\documentclass[11pt]{article}
\usepackage[utf8]{inputenc}
\usepackage[a4paper, total={6in, 8in}]{geometry}
\usepackage{mathrsfs}
\usepackage{amssymb}
\usepackage{braket}
\usepackage{amsmath}
\usepackage{verbatim}
\usepackage{appendix}
\usepackage{bbold}
\usepackage{graphicx}
\usepackage{dsfont}
\usepackage{bbm}
\usepackage{multicol}
\usepackage[T1]{fontenc}
\usepackage{lmodern}
\usepackage{morefloats}
\usepackage[svgnames]{xcolor}
\usepackage{tikz}
\usepackage{pbox}
\usepackage{ytableau}
\usetikzlibrary{calc,decorations.markings, shapes.misc, decorations.pathmorphing}
\tikzset{cross/.style={cross out, draw, 
		minimum size=2*(#1-\pgflinewidth), 
		inner sep=0pt, outer sep=0pt}} 
\definecolor{lstbgcolor}{rgb}{0.9,0.9,0.9} 
\usepackage{listings} 
\usepackage[vcentermath]{youngtab}
\usepackage{mathtools}

\usepackage[breakable]{tcolorbox}
\usepackage{caption}
 \usepackage{fancyvrb} 
\input{code_preamble.tex}

\newcommand{\bea}{\begin{eqnarray}}
\newcommand{\eea}{\end{eqnarray}}
\newcommand{\bg}{\begin{gathered}}
\newcommand{\eg}{\end{gathered}}

\numberwithin{equation}{section}
\allowdisplaybreaks
\newcommand{\less}{<}

\newcommand{\idn}{\mathbbm{1}}

\DeclareMathOperator{\End}{End}

\newcommand{\DimSD}[1]{\Dim V^{S_D}_{#1}}

\def\Dim{ {\rm Dim } }
 
\def\ls[#1]{ {}_{#1}}
\def\cYS{ \mathcal{Y_S}} 
\def\mC{ \mathbb{C} }

\usepackage{subcaption}

\usepackage{amsthm}
\usepackage{rotating}
\usepackage{physics}
\makeatletter
\newcommand{\bigtimes}{%
	\DOTSB\mathop{\mathpalette\mattos@bigtimes\relax}\slimits@
}
\newcommand\mattos@bigtimes[2]{%
	\vcenter{\hbox{%
			\sbox\z@{$#1\sum$}%
			\resizebox{!}{0.9\dimexpr\ht\z@+\dp\z@}{\raisebox{\depth}{$\m@th#1\times$}}%
	}}%
	\vphantom{\sum}%
}

\newcommand{\mytikz}[1]{
	\pbox{\textwidth}{\begin{tikzpicture}[>=stealth,decoration={
    markings,
    mark=at position 0.5 with {\arrow{>}}}]	
		#1
	\end{tikzpicture}}
}

\newcommand{\PAdiagrambig}[3][]{\;\begin{tikzpicture}[baseline={([yshift=-.5ex]current bounding box.center)}]
	\def \n {#2};
	\def \edges {#3};
	\def \arcs {#1}
	\def \sep {0.5};
	\foreach \v in {1,...,\n}
	{
		\pgfmathparse{(\v-1)*\sep};
		\coordinate (v\v) at (\pgfmathresult,0.25);
		\node[circle,fill,inner sep=1pt] at (v\v) {};
	}
	\foreach \v in {1,...,\n}
	{
		\pgfmathparse{(\v-1)*\sep};
		\coordinate (v-\v) at (\pgfmathresult,-0.25);
		\node[circle,fill,inner sep=1pt] at (v-\v) {};
	}
	\foreach \endpointOne/\endpointTwo in \edges
	{
		\draw[] (v\endpointOne) -- (v\endpointTwo);
	}
	\foreach \endpointOne/\endpointTwo in \arcs
	{
		\draw[] (v\endpointOne) to[bend left] (v\endpointTwo);
	}
	\end{tikzpicture}\;}

\usepackage{hyperref}
\hypersetup{
  colorlinks = true,
  linkcolor  = black
}

\makeatother

\newcommand{\Span}{\mathrm{Span}}

\definecolor{GREEN}{rgb}{0.0,0.70,0.24}
\definecolor{BLUE}{rgb}{0.0,0.24,0.70}

\usepackage[numbers]{natbib}
\begin{document}

\begin{flushright}
QMUL-PH-23-35
\end{flushright}

\bigskip

\begin{center}

{\Large \bf Permutation invariant tensor models and partition algebras}

\bigskip

George Barnes$^{a, *}$, Adrian Padellaro$^{b, \ddagger}$,
 Sanjaye Ramgoolam$^{a , c ,\dag}  $

\bigskip
$^{a}${\em School of Physics and Astronomy} , {\em  Centre for Research in String Theory}\\
{\em Queen Mary University of London, London E1 4NS, United Kingdom }\\
\medskip
$^{b}${\em Faculty of Physics} , {\em  Bielefeld University}\\
{\em PO-Box 100131, D-33501 Bielefeld, German}\\
\medskip
$^{c}${\em  School of Physics and Mandelstam Institute for Theoretical Physics,} \\   
{\em University of Witwatersrand, Wits, 2050, South Africa} \\
\medskip
E-mails:  $^{*}$g.barnes@qmul.ac.uk, \\ $^{\ddagger}$apadellaro@physik.uni-bielefeld.de
\quad $^{\dag}$s.ramgoolam@qmul.ac.uk

\begin{abstract} \noindent
Matrix models with continuous symmetry are powerful tools for studying quantum gravity and holography. 
Tensor models have also found applications in holographic quantum gravity.  
Matrix models with discrete permutation symmetry  have been shown to satisfy large $N$ factorisation  properties relevant to holography, while also having applications to the statistical analysis of ensembles of real-world matrices. 
Here we develop 3-index tensor models in dimension $D$ with a discrete symmetry of permutations in the symmetric group $S_D$. We construct the most general permutation invariant Gaussian tensor model using  the representation theory of symmetric groups and associated partition algebras. We define a representation basis  for the 3-index tensors,  where the two-point function is diagonalised. Inverting the change of basis gives an explicit formula for the two-point function 
in the tensor basis for general $D$. 
\end{abstract}

\end{center}


\newpage 

\tableofcontents

\newpage
\section{Introduction}
Since its introduction by Wigner and Dyson \cite{Wigner1955, Dyson1962} random matrix theory, based on matrix models having continuous symmetries,  has been fruitfully applied to diverse areas of science including nuclear physics, chaos, condensed matter physics, biological networks, feature-matrices in bio-statistics, data science, financial correlations and quantum gravity \cite{Mehta2004, Guhr1998, Beenakker1997, Edelman2013, Klebanov1991, GinsMoore}. Permutation invariant matrix models have been developed and applied to real-world matrix ensembles \cite{LMT, PIGMM,RSS}. Permutation invariant random  matrix theory follows a similar approach to traditional random matrix theory  but enriches and extends traditional areas of application as the matrix model observables include more general observables than those invariant under continuous symmetries. This has been demonstrated in the case of
permutation invariant Gaussian matrix (PIGM) models already in a variety of computational linguistic applications: the Gaussianity analysed in \cite{RSS} was based on the construction of matrices by linear regression in \cite{LMT} while \cite{HCRS} extended the analysis of \cite{RSS} and also analysed the matrices constructed by neural network methods in \cite{wijnholds2020}. It is natural to consider the generalisation of these matrix models to models containing higher index, tensor, objects. 

In quantum gravity, tensor models with continuous group symmetry generalise the connection between matrix models and random two-dimensional geometries to higher dimensions. The Feynman diagram expansion of tensor models with $d$ indices can be organised in terms of $d$-dimensional geometries \cite{Ambjorn:1990ge,Sasakura:1990fs,Gross:1991hx,Oriti:2006se,Gurau:2010ba,Gurau:2011aq}. More recently, quantum mechanical tensor models have been studied as toy models of black hole holography \cite{Witten2016}. This connection exploits similarities between tensor models and SYK models, which are believed to be dual to black holes in 1+1 dimensions \cite{Kitaev_2015,Sachdev_2015}. In this paper, we present a framework for studying tensor models with discrete permutation symmetry.

The Linguistic Matrix Theory programme \cite{LMT, PIGMM} proposed  random matrix theory  with discrete permutation symmetries as a way to model the statistics of ensembles of matrices arising within the context of compositional distributional semantics \cite{Coecke2010, Baroni2014}. Distributional models of meaning in natural language are based on the simple idea that the meaning of a word can be deduced from its co-occurrence with other context words \cite{Firth, Harris}. The use of vectors to represent words is a well-established technique in computational linguistics. Compositional distributional semantics  assigns words  to vectors, matrices or higher rank tensors, depending on their grammatical role and the grammatical  composition of words corresponds to the contraction of indices.

In \cite{LMT, PIGMM} a permutation invariant statistical model of $D \times D$ matrices $M$ was defined through a quadratic polynomial function $\mathcal{S}(M_{ij})$ satisfying
\begin{equation}
	\mathcal{S}(M_{\sigma(i)\sigma(j)}) = \mathcal{S}(M_{ij}) \qquad \forall \, \sigma \in S_D \, .
\end{equation}
In physics parlance, such a polynomial function is called an action.
The most general quadratic permutation invariant action has $13$ free parameters. It defines a statistical model with partition function
\begin{equation}
	Z = \int \dd{M} \exp{-\mathcal{S}(M)} \, .
\end{equation}
Conventionally, solving a model of this type involves giving an algorithm for computing expectation values
\begin{equation}
	\expval{f(M_{ij})} = \frac{1}{Z}\int \dd{M}  \exp{-\mathcal{S}(M)} f(M)
\end{equation}
of polynomial functions $f(M_{ij})$. For models defined by a quadratic action general expectation values can be computed from one-point and two-point functions
\begin{equation}
	\expval{M_{ij}},  \quad \expval{M_{ij} M_{kl}}. \label{eq: matrix model 1pt 2pt}
\end{equation}
This uses Wick's theorem, which expresses higher-order expectation values/moments as sums of products of one and two-point functions. Elegant expressions for \eqref{eq: matrix model 1pt 2pt} were given in \cite{PIGMM} in terms of $S_D$ representation theoretic quantities. Observables in this model were defined to be general polynomial functions $f(M)$ satisfying
\begin{equation}
	f(M_{\sigma(i)\sigma(j)}) = f(M_{ij}) \qquad \forall \, \sigma \in S_D \, .
\end{equation}
Closed form expressions, as functions of $D$, for a selection of expectation values of observables were given. 

In \cite{PIG2MM} the theoretical results in \cite{PIGMM} were extended to 2-matrix models, and a combinatorial algorithm was developed for computing expectation values of general observables. In \cite{PIMO_Factor}, permutation invariant matrix observables were described in terms of 
partition algebras which are actively studied in representation theory (e.g. \cite{PartitionAlgebras}) and this was exploited to derive a large N factorisation property of expectation values in permutation invariant matrix models  analogous to similar large N factorisation for matrix models with continuous symmetry.  The factorisation property in the continuous symmetry case which has known  applications in  the AdS/CFT correspondence \cite{Aharony:1999ti}, see e.g \cite{BBNS,CJR,GRW}.   Partition algebras were discovered by Jones \cite{Jones} and Martin \cite{Martin} in the context of classical statistical mechanics. Recently, permutation invariant matrix models and partition algebras were connected to quantum many-body physics and holography \cite{PIMQM}, using mechanisms of hidden symmetry for large $N$ matrix/tensor systems based on Schur-Weyl duality \cite{SWrev2008,CombfiniteN}. Permutation invariant matrix models were developed for the case of matrices obeying constraints $M^T = M $ and having vanishing (or constant)  diagonal entries. This development was motivated by applications to correlation matrices in statistical finance \cite{LMTFinance}.

Building on these developments in matrix systems of general size $N$ with  discrete permutation symmetries, we initiate the study of tensor systems for tensors of general size $N$ with discrete permutation symmetry. We will develop  a statistical model for a three-index tensor $\Phi_{ijk}$ with $i,j,k =1,\dots,D$ (we are using $D$ instead of $N$ here).  In compositional distributional semantics setting three-index tensors are needed to encode the grammatical role of transitive verbs, while along the lines of \cite{PIMQM} we expect this study will have implications for solvable quantum mechanical tensor systems. 

In section \ref{sec: the model} we describe the general structure of the permutation invariant Gaussian tensor model and describe the permutation invariant observables formed from $\Phi_{ijk}$. Representation theory of the symmetric group is used to find expressions for the one-point and two-point functions
\begin{equation}
	\expval{\Phi_{ijk}}, \quad \expval{\Phi_{ijk} \Phi_{pqr}},
\end{equation}
in terms of $S_D$ invariant endomorphism tensors which are defined in equation \eqref{eq: Q clebschs} in terms of symmetric group Clebsch-Gordan coefficients.

Calculating the invariant endomorphism tensors by explicitly constructing the Clebsch-Gordan coefficients for general $D$ is highly complex. Section \ref{sec: the Qs} is devoted to explicitly constructing the invariant endomorphism tensors without having to calculate Clebsch-Gordan coefficients. These invariant endomorphism tensors are elements of a partition algebra, and admit a labelling in terms of graphs with $S_D$ representation theory data. The construction involves diagonalizing matrices made out of partition algebra structure constants and the size of the matrices is independent of $D$. This can be done with the help of open-source software SageMath for partition algebra calculations \cite{sagemath} (see appendix \ref{app: code}). The construction works efficiently for general large $D$ of interest (there is a minor technical restriction to $D \geq 6$). For concreteness we end the section with some explicit examples of invariant endomorphism tensors.

In section \ref{sec: counting} we give a general formula for the counting of permutation invariant tensor observables using characters of permutations in the natural representation $V_D$. We then show that there is a basis of invariant tensor observables which is in one-to-one correspondence with a family of bi-partite 3-colored graphs. We prove that the counting of these graphs agrees with the representation theory counting. We give a summary in section \ref{sec: summary and outlook} along with a discussion of future directions.

\section{Permutation invariant Gaussian tensor model} \label{sec: the model}
In this section we will define and construct the permutation invariant Gaussian tensor model. We solve the model using a representation theoretic change of basis which diagonalises the two point function. We also give formulas for the one point and two point function in the tensor basis.

The symmetric group $S_D$ can be defined  to be the set of bijective maps 
\begin{equation}
	\sigma: \{1,\dots,D\} \rightarrow \{1,\dots,D\}
\end{equation}
 with multiplication given by composition of maps. The degrees of freedom in the problem are packaged into a tensor $\Phi_{ijk}$ with $i,j,k=1,\dots,D$. The symmetric groups acts diagonally on $\Phi_{ijk}$
\begin{align} \label{eq: sigma on V_D^3}
	\Phi_{ijk} \rightarrow \Phi_{\sigma(i) \sigma(j) \sigma(k)} \, , \qquad \forall \, \sigma \in S_D \, , \qquad i, j, k \in \{ 1, 2, \dots , D \} \, .
\end{align}
A permutation invariant Gaussian tensor model is defined by a Gaussian/quadratic action $\mathcal{S}(\Phi_{ijk})$ satisfying
\begin{equation}
	\mathcal{S}(\Phi_{ijk}) = \mathcal{S}(\Phi_{\sigma(i) \sigma(j) \sigma(k)}), \qquad \forall \, \sigma \in S_D \, . \label{eq: SD invariant action}
\end{equation}
The corresponding partition function is
\begin{equation}
	Z = \int \dd{\Phi} \exp(- \mathcal{S}(\Phi)) .
\end{equation}
We take observables of the model to be $S_D$-invariant polynomial functions $f(\Phi_{ijk})$ of general degree
\begin{equation} \label{eqn: observables definition}
	f(\Phi_{ijk}) = f(\Phi_{\sigma(i) \sigma(j) \sigma(k)}), \quad \forall \sigma \in S_D.
\end{equation}
Expectation values of these observables are then given by
\begin{equation} \label{eqn: expectation values}
	\langle f(\Phi) \rangle \equiv \frac{1}{Z}\int d \Phi f(\Phi) e^{- \mathcal{S}(\Phi)}.
\end{equation}
They can be computed using Wick's theorem because the action is Gaussian. To use Wick's theorem we need expressions for
\begin{equation}
	\expval{\Phi_{ijk}}, \quad \expval{\Phi_{ijk} \Phi_{pqr}}. \label{eq: 1pt 2pt tensor}
\end{equation}
For a Gaussian action, computing \eqref{eq: 1pt 2pt tensor} requires the inversion of a $D^3 \times D^3$ matrix of quadratic couplings.

Given this definition, the first hurdle in the construction and solution of the permutation invariant Gaussian tensor model is to give a parametrisation of $\mathcal{S}(\Phi_{ijk})$. Note that the condition \eqref{eq: SD invariant action} implies that $\mathcal{S}(\Phi_{ijk})$ is a linear combination of invariant polynomials of degree one and two. 
In the language of \cite{LMT, PIGMM}, there exists a graph basis of invariant polynomials. A basis of invariant polynomials of degree $m$ is labelled by set partitions of $3m$ objects. For example there are a total of five degree one invariant combinations 
\begin{align} \label{eq: linear graph basis invariants}
\sum_{i = 1}^D \Phi_{iii}, \quad \sum_{i,j = 1}^D \Phi_{iij}, \quad \sum_{i,j = 1}^D \Phi_{iji}, \quad \sum_{i,j = 1}^D \Phi_{jii}, \quad \sum_{i,j, k = 1}^D \Phi_{ijk}. 
\end{align}
They correspond to bipartite graphs (that is graphs with black and white vertices, edges joining black vertices to white), with edges having three colours which we will draw as red, blue and green. The tensors $\Phi$ correspond to white vertices, each index is a coloured edge and black vertices correspond to the identification of indices. This description is used in section \ref{sec: counting} to count observables of general degree, for general $D$. For example, three of the linear invariants correspond to the following graphs
\begin{align}
 &\sum_{i=1}^D \Phi_{iii} \Longleftrightarrow \vcenter{\hbox{
\tikzset{every picture/.style={line width=0.75pt}} 
\begin{tikzpicture}[x=0.75pt,y=0.75pt,yscale=-1,xscale=1]
\draw [color={rgb, 255:red, 65; green, 117; blue, 5 }  ,draw opacity=1 ]   (380,75) -- (395,95) ;
\draw [color={rgb, 255:red, 208; green, 2; blue, 27 }  ,draw opacity=1 ]   (380,75) -- (365,100) ;
\draw [color={rgb, 255:red, 42; green, 116; blue, 197 }  ,draw opacity=1 ]   (380,75) -- (380,100) ;
\draw  [fill={rgb, 255:red, 255; green, 255; blue, 255 }  ,fill opacity=1 ] (385,75) .. controls (385,72.24) and (382.76,70) .. (380,70) .. controls (377.24,70) and (375,72.24) .. (375,75) .. controls (375,77.76) and (377.24,80) .. (380,80) .. controls (382.76,80) and (385,77.76) .. (385,75) -- cycle ;
\draw [color={rgb, 255:red, 208; green, 2; blue, 27 }  ,draw opacity=1 ]   (365,100) -- (365,130) ;
\draw [color={rgb, 255:red, 208; green, 2; blue, 27 }  ,draw opacity=1 ]   (380,160) -- (365,130) ;
\draw [color={rgb, 255:red, 42; green, 116; blue, 197 }  ,draw opacity=1 ]   (380,100) -- (380,130) ;
\draw [color={rgb, 255:red, 42; green, 116; blue, 197 }  ,draw opacity=1 ]   (380,130) -- (380,160) ;
\draw  [fill={rgb, 255:red, 0; green, 0; blue, 0 }  ,fill opacity=1 ] (385,160) .. controls (385,157.24) and (382.76,155) .. (380,155) .. controls (377.24,155) and (375,157.24) .. (375,160) .. controls (375,162.76) and (377.24,165) .. (380,165) .. controls (382.76,165) and (385,162.76) .. (385,160) -- cycle ;
\draw [color={rgb, 255:red, 65; green, 117; blue, 5 }  ,draw opacity=1 ]   (395,95) -- (395,135) ;
\draw [color={rgb, 255:red, 65; green, 117; blue, 5 }  ,draw opacity=1 ]   (395,135) -- (380,160) ;
\end{tikzpicture}}}\quad,\quad \sum_{i,j = 1}^D \Phi_{iij} \Longleftrightarrow \vcenter{\hbox{
\begin{tikzpicture}[x=0.75pt,y=0.75pt,yscale=-1,xscale=1]
\draw [color={rgb, 255:red, 65; green, 117; blue, 5 }  ,draw opacity=1 ]   (435,75) -- (450,95) ;
\draw [color={rgb, 255:red, 208; green, 2; blue, 27 }  ,draw opacity=1 ]   (435,75) -- (420,100) ;
\draw [color={rgb, 255:red, 42; green, 116; blue, 197 }  ,draw opacity=1 ]   (435,75) -- (435,100) ;
\draw  [fill={rgb, 255:red, 255; green, 255; blue, 255 }  ,fill opacity=1 ] (440,75) .. controls (440,72.24) and (437.76,70) .. (435,70) .. controls (432.24,70) and (430,72.24) .. (430,75) .. controls (430,77.76) and (432.24,80) .. (435,80) .. controls (437.76,80) and (440,77.76) .. (440,75) -- cycle ;
\draw [color={rgb, 255:red, 208; green, 2; blue, 27 }  ,draw opacity=1 ]   (420,100) -- (420,130) ;
\draw [color={rgb, 255:red, 208; green, 2; blue, 27 }  ,draw opacity=1 ]   (435,160) -- (420,130) ;
\draw [color={rgb, 255:red, 42; green, 116; blue, 197 }  ,draw opacity=1 ]   (435,100) -- (435,130) ;
\draw [color={rgb, 255:red, 42; green, 116; blue, 197 }  ,draw opacity=1 ]   (435,130) -- (435,160) ;
\draw  [fill={rgb, 255:red, 0; green, 0; blue, 0 }  ,fill opacity=1 ] (440,160) .. controls (440,157.24) and (437.76,155) .. (435,155) .. controls (432.24,155) and (430,157.24) .. (430,160) .. controls (430,162.76) and (432.24,165) .. (435,165) .. controls (437.76,165) and (440,162.76) .. (440,160) -- cycle ;
\draw [color={rgb, 255:red, 65; green, 117; blue, 5 }  ,draw opacity=1 ]   (450,95) -- (465,160) ;
\draw  [fill={rgb, 255:red, 0; green, 0; blue, 0 }  ,fill opacity=1 ] (470,160) .. controls (470,157.24) and (467.76,155) .. (465,155) .. controls (462.24,155) and (460,157.24) .. (460,160) .. controls (460,162.76) and (462.24,165) .. (465,165) .. controls (467.76,165) and (470,162.76) .. (470,160) -- cycle ;
\end{tikzpicture}}},\quad \quad \sum_{i,j, k = 1}^D \Phi_{ijk} \Longleftrightarrow \vcenter{\hbox{
\begin{tikzpicture}[x=0.75pt,y=0.75pt,yscale=-1,xscale=1]
\draw [color={rgb, 255:red, 65; green, 117; blue, 5 }  ,draw opacity=1 ]   (525,75) -- (540,95) ;
\draw [color={rgb, 255:red, 208; green, 2; blue, 27 }  ,draw opacity=1 ]   (525,75) -- (510,100) ;
\draw [color={rgb, 255:red, 42; green, 116; blue, 197 }  ,draw opacity=1 ]   (525,75) -- (525,100) ;
\draw  [fill={rgb, 255:red, 255; green, 255; blue, 255 }  ,fill opacity=1 ] (530,75) .. controls (530,72.24) and (527.76,70) .. (525,70) .. controls (522.24,70) and (520,72.24) .. (520,75) .. controls (520,77.76) and (522.24,80) .. (525,80) .. controls (527.76,80) and (530,77.76) .. (530,75) -- cycle ;
\draw [color={rgb, 255:red, 208; green, 2; blue, 27 }  ,draw opacity=1 ]   (510,100) -- (495,160) ;
\draw [color={rgb, 255:red, 42; green, 116; blue, 197 }  ,draw opacity=1 ]   (525,100) -- (525,130) ;
\draw [color={rgb, 255:red, 42; green, 116; blue, 197 }  ,draw opacity=1 ]   (525,130) -- (525,160) ;
\draw  [fill={rgb, 255:red, 0; green, 0; blue, 0 }  ,fill opacity=1 ] (530,160) .. controls (530,157.24) and (527.76,155) .. (525,155) .. controls (522.24,155) and (520,157.24) .. (520,160) .. controls (520,162.76) and (522.24,165) .. (525,165) .. controls (527.76,165) and (530,162.76) .. (530,160) -- cycle ;
\draw [color={rgb, 255:red, 65; green, 117; blue, 5 }  ,draw opacity=1 ]   (540,95) -- (555,160) ;
\draw  [fill={rgb, 255:red, 0; green, 0; blue, 0 }  ,fill opacity=1 ] (560,160) .. controls (560,157.24) and (557.76,155) .. (555,155) .. controls (552.24,155) and (550,157.24) .. (550,160) .. controls (550,162.76) and (552.24,165) .. (555,165) .. controls (557.76,165) and (560,162.76) .. (560,160) -- cycle ;
\draw  [fill={rgb, 255:red, 0; green, 0; blue, 0 }  ,fill opacity=1 ] (500,160) .. controls (500,157.24) and (497.76,155) .. (495,155) .. controls (492.24,155) and (490,157.24) .. (490,160) .. controls (490,162.76) and (492.24,165) .. (495,165) .. controls (497.76,165) and (500,162.76) .. (500,160) -- cycle ;
\end{tikzpicture}}} \, .
\end{align}
The invariants in \eqref{eq: linear graph basis invariants} also correspond to the set partitions
\begin{equation}
  \{\{1,2,3\}\}, \quad \{\{1,2\}, \{3\}\}, \quad \{\{1,3\}, \{2\}\}, \quad \{\{2,3\}, \{1\}\}, \quad \{\{1\},\{2\}, \{3\}\}.
\end{equation}

A subset of degree two invariants is
\begin{align} \label{eq: quadratic graph basis invariant examples}
\sum_{i = 1}^D \Phi_{iii} \Phi_{iii}, \quad \sum_{i, j = 1}^D \Phi_{iii} \Phi_{iij}, \quad \sum_{i, j = 1}^D \Phi_{iii} \Phi_{jjj}, \quad \sum_{i,j,k = 1}^D \Phi_{ijk} \Phi_{jkk}, \quad \sum_{i,j,k,p,q,r = 1}^D \Phi_{ijk} \Phi_{pqr} \, .
\end{align}
Some examples of the corresponding graphs are
\begin{align}
	\sum_{i = 1}^D \Phi_{iii} \Phi_{iii} &\Longleftrightarrow \vcenter{\hbox{
\begin{tikzpicture}[x=0.75pt,y=0.75pt,yscale=-1,xscale=1]
\draw [color={rgb, 255:red, 65; green, 117; blue, 5 }  ,draw opacity=1 ]   (115,195) -- (40,225) ;
\draw [color={rgb, 255:red, 65; green, 117; blue, 5 }  ,draw opacity=1 ]   (40,140) -- (50,160) ;
\draw [color={rgb, 255:red, 208; green, 2; blue, 27 }  ,draw opacity=1 ]   (40,140) -- (25,165) ;
\draw [color={rgb, 255:red, 42; green, 116; blue, 197 }  ,draw opacity=1 ]   (40,140) -- (40,165) ;
\draw  [fill={rgb, 255:red, 255; green, 255; blue, 255 }  ,fill opacity=1 ] (45,140) .. controls (45,137.24) and (42.76,135) .. (40,135) .. controls (37.24,135) and (35,137.24) .. (35,140) .. controls (35,142.76) and (37.24,145) .. (40,145) .. controls (42.76,145) and (45,142.76) .. (45,140) -- cycle ;
\draw [color={rgb, 255:red, 65; green, 117; blue, 5 }  ,draw opacity=1 ]   (100,140) -- (115,165) ;
\draw [color={rgb, 255:red, 208; green, 2; blue, 27 }  ,draw opacity=1 ]   (100,140) -- (75,170) ;
\draw [color={rgb, 255:red, 42; green, 116; blue, 197 }  ,draw opacity=1 ]   (100,140) -- (100,190) ;
\draw  [fill={rgb, 255:red, 255; green, 255; blue, 255 }  ,fill opacity=1 ] (105,140) .. controls (105,137.24) and (102.76,135) .. (100,135) .. controls (97.24,135) and (95,137.24) .. (95,140) .. controls (95,142.76) and (97.24,145) .. (100,145) .. controls (102.76,145) and (105,142.76) .. (105,140) -- cycle ;
\draw [color={rgb, 255:red, 208; green, 2; blue, 27 }  ,draw opacity=1 ]   (25,165) -- (25,195) ;
\draw [color={rgb, 255:red, 208; green, 2; blue, 27 }  ,draw opacity=1 ]   (40,225) -- (25,195) ;
\draw [color={rgb, 255:red, 42; green, 116; blue, 197 }  ,draw opacity=1 ]   (40,165) -- (40,195) ;
\draw [color={rgb, 255:red, 42; green, 116; blue, 197 }  ,draw opacity=1 ]   (40,195) -- (40,225) ;
\draw [color={rgb, 255:red, 208; green, 2; blue, 27 }  ,draw opacity=1 ]   (75,170) -- (40,225) ;
\draw [color={rgb, 255:red, 65; green, 117; blue, 5 }  ,draw opacity=1 ]   (115,165) -- (115,195) ;
\draw [color={rgb, 255:red, 65; green, 117; blue, 5 }  ,draw opacity=1 ]   (50,160) -- (50,205) ;
\draw [color={rgb, 255:red, 65; green, 117; blue, 5 }  ,draw opacity=1 ]   (50,205) -- (40,230) ;
\draw [color={rgb, 255:red, 42; green, 116; blue, 197 }  ,draw opacity=1 ]   (100,190) -- (40,225) ;
\draw  [fill={rgb, 255:red, 0; green, 0; blue, 0 }  ,fill opacity=1 ] (45,225) .. controls (45,222.24) and (42.76,220) .. (40,220) .. controls (37.24,220) and (35,222.24) .. (35,225) .. controls (35,227.76) and (37.24,230) .. (40,230) .. controls (42.76,230) and (45,227.76) .. (45,225) -- cycle ;
\end{tikzpicture}}} \, , \\
	\sum_{i, j = 1}^D \Phi_{iii} \Phi_{iij} &\Longleftrightarrow \vcenter{\hbox{
\begin{tikzpicture}[x=0.75pt,y=0.75pt,yscale=-1,xscale=1]
\draw [color={rgb, 255:red, 65; green, 117; blue, 5 }  ,draw opacity=1 ]   (105,310) -- (105,340) ;
\draw [color={rgb, 255:red, 65; green, 117; blue, 5 }  ,draw opacity=1 ]   (30,255) -- (40,275) ;
\draw [color={rgb, 255:red, 208; green, 2; blue, 27 }  ,draw opacity=1 ]   (30,255) -- (15,280) ;
\draw [color={rgb, 255:red, 42; green, 116; blue, 197 }  ,draw opacity=1 ]   (30,255) -- (30,280) ;
\draw  [fill={rgb, 255:red, 255; green, 255; blue, 255 }  ,fill opacity=1 ] (35,255) .. controls (35,252.24) and (32.76,250) .. (30,250) .. controls (27.24,250) and (25,252.24) .. (25,255) .. controls (25,257.76) and (27.24,260) .. (30,260) .. controls (32.76,260) and (35,257.76) .. (35,255) -- cycle ;
\draw [color={rgb, 255:red, 65; green, 117; blue, 5 }  ,draw opacity=1 ]   (90,255) -- (105,280) ;
\draw [color={rgb, 255:red, 208; green, 2; blue, 27 }  ,draw opacity=1 ]   (90,255) -- (65,285) ;
\draw [color={rgb, 255:red, 42; green, 116; blue, 197 }  ,draw opacity=1 ]   (90,255) -- (90,305) ;
\draw  [fill={rgb, 255:red, 255; green, 255; blue, 255 }  ,fill opacity=1 ] (95,255) .. controls (95,252.24) and (92.76,250) .. (90,250) .. controls (87.24,250) and (85,252.24) .. (85,255) .. controls (85,257.76) and (87.24,260) .. (90,260) .. controls (92.76,260) and (95,257.76) .. (95,255) -- cycle ;
\draw [color={rgb, 255:red, 208; green, 2; blue, 27 }  ,draw opacity=1 ]   (15,280) -- (15,310) ;
\draw [color={rgb, 255:red, 208; green, 2; blue, 27 }  ,draw opacity=1 ]   (30,340) -- (15,310) ;
\draw [color={rgb, 255:red, 42; green, 116; blue, 197 }  ,draw opacity=1 ]   (30,280) -- (30,310) ;
\draw [color={rgb, 255:red, 42; green, 116; blue, 197 }  ,draw opacity=1 ]   (30,310) -- (30,340) ;
\draw [color={rgb, 255:red, 208; green, 2; blue, 27 }  ,draw opacity=1 ]   (65,285) -- (30,340) ;
\draw [color={rgb, 255:red, 65; green, 117; blue, 5 }  ,draw opacity=1 ]   (105,280) -- (105,310) ;
\draw [color={rgb, 255:red, 65; green, 117; blue, 5 }  ,draw opacity=1 ]   (40,275) -- (40,320) ;
\draw [color={rgb, 255:red, 65; green, 117; blue, 5 }  ,draw opacity=1 ]   (40,320) -- (30,340) ;
\draw [color={rgb, 255:red, 42; green, 116; blue, 197 }  ,draw opacity=1 ]   (90,305) -- (30,340) ;
\draw  [fill={rgb, 255:red, 0; green, 0; blue, 0 }  ,fill opacity=1 ] (110,340) .. controls (110,337.24) and (107.76,335) .. (105,335) .. controls (102.24,335) and (100,337.24) .. (100,340) .. controls (100,342.76) and (102.24,345) .. (105,345) .. controls (107.76,345) and (110,342.76) .. (110,340) -- cycle ;
\draw  [fill={rgb, 255:red, 0; green, 0; blue, 0 }  ,fill opacity=1 ] (35,340) .. controls (35,337.24) and (32.76,335) .. (30,335) .. controls (27.24,335) and (25,337.24) .. (25,340) .. controls (25,342.76) and (27.24,345) .. (30,345) .. controls (32.76,345) and (35,342.76) .. (35,340) -- cycle ;
\end{tikzpicture}}} \, , \\
\sum_{i,j,k,p,q,r = 1}^D \Phi_{ijk} \Phi_{pqr}&\Longleftrightarrow \vcenter{\hbox{
\begin{tikzpicture}[x=0.75pt,y=0.75pt,yscale=-1,xscale=1]
\draw [color={rgb, 255:red, 65; green, 117; blue, 5 }  ,draw opacity=1 ]   (45,20) -- (60,45) ;
\draw [color={rgb, 255:red, 65; green, 117; blue, 5 }  ,draw opacity=1 ]   (120,75) -- (120,105) ;
\draw [color={rgb, 255:red, 208; green, 2; blue, 27 }  ,draw opacity=1 ]   (45,20) -- (30,45) ;
\draw [color={rgb, 255:red, 42; green, 116; blue, 197 }  ,draw opacity=1 ]   (45,20) -- (45,45) ;
\draw  [fill={rgb, 255:red, 255; green, 255; blue, 255 }  ,fill opacity=1 ] (50,20) .. controls (50,17.24) and (47.76,15) .. (45,15) .. controls (42.24,15) and (40,17.24) .. (40,20) .. controls (40,22.76) and (42.24,25) .. (45,25) .. controls (47.76,25) and (50,22.76) .. (50,20) -- cycle ;
\draw [color={rgb, 255:red, 65; green, 117; blue, 5 }  ,draw opacity=1 ]   (105,20) -- (120,45) ;
\draw [color={rgb, 255:red, 208; green, 2; blue, 27 }  ,draw opacity=1 ]   (105,20) -- (90,45) ;
\draw [color={rgb, 255:red, 42; green, 116; blue, 197 }  ,draw opacity=1 ]   (105,20) -- (105,105) ;
\draw  [fill={rgb, 255:red, 255; green, 255; blue, 255 }  ,fill opacity=1 ] (110,20) .. controls (110,17.24) and (107.76,15) .. (105,15) .. controls (102.24,15) and (100,17.24) .. (100,20) .. controls (100,22.76) and (102.24,25) .. (105,25) .. controls (107.76,25) and (110,22.76) .. (110,20) -- cycle ;
\draw [color={rgb, 255:red, 208; green, 2; blue, 27 }  ,draw opacity=1 ]   (30,45) -- (30,75) ;
\draw [color={rgb, 255:red, 208; green, 2; blue, 27 }  ,draw opacity=1 ]   (90,45) -- (90,105) ;
\draw [color={rgb, 255:red, 208; green, 2; blue, 27 }  ,draw opacity=1 ]   (30,105) -- (30,75) ;
\draw [color={rgb, 255:red, 42; green, 116; blue, 197 }  ,draw opacity=1 ]   (45,45) -- (45,75) ;
\draw [color={rgb, 255:red, 42; green, 116; blue, 197 }  ,draw opacity=1 ]   (45,75) -- (45,105) ;
\draw  [fill={rgb, 255:red, 0; green, 0; blue, 0 }  ,fill opacity=1 ] (35,105) .. controls (35,102.24) and (32.76,100) .. (30,100) .. controls (27.24,100) and (25,102.24) .. (25,105) .. controls (25,107.76) and (27.24,110) .. (30,110) .. controls (32.76,110) and (35,107.76) .. (35,105) -- cycle ;
\draw  [fill={rgb, 255:red, 0; green, 0; blue, 0 }  ,fill opacity=1 ] (125,105) .. controls (125,102.24) and (122.76,100) .. (120,100) .. controls (117.24,100) and (115,102.24) .. (115,105) .. controls (115,107.76) and (117.24,110) .. (120,110) .. controls (122.76,110) and (125,107.76) .. (125,105) -- cycle ;
\draw [color={rgb, 255:red, 65; green, 117; blue, 5 }  ,draw opacity=1 ]   (120,45) -- (120,75) ;
\draw  [fill={rgb, 255:red, 0; green, 0; blue, 0 }  ,fill opacity=1 ] (50,105) .. controls (50,102.24) and (47.76,100) .. (45,100) .. controls (42.24,100) and (40,102.24) .. (40,105) .. controls (40,107.76) and (42.24,110) .. (45,110) .. controls (47.76,110) and (50,107.76) .. (50,105) -- cycle ;
\draw  [fill={rgb, 255:red, 0; green, 0; blue, 0 }  ,fill opacity=1 ] (65,105) .. controls (65,102.24) and (62.76,100) .. (60,100) .. controls (57.24,100) and (55,102.24) .. (55,105) .. controls (55,107.76) and (57.24,110) .. (60,110) .. controls (62.76,110) and (65,107.76) .. (65,105) -- cycle ;
\draw  [fill={rgb, 255:red, 0; green, 0; blue, 0 }  ,fill opacity=1 ] (95,105) .. controls (95,102.24) and (92.76,100) .. (90,100) .. controls (87.24,100) and (85,102.24) .. (85,105) .. controls (85,107.76) and (87.24,110) .. (90,110) .. controls (92.76,110) and (95,107.76) .. (95,105) -- cycle ;
\draw  [fill={rgb, 255:red, 0; green, 0; blue, 0 }  ,fill opacity=1 ] (110,105) .. controls (110,102.24) and (107.76,100) .. (105,100) .. controls (102.24,100) and (100,102.24) .. (100,105) .. controls (100,107.76) and (102.24,110) .. (105,110) .. controls (107.76,110) and (110,107.76) .. (110,105) -- cycle ;
\draw [color={rgb, 255:red, 65; green, 117; blue, 5 }  ,draw opacity=1 ]   (60,75) -- (60,105) ;
\draw [color={rgb, 255:red, 65; green, 117; blue, 5 }  ,draw opacity=1 ]   (60,45) -- (60,75) ;
\end{tikzpicture}}} \, .
\end{align}
Note that bosonic symmetry implies a redundancy in the set partition description corresponding to the permutation symmetry $1\leftrightarrow 4, 2\leftrightarrow 5, 3\leftrightarrow 6$.
Quadratic invariants also have a corresponding set partition description
\begin{align} \nonumber
  &\{\{1,2,3,4,5,6\}\}, \quad \{\{1,2,3,4,5\}, \{6\}\}, \quad \{\{1,2,3\}, \{4,5,6\}\}, \\
  &\{\{1\}, \{2,4\}, \{3,5,6\}\}, \quad \{\{1\},\{2\}, \{3\},\{4\},\{5\},\{6\}\}.
\end{align}

Each of the  invariant polynomials in \eqref{eq: linear graph basis invariants} and \eqref{eq: quadratic graph basis invariant examples} 
 will appear in the Gaussian action, along with all other possible degree two invariant combinations, each with an independent coupling parameter. However, \eqref{eq: 1pt 2pt tensor} are difficult to compute, for large $D$, in the graph basis used in \eqref{eq: linear graph basis invariants} and \eqref{eq: quadratic graph basis invariant examples} because the calculation involves the inversion of a $D^3 \times D^3$ matrix of couplings with no obvious structure that is useful in computing the inverse.

This motivates us to find a better parametrisation of the action and we will now build towards a representation theoretic construction of $\mathcal{S}(\Phi_{ijk})$.  This will lead to a parametrisation where the quadratic coupling matrix is block diagonal, making the computation of \eqref{eq: 1pt 2pt tensor} much simpler -- involving the inverse of seven symmetric matrices of dimensions $5, 10, 6, 6, 1, 2, 1$, respectively.

\subsection{The natural representation of $S_D$ and its tensor products}
To describe the construction we will need some facts about the representation theory of symmetric groups.
The natural representation of $S_D$ is given by a $D$-dimensional vector space
\begin{equation}
	V_D = \Span( \, e_1, e_2, \dots, e_D\,),
\end{equation}
together with the following map $\rho$ from $S_D$ to the set of linear maps on $V_D$
\begin{equation}
	\rho(\sigma)e_{i} = e_{\sigma^{-1}(i)}.
\end{equation}
It can be verified that this is a homomorphism: $\rho$ satisfies
\begin{equation}
	\rho(\sigma)\rho(\sigma')e_i = \rho(\sigma \sigma')e_i.
\end{equation}
for all $\sigma, \sigma' \in S_D$.

The natural representation is a reducible representation, it contains two subspaces that are closed under the action of $\rho(\sigma)$
\begin{equation}
	V_D \cong V_{[D]} \oplus V_{[D-1,1]}.
\end{equation}
We have labelled irreducible representations of $S_D$ by integer partitions of $D$, as is common in the literature on representation theory of symmetric groups.
See \cite{PIGMM} for further details about the decomposition of $V_D$, and \cite{Hamermesh} or \cite{Sagan} for further details on the general representation theory of $S_D$.

The subspace $V_{[D]}$ is the trivial representation
\begin{equation}
	V_{[D]} = \Span( E_0 = e_1 + e_2 + \dots + e_D ).
\end{equation}
Note that
\begin{equation}
	\rho(\sigma)E_0 = E_0, \quad \forall \sigma \in S_D.
\end{equation}
The subspace $V_{[D-1,1]}$ is a $(D-1)$-dimensional irreducible representation with basis
\begin{equation}
	E_a = \sum_{i=1}^D C_{a,i} e_i
\end{equation}
where
\begin{equation}
	C_{a,i} = \frac{1}{\sqrt{a(a+1)}} \qty(-a \delta_{i,a+1} + \sum_{j=1}^{a} \delta_{ji}).
\end{equation}
This basis is orthonormal with respect to the inner product
\begin{equation}
  (e_i, e_j) = \delta_{ij}.
\end{equation}

From equation \eqref{eq: sigma on V_D^3}, we have the isomorphism
\begin{equation}
	\Span(\, \Phi_{ijk}\, ) \cong V_D \otimes V_D \otimes V_D.
\end{equation}
The representation $V_D \otimes V_D \otimes V_D$ has the following decomposition into irreducible representations
\begin{align} \label{eq: VD x VD x VD Decomposition} \nonumber
V_D& \otimes V_D \otimes V_D \\
&\cong 5 V_{[D]} \oplus 10 V_{[D-1,1]} \oplus 6 V_{[D-2,2]} \oplus 6 V_{[D-2,1,1]} \oplus V_{[D-3,3]} \oplus 2 V_{[D-3,2,1]} \oplus V_{[D-3,1,1,1]},
\end{align}
where we have, again, used partitions of $D$ to label the irreducible representations on the RHS. This decomposition is derived in appendix \ref{apx: V_D3 decomp} and the multiplicities are represented in terms of graphs with edges labelled by irreducible representations. For the familiar reader, they are equivalently thought of as Young diagrams with $D$ boxes. In \eqref{eq: VD x VD x VD Decomposition} we used the rule \cite[Section 7.13]{Hamermesh} for decomposing tensor products of the form $V_{R} \otimes V_{[D-1,1]}$ (also see \cite{Quasi-partition}).
Equation \eqref{eq: VD x VD x VD Decomposition} says that there exists a basis for $V_D \otimes V_D \otimes V_D$ spanned by elements
\begin{align} \nonumber
	&S^{V_{\Lambda}; \alpha}_a \quad \text{with} \\
	&\Lambda \in \{ [D], [D-1,1], [D-2,2], [D-2,1,1], [D-3,3], [D-3,2,1], [D-3,1,1,1] \},
\end{align}
with $\alpha$ ranging over the multiplicity of $V_{\Lambda}$, and the $a$'s label an orthonormal basis in each irreducible subspace. For the sake of brevity we define the following short-hand for each of the irreducible representations appearing on the right hand side of \eqref{eq: VD x VD x VD Decomposition} 
\begin{align} \label{eq: Young diagram map} \nonumber
&V_{[D]} \equiv V_0 \, , \quad V_{[D-1,1]} \equiv V_H \, , \quad V_{[D-2,2]} \equiv V_2 \, , \quad V_{[D-2,1,1]} \equiv V_3 \, , \quad  \\ &V_{[D-3,3]} \equiv V_4 \, , \quad  V_{[D-3,2,1]} \equiv V_5 \, , \quad  V_{[D-3,1,1,1]} \equiv V_6 \, . 
\end{align}

An explicit change of basis,
\begin{align}
	S^{\Lambda, \alpha}_a = \sum_{i,j,k = 1}^D C^{\Lambda, \alpha}_{a, ijk} \Phi^{ijk},
\end{align}
implementing the isomorphism \eqref{eq: VD x VD x VD Decomposition}, defines a set of coefficients $C^{\Lambda, \alpha}_{a, ijk}$ known as Clebsch-Gordan coefficients. The statement that they implement an isomorphism of representations implies that the coefficients satisfy the equivariance property
\begin{equation}
	C^{\Lambda, \alpha}_{a, \sigma(i)\sigma(j)\sigma(k)} = \sum_{b=1}^{\dim V_\Lambda}D^{\Lambda}_{ab}(\sigma)C^{\Lambda, \alpha}_{b, ijk},
\end{equation}
where $D^{\Lambda}_{ab}(\sigma)$ is an irreducible representation of $S_D$ labelled by $\Lambda$.

\subsection{Representation basis for linear part of action}
Having defined the representation basis we can now give a good parametrisation of the action, starting with the linear part.
According to \eqref{eq: VD x VD x VD Decomposition}, the vectors
\begin{equation}
	S^{V_{[D]}, 1}, S^{V_{[D]}, 2}, S^{V_{[D]}, 3}, S^{V_{[D]}, 4}, S^{V_{[D]}, 5},
\end{equation}
form a basis for the 5-dimensional subspace of  invariant vectors in $V_D \otimes V_D \otimes V_D$,
\begin{equation}
	\rho(\sigma)S^{V_{[D]}, \alpha} = S^{V_{[D]}, \alpha}. \label{eq: rep invariants}
\end{equation}
The action decomposes into a linear and quadratic part
\begin{equation}
	\mathcal{S}(\Phi) = \mathcal{S}_{\text{linear}}(\Phi) + \mathcal{S}_{\text{quadratic}}(\Phi),
\end{equation}
and from \eqref{eq: rep invariants}
\begin{equation}
	\mathcal{S}_{\text{linear}}(\Phi) = \sum_{\alpha=1}^5 \mu_{[D], \alpha} S^{[D], \alpha} = \sum_{\alpha} \mu_{[D], \alpha} C^{[D], \alpha}_{ijk}\Phi^{ijk}
\end{equation}
where $\mu_{[D], \alpha}$ are five independent parameters.

\subsection{Representation basis for quadratic part of action}
The quadratic invariants are in one-to-one correspondence with the trivial representations in the irreducible decomposition of
\begin{align} \label{Eqn: T squared decomp}
	\text{Sym}^2(V_D^{\otimes 3})
\end{align}
which is the symmetric part of $V_D^{\otimes 3} \otimes V_D^{\otimes 3}$. Only the symmetric part is relevant, due to the commuting nature of $\Phi$ -- they satisfy
\begin{equation}
	\Phi_{ijk} \Phi_{pqr} =  \Phi_{pqr}\Phi_{ijk}.
\end{equation}

To find the multiplicity of the trivial representation $V_{0}$ in the decomposition of \eqref{Eqn: T squared decomp} we first  note that all irreducible representations of $S_D$ can be chosen to be real and unitary, or equivalently orthogonal. Therefore, Schur's lemma implies that the trivial representation $V_0$ appears in the tensor product $V_R \otimes V_S$ of two irreducible representations $V_R, V_S$, if and only if the irreducible representations are isomorphic. If they are isomorphic, the multiplicity of the trivial representation in the tensor product decomposition is one. That is, we have
\begin{equation}
  \dim \mathrm{Hom}_{S_D}(V_R \otimes V_S, V_0) = \delta_{RS},
\end{equation}
where $\delta_{RS}$ is one if $V_R \cong V_S$ and zero otherwise.

From this observation, the construction of quadratic invariants is simple in the representation basis. In general they take the form
\begin{align}
	\sum_a S^{\Lambda, \alpha}_a S^{\Lambda, \beta}_a = \sum_{\substack{i,j,k, \\ p,q,r}} \sum_a C^{\Lambda, \alpha}_{a, ijk} \Phi^{ijk} C^{\Lambda, \beta}_{a, pqr} \Phi^{pqr}.
\end{align}
This follows from the fact that the representation of $S_D$ acts with an orthogonal matrix in this basis. Therefore, summing over the index $a$ gives the unique invariant vector in $V_\Lambda \otimes V_\Lambda$, up to normalisation.

Defining the invariant endomorphism tensors $Q$ as
\begin{align} \label{eq: Q clebschs}
	Q^{\Lambda, \alpha \beta}_{ijk ; pqr} \equiv \sum_a C^{\Lambda, \alpha}_{a, ijk} C^{\Lambda, \beta}_{a, pqr},
\end{align}
we have
\begin{equation}
	\sum_a S^{\Lambda, \alpha}_a S^{\Lambda, \beta}_a = \sum_{\substack{i,j,k, \\ p,q,r}} \Phi^{ijk} Q^{\Lambda, \alpha \beta}_{ijk ; pqr} \Phi^{pqr}.
\end{equation}
It follows from the definition \eqref{eq: Q clebschs} that they satisfy
\begin{equation}
	Q^{\Lambda, \alpha \beta}_{\sigma(i)\sigma(j)\sigma(k) ; \sigma(p)\sigma(q)\sigma(r)} = Q^{\Lambda, \alpha \beta}_{ijk ; pqr}  \, , \label{eq: Q invariance}
\end{equation}
and the transposition condition
\begin{equation}
	Q^{\Lambda, \alpha \beta}_{ijk ; pqr} = Q^{\Lambda, \beta \alpha}_{pqr ; ijk} \, , 
\end{equation}
and therefore span the $S_D$ invariants of \eqref{Eqn: T squared decomp}. 

Counting the number of such combinations, we find 117 quadratic terms
\begin{align}
\frac{5 \cdot 6}{2} + \frac{10 \cdot 11}{2} + \frac{6 \cdot 7}{2} + \frac{6 \cdot 7}{2} + \frac{1 \cdot 2}{2} + \frac{2 \cdot 3}{2} + \frac{1 \cdot 2}{2} = 117,
\end{align}
where the numbers $5,10,6,6,1,2,1$ correspond to the multiplicities in \eqref{eq: VD x VD x VD Decomposition}.

To summarize, we can write the quadratic action, in the representation basis, as
\begin{align}
	\mathcal{S}_{\text{quadratic}}(\Phi) = \sum_{\Lambda, \alpha, \beta} g^{\Lambda}_{\alpha \beta} Q^{\Lambda , \alpha \beta}_{ijk ; pqr} \Phi^{ijk} \Phi^{pqr},
\end{align}
where $g^{\Lambda}$ are symmetric matrices of parameters labelled by irreps $\Lambda$ and indexed by the multiplicity indices $\alpha, \beta$. The matrices $g^{\Lambda}_{\alpha \beta}$ must have non-negative eigenvalues to define a convergent integral. Including the linear terms we can write the full partition function
\begin{align} \label{eq: action graph basis} \nonumber
Z &= \int \dd \Phi  \exp(- \mathcal{S}(\Phi)) \\
&= \int \dd \Phi \exp(  \sum_{\alpha=1}^5 \sum_{i, j, k} \mu_{[D], \alpha} C^{[D], \alpha}_{ijk}\Phi^{ijk} - \sum_{\Lambda, \alpha, \beta} \sum_{\substack{i,j,k \\ p,q,r}} g^{\Lambda}_{\alpha \beta} Q^{\Lambda , \alpha \beta}_{ijk ; pqr} \Phi^{ijk} \Phi^{pqr} ).
\end{align}

\subsection{One-point and two-point functions in representation basis}
Given the above expression for the partition function, we can now derive the one-point and two-point functions using standard techniques from quantum field theory.
To derive the one and two-point functions of $\Phi$ we first compute
\begin{equation}
	\langle S^{\Lambda, \alpha}_a \rangle, \quad \langle S^{\Lambda, \alpha}_a S^{\Lambda', \beta}_b \rangle,
\end{equation}
defined in \eqref{eqn: expectation values}. To this end, it is useful to write the partition function entirely in terms of the representation basis elements. Further, we introduce auxiliary terms in the linear part of the action, multiplying non-invariant representation variables, with the understanding that the coupling of these auxiliary terms should be set to zero in order to recover the permutation invariant model. This object, 
\begin{align} \label{eq: generating function}
	Z[\mu] = \int \dd S \exp( \sum_{\Lambda, \alpha, a} \mu^a_{\Lambda, \alpha} S^{\Lambda, \alpha}_a - \frac{1}{2} \sum_{\Lambda, \alpha, \beta, a} S^{\Lambda, \alpha}_a g^{\Lambda}_{\alpha \beta} S^{\Lambda, \beta}_a),
\end{align}
with
\begin{align}
	\dd{S} = \prod_{\Lambda, \alpha, a} \dd{S_a^{\Lambda, \alpha}},
\end{align}
defines a generating function of expectation values in the representation basis.
In this form, the integral can be computed using standard Gaussian integration techniques. For notational convenience we define
\begin{align}
\tilde{g}_{\Lambda} = g_{\Lambda}^{-1}.
\end{align}
Performing the integral on the right hand side of \eqref{eq: generating function} gives
\begin{align} \label{eq: partition function expression}
	Z[\mu] =\frac{(2 \pi)^{\frac{D^3}{2}}}{(\text{det}g)^{\frac{1}{2}}} \exp(\frac{1}{2} \sum_{\Lambda, \alpha, \beta, a} \mu_{\Lambda, \alpha}^a \tilde{g}_{\Lambda}^{\alpha \beta} \mu_{\Lambda, \beta}^a ),
\end{align}

In order to calculate expectation values of the $S$ variables we differentiate $Z[\mu]$ with respect to $\mu$. For example
\begin{align} \nonumber
\langle S^{\Lambda, \alpha}_a \rangle &= \frac{1}{Z} \int \dd S S^{\Lambda, \alpha}_a e^{-\mathcal{S}} = \frac{1}{Z} \frac{\partial Z}{\partial \mu_{\Lambda, \alpha}^a} \Big|_{\mu_{\Lambda} \neq 0\, \text{iff} \, \Lambda = [D]} \\ \nonumber
&= \frac{1}{Z} \tilde{g}_{\Lambda}^{\alpha \beta} \mu_{\Lambda, \beta} \delta(\Lambda, [D]) Z \\ 
&= \tilde{g}_{\Lambda}^{\alpha \beta} \mu_{\Lambda, \beta} \delta(\Lambda, [D]). \label{eq: S 1pt}
\end{align}
In going to the second line we have differentiated the expression for $Z[\mu]$ given on the right hand side of \eqref{eq: partition function expression}. The $\delta$-function in the last line follows due to the fact that $\mu_{\Lambda, \beta} = 0$ unless $\Lambda = [D]$. We have also dropped the state index on the linear couplings from the second line as they do not depend on the state index - the only non-zero linear couplings are those for the trivial irrep which has dimension one.
Writing this out explicitly, the non-zero linear expectation values are given by
\begin{align}
\langle S^{[D], \alpha} \rangle &= \sum_{\beta = 1}^5 \tilde{g}_{[D]}^{\alpha \beta} \mu_{[D], \beta},
\end{align}
with all other linear expectation values equal to zero.

In order to calculate quadratic expectation values of the $S$ variables we differentiate the partition function twice, giving
\begin{align} \label{eq: 2pt of S vars} \nonumber
\langle S^{\Lambda_1, \alpha}_a S^{\Lambda_2, \beta}_b \rangle &= \frac{1}{Z} \int \dd S S^{\Lambda_1, \alpha}_a S^{\Lambda_2, \beta}_b e^{-\mathcal{S}} = \frac{1}{Z} \frac{\partial }{\partial \mu_{\Lambda_2, \beta}^b} \frac{\partial Z}{\partial \mu_{\Lambda_1, \alpha}^a} \Big|_{\mu_{\Lambda} \neq 0\, \text{iff} \, \Lambda = [D]} \\ \nonumber
&= \tilde{g}_{\Lambda_1}^{\alpha \beta} \delta(\Lambda_1, \Lambda_2) \delta_{ab} + \langle S^{\Lambda_1, \alpha}_a \rangle \langle S^{\Lambda_2, \beta}_b \rangle \\ 
&= \tilde{g}_{\Lambda_1}^{\alpha \beta} \delta(\Lambda_1, \Lambda_2) \delta_{ab} + \sum_{\gamma, \rho} \tilde{g}_{\Lambda_1}^{\alpha \gamma} \mu_{\Lambda_1, \gamma} \tilde{g}_{\Lambda_2}^{\alpha \rho} \mu_{\Lambda_2, \rho} \delta(\Lambda_1, [D]) \delta(\Lambda_2, [D]).
\end{align}
For simplicity we define the connected piece of the two-point function as
\begin{align} \label{eq: rep basis connected two-point}
\langle S^{\Lambda_1, \alpha}_a S^{\Lambda_2, \beta}_b \rangle_{\text{conn}} \equiv \langle S^{\Lambda_1, \alpha}_a S^{\Lambda_2, \beta}_b \rangle -\langle S^{\Lambda_1, \alpha}_a \rangle \langle S^{\Lambda_2, \beta}_b \rangle = \tilde{g}_{\Lambda_1}^{\alpha \beta} \delta(\Lambda_1, \Lambda_2) \delta_{ab}.
\end{align}
We will now use these results to compute expressions for the one and two-point functions in the tensor basis.

\subsection{One-point and two-point functions in tensor basis}
The one-point function in the tensor basis is given by
\begin{align} \nonumber
	\expval{\Phi_{ijk}}  &= \sum_{\Lambda} \sum_{\alpha} \sum_{a=1}^{\text{dim} \Lambda}C^{\Lambda, \alpha}_{a, ijk} \langle S^{\Lambda, \alpha}_a \rangle \\ \nonumber
	&=\sum_{\Lambda} \sum_{\alpha} \sum_{a=1}^{\text{dim} \Lambda}C^{\Lambda, \alpha}_{a, ijk} \tilde{g}_{\Lambda}^{\alpha \beta} \mu_{\Lambda, \beta} \delta(\Lambda, [D]) \\
	&= \sum_{\alpha} C^{V_{[D]}, \alpha}_{ijk} \tilde{g}_{[D]}^{\alpha \beta} \mu_{[D], \beta}. \label{eq: tensor 1pt}
\end{align}
Expressions for the above Clebsch-Gordan coefficents are found in appendix \ref{apx: CGC}. We can write the $S_D$ invariant two-point function of two tensors as a sum over the $S$-variable two-point functions. In turn, this can be written as a sum of invariant endomorphism tensors $Q$
\begin{align} \label{eq: 2pt schematic} \nonumber
\langle \Phi_{ijk} \Phi_{pqr} \rangle &= \sum_{\Lambda_1, \Lambda_2} \sum_{\alpha, \beta} \sum_{a=1}^{\text{dim} \Lambda_1} \sum_{b=1}^{\text{dim} \Lambda_2} C^{\Lambda_1, \alpha}_{a, ijk} C^{\Lambda_2, \beta}_{b, pqr} \langle S^{\Lambda_1, \alpha}_a S^{\Lambda_2, \beta}_b \rangle \\ \nonumber
&= \sum_{\Lambda_1, \Lambda_2} \sum_{\alpha, \beta} \sum_{a=1}^{\text{dim} \Lambda_1} \sum_{b=1}^{\text{dim} \Lambda_2} C^{\Lambda_1, \alpha}_{a, ijk} C^{\Lambda_2, \beta}_{b, pqr} \Big( \tilde{g}_{\Lambda_1}^{\alpha \beta} \delta(\Lambda_1, \Lambda_2) \delta_{ab} + \langle S^{\Lambda_1 \alpha}_a \rangle \langle S^{\Lambda_2, \beta}_b \rangle \Big) \\ \nonumber
&= \sum_{\Lambda_1} \sum_{\alpha, \beta} \sum_{a=1}^{\text{dim} \Lambda_1} C^{\Lambda_1, \alpha}_{a, ijk} C^{\Lambda_1, \beta}_{a, pqr}  \tilde{g}_{\Lambda_1}^{\alpha \beta} \\ \nonumber 
&\hspace{1cm} + \sum_{\Lambda_1, \Lambda_2} \sum_{\alpha, \beta, \gamma, \rho} \sum_{a,b=1}^{\text{dim} \Lambda_1} C^{\Lambda_1, \alpha}_{a, ijk} C^{\Lambda_2, \beta}_{b, pqr}  \tilde{g}_{\Lambda_1}^{\alpha \gamma} \mu_{\Lambda_1, \gamma} \delta(\Lambda_1, [D]) \tilde{g}_{\Lambda_2}^{\beta \rho} \mu_{\Lambda_2, \rho} \delta(\Lambda_2, [D]) \\ 
&= \sum_{\Lambda_1} \sum_{\alpha, \beta} Q^{\Lambda_1, \alpha \beta}_{ijk ; pqr} \tilde{g}_{\Lambda_1}^{\alpha \beta} + \sum_{\alpha, \beta, \gamma, \rho} Q^{  [D], \alpha \beta }_{ijk ; pqr} \tilde{g}_{[D]}^{\alpha \gamma} \mu_{[D], \gamma} \tilde{g}_{[D]}^{\beta \rho} \mu_{[D], \rho}
\end{align}
In the second line we have used the result \eqref{eq: 2pt of S vars} and in the final line we have used the definition \eqref{eq: Q clebschs}. Akin to \eqref{eq: rep basis connected two-point}, for simplicity we again write the connected piece of the two-point function in the tensor basis
\begin{align}
\langle \Phi_{ijk} \Phi_{pqr} \rangle_{\text{conn}} \equiv \langle \Phi_{ijk} \Phi_{pqr} \rangle - \langle \Phi_{ijk} \rangle \langle \Phi_{pqr} \rangle = \sum_{\Lambda_1} \sum_{\alpha, \beta} Q^{\Lambda_1, \alpha \beta}_{ijk ; pqr} \tilde{g}_{\Lambda_1}^{\alpha \beta}. \label{eq: tensor 2pt}
\end{align}
Given \eqref{eq: tensor 2pt}, our task in the next section is to find explicit expressions for the invariant endomorphism tensors $Q^{\Lambda, \alpha \beta}_{ijk ; pqr}$ that determine the tensor two-point function. 

\section{Explicit computation of the two-point functions in the tensor basis}
\label{sec: the Qs}
The aim of this section is to develop an algorithm for constructing the invariant endomorphism tensors \eqref{eq: Q clebschs}, appearing in the two-point function in the tensor basis \eqref{eq: 2pt schematic}, for all $D \geq 6$. The algorithm involves resolving the labels $\Lambda, \alpha, \beta$ on the tensors $Q^{\Lambda, \alpha \beta}_{ijk ; pqr}$. We show that the labels are described by a pair of graphs labelled by irreducible representations of $S_D$. Each pair of graphs is uniquely determined by a set of eigenvalue equations related to central elements in the $\mC ( S_D) $ group algebras. We solve the eigenvalue equations using a correspondence between central elements in $\mathbb{C}(S_D)$ and elements of partition algebras. This approach bypasses the explicit construction of the  Clebsch-Gordan coefficients appearing in the definition of $Q^{\Lambda, \alpha \beta}_{ijk ; pqr}$ \eqref{eq: Q clebschs}. Specifying bases in irreps of $S_N$ at general $N$, working out the corresponding Clebsch's and summing them are the highly non-trivial steps which are bypassed. We find that the  $Q^{\Lambda, \alpha \beta}_{ijk ; pqr}$ themselves are elements of the partition algebra and the eigenvalue equations involve matrices whose elements are structure constants of the partition algebra. Combinatorially constructed eigenvalue systems for central elements in symmetric group algebras have also been used in the identification of representation theoretic labels with motivations coming from holography, quantum information and combinatorial representation theory \cite{KRkst,JBGSR1,JBGSR2}.

In the construction, it is useful to view the invariant tensors $Q^{\Lambda, \alpha \beta}_{ijk ; pqr}$ as equivariant maps
\begin{equation} \label{eq: Q as equivariant map}
	Q^{\Lambda, \alpha \beta}: V_D^{\otimes 3} \rightarrow V_D^{\otimes 3}
\end{equation}
where
\begin{equation}
	Q^{\Lambda, \alpha \beta}(\Phi_{ijk}) = \sum_{p,q,r=1}^D Q^{\Lambda, \alpha \beta}_{pqr; ijk} \Phi_{pqr} \, ,
\end{equation}
and from equation \eqref{eq: Q invariance},
\begin{equation}
	Q^{\Lambda, \alpha \beta}\rho( \sigma )= \rho(\sigma ) Q^{\Lambda, \alpha \beta} \, .
\end{equation}
They form a basis for the vector space of equivariant maps of the kind in \eqref{eq: Q as equivariant map}, commonly denoted
\begin{equation}
	\End_{S_D}(V_D^{\otimes 3}) = \Span(Q^{\Lambda, \alpha \beta}) \, .
\end{equation}
This vector space is an algebra, with multiplication given by composition of maps. In subsection \ref{Q and graphs} we will describe how the multiplicity labels $\alpha, \beta$ can be understood as graphs $G_{\vec{R}}^{\Lambda}$ decorated with irreducible representations ${\vec{R}}=(R_1, R_2, R_3, R_4)$ and $\Lambda$
\begin{equation}
	G_{\vec{R}}^{\Lambda}= \mytikz{	
		\node (t1) at (0,0) [circle,fill,inner sep=0.5mm,label=above:] {};	
		\node (t2) at (1.5,0) [circle,fill,inner sep=0.5mm,label=above:] {};	
		\node (i1) at (-1.5,0.9) {};
		\node (i2) at (-1.5,0) {};
		\node (i3) at (-1.5,-0.9) {};
		\node (m) at (3,0) {};
		\draw [postaction={decorate}] (t2) to node[above]{$\Lambda$} (m);
		\draw [postaction={decorate}] (t1) to node[above]{$R_4$} (t2);
		\draw [postaction={decorate}] (i1) to node[above]{$R_1$} (t1);
		\draw [postaction={decorate}] (i2) to node[below]{$R_2$} (t1);
		\draw [postaction={decorate}] (i3) to node[below]{$R_3$} (t2);
	}.
\end{equation}
Sections \ref{subsec: T2}-\ref{subsec: T2 dual} build towards the construction of a set of commuting operators whose eigenvalues distinguish the pairs of graphs, and consequently, have simultaneous eigenvectors $Q^{\Lambda, \alpha \beta}$. The last subsection gives an algorithm for solving these eigenvector equations as analytic functions of $D$.

\subsection{Resolving multiplicity labels using graphs} \label{Q and graphs}

As we now explain, the multiplicity indices $\alpha, \beta$ are in correspondence with decorated graphs. It will be useful to introduce some diagrammatic notation for Clebsch-Gordan coefficients used in \cite{Quiverscalc}. Clebsch-Gordan coefficients for the decomposition $V_{R_1} \otimes V_{R_2}$ can be represented by a graph
\begin{equation}
\label{eq:CG_diag}
C^{R_1 R_2,\Lambda \,\tau}_{\,a_1\;a_2,\,m}
\; = \;
\mytikz{	
	\node (t) at (0,0) [circle,fill,inner sep=0.5mm,label=below:$\tau$] {};	
	\node (i1) at (-1.5,0.7) {$a_1$};
	\node (i2) at (-1.5,-0.7) {$a_2$};
	\node (m) at (1.5,0) {$m$};
	\draw [postaction={decorate}] (t) to node[above]{$\Lambda$} (m);		
	\draw [postaction={decorate}] (i1) to node[above]{$R_1$} (t);
	\draw [postaction={decorate}] (i2) to node[below]{$R_2$} (t);
},
\end{equation}
where $a_1, a_2, m$ label orthonormal bases for $V_{R_1}, V_{R_2}, V_\Lambda$, respectively and $\tau$ is a further index ranging over the multiplicity of $\Lambda$ in the decomposition of $V_{R_1} \otimes V_{R_2}$.
The equivariance of Clebsch-Gordan coefficients
\begin{equation}
\sum_{b_1 , b_2} D^{R_1}_{a_1 b_1}(\sigma) D^{R_2}_{a_2 b_2}(\sigma) 
C^{R_1 R_2,\Lambda \,\tau}_{\,b_1\;b_2,\,m} = \sum_{l}
C^{R_1 R_2,\Lambda \,\tau}_{\,a_1\;a_2,\,l}
D^{\Lambda}_{l m}(\sigma) \, , \quad \forall \sigma \in S_D \, ,
\end{equation}
can be written diagrammatically as
\begin{align}
\label{eq:CG_gamma_pull}
	\mytikz{
		\node (m) at (0,0) [circle,fill,inner sep=0.5mm] {};	
		\node (g1) at (-1,0.7) [rectangle,draw] {$\sigma$};
		\node (g2) at (-1,-0.7) [rectangle,draw] {$\sigma$};
		\draw [postaction={decorate}] (m) to node[above]{$\Lambda$} +(1,0);	
		\draw [postaction={decorate}] (g1) to node[above]{$R_1$} (m);
		\draw [postaction={decorate}] (g2) to node[below]{$R_2$} (m);		
		\draw [postaction={decorate}] ($(g1)+(-1.0,0)$) to (g1);
		\draw [postaction={decorate}] ($(g2)+(-1.0,0)$) to (g2);
	}
\; = \;
	\mytikz{
		\node (s) at (-0.2,0) [rectangle,draw] {$\sigma$};		
		\node (m) at (-1.5,0) [circle,fill,inner sep=0.5mm] {};	
		\draw [postaction={decorate}] (s) to +(0.7,0);	
		\draw [postaction={decorate}] (m) to node[above]{$\Lambda$} (s);		
		\draw [postaction={decorate}] ($(m)+(-1,0.7)$) to node[above]{$R_1$} (m);
		\draw [postaction={decorate}] ($(m)+(-1,-0.7)$) to node[below]{$R_2$} (m);
	} \, , \quad \forall \sigma \in S_D \, .
\end{align}
The coefficients relevant to the decomposition of $V_D \otimes V_D \otimes V_D$ into irreducible representations are composed of two Clebsch-Gordan coefficients
\begin{equation}
\label{eq:CG_CG_diag}
C^{G^{\Lambda}_{\vec{R}}}_{i_1 i_2 i_3 \rightarrow m} = 
\sum_{m^{\prime}}C^{R_1 R_2,R_4\, }_{\,i_1\;i_2,\,m^{\prime}} C ^{R_4 R_3,\Lambda }_{\,m^{\prime}\;i_3,\,m}
\; = \;
\mytikz{	
	\node (t1) at (0,0) [circle,fill,inner sep=0.5mm,label=above:] {};	
	\node (t2) at (1.5,0) [circle,fill,inner sep=0.5mm,label=above:] {};	
	\node (i1) at (-1.5,0.9) {$i_1$};
	\node (i2) at (-1.5,0) {$i_2$};
	\node (i3) at (-1.5,-0.9) {$i_3$};
	\node (m) at (3,0) {$m$};
	\draw [postaction={decorate}] (t2) to node[above]{$\Lambda$} (m);
	\draw [postaction={decorate}] (t1) to node[above]{$R_4$} (t2);
	\draw [postaction={decorate}] (i1) to node[above]{$R_1$} (t1);
	\draw [postaction={decorate}] (i2) to node[below]{$R_2$} (t1);
	\draw [postaction={decorate}] (i3) to node[below]{$R_3$} (t2);
}.
\end{equation}
An important feature of the relevant Clebsch-Gordan coefficients appearing above is that they are all multiplicity free, that is the $\tau$ appearing in \eqref{eq:CG_diag} is 1 for any given $\vec{R}$ and $\Lambda$. We can therefore drop the multiplicity labels. We refer to the content of this graph as $G^{\Lambda}_{\vec{R}}$, $\Lambda$ being the final irreducible representation and ${\vec{R}}={(R_1, R_2, R_3, R_4)}$ specifying the intermediate irreducible representations.
It follows that the multiplicities in \eqref{eq: VD x VD x VD Decomposition} are uniquely specified by the content of such graphs.

Given the form of $Q^{\Lambda, \alpha \beta}_{ijk, pqr}$ in \eqref{eq: Q clebschs}
we have the following diagrammatic expression
\begin{equation}
	Q^{G^{\Lambda}_{\vec{R}} G^{\Lambda}_{\vec{S}}}_{ijk, pqr} = 
	\vcenter{\hbox{
			\tikzset{every picture/.style={line width=0.75pt}} 
			
			\begin{tikzpicture}[x=0.75pt,y=0.75pt,yscale=-1,xscale=1]
				
				\draw [color={rgb, 255:red, 0; green, 0; blue, 0 }  ,draw opacity=1 ]   (120.6,77.39) -- (37.92,120.46) ;
				\draw [color={rgb, 255:red, 0; green, 0; blue, 0 }  ,draw opacity=1 ]   (94.76,77.39) -- (68.93,51.55) ;
				\draw [color={rgb, 255:red, 0; green, 0; blue, 0 }  ,draw opacity=1 ]   (94.76,77.39) -- (68.93,77.39) ;
				\draw [color={rgb, 255:red, 0; green, 0; blue, 0 }  ,draw opacity=1 ]   (68.93,51.55) -- (37.92,51.55) ;
				\draw [color={rgb, 255:red, 0; green, 0; blue, 0 }  ,draw opacity=1 ]   (68.93,77.39) -- (37.92,77.39) ;
				\draw    (120.6,77.39) -- (94.76,77.39) ;
				\draw [shift={(94.76,77.39)}, rotate = 180] [color={rgb, 255:red, 0; green, 0; blue, 0 }  ][fill={rgb, 255:red, 0; green, 0; blue, 0 }  ][line width=0.75]      (0, 0) circle [x radius= 1.34, y radius= 1.34]   ;
				\draw [shift={(120.6,77.39)}, rotate = 180] [color={rgb, 255:red, 0; green, 0; blue, 0 }  ][fill={rgb, 255:red, 0; green, 0; blue, 0 }  ][line width=0.75]      (0, 0) circle [x radius= 1.34, y radius= 1.34]   ;
				\draw    (130.94,77.39) -- (120.6,77.39) ;
				\draw [shift={(120.6,77.39)}, rotate = 180] [color={rgb, 255:red, 0; green, 0; blue, 0 }  ][fill={rgb, 255:red, 0; green, 0; blue, 0 }  ][line width=0.75]      (0, 0) circle [x radius= 1.34, y radius= 1.34]   ;
				\draw [color={rgb, 255:red, 0; green, 0; blue, 0 }  ,draw opacity=1 ]   (120.6,170.41) -- (37.92,213.48) ;
				\draw [color={rgb, 255:red, 0; green, 0; blue, 0 }  ,draw opacity=1 ]   (94.76,170.41) -- (68.93,144.57) ;
				\draw [color={rgb, 255:red, 0; green, 0; blue, 0 }  ,draw opacity=1 ]   (94.76,170.41) -- (68.93,170.41) ;
				\draw [color={rgb, 255:red, 0; green, 0; blue, 0 }  ,draw opacity=1 ]   (68.93,144.57) -- (37.92,144.57) ;
				\draw [color={rgb, 255:red, 0; green, 0; blue, 0 }  ,draw opacity=1 ]   (68.93,170.41) -- (37.92,170.41) ;
				\draw    (120.6,170.41) -- (94.76,170.41) ;
				\draw [shift={(94.76,170.41)}, rotate = 180] [color={rgb, 255:red, 0; green, 0; blue, 0 }  ][fill={rgb, 255:red, 0; green, 0; blue, 0 }  ][line width=0.75]      (0, 0) circle [x radius= 1.34, y radius= 1.34]   ;
				\draw [shift={(120.6,170.41)}, rotate = 180] [color={rgb, 255:red, 0; green, 0; blue, 0 }  ][fill={rgb, 255:red, 0; green, 0; blue, 0 }  ][line width=0.75]      (0, 0) circle [x radius= 1.34, y radius= 1.34]   ;
				\draw    (130.94,170.41) -- (120.6,170.41) ;
				\draw [shift={(120.6,170.41)}, rotate = 180] [color={rgb, 255:red, 0; green, 0; blue, 0 }  ][fill={rgb, 255:red, 0; green, 0; blue, 0 }  ][line width=0.75]      (0, 0) circle [x radius= 1.34, y radius= 1.34]   ;
				\draw    (130.94,77.39) .. controls (156.13,76.57) and (156.57,170.03) .. (130.94,170.41) ;
				
				\draw (65.37,57.59) node [anchor=north west][inner sep=0.75pt]  [font=\tiny]  {$R_{1}$};
				\draw (65.37,81.7) node [anchor=north west][inner sep=0.75pt]  [font=\tiny]  {$R_{2}$};
				\draw (65.37,109.26) node [anchor=north west][inner sep=0.75pt]  [font=\tiny]  {$R_{3}$};
				\draw (97.41,57.59) node [anchor=north west][inner sep=0.75pt]  [font=\tiny]  {$R_{4}$};
				\draw (24.75,49) node [anchor=north west][inner sep=0.75pt]  [font=\tiny]  {$i$};
				\draw (24.75,76.56) node [anchor=north west][inner sep=0.75pt]  [font=\tiny]  {$j$};
				\draw (24.75,117.9) node [anchor=north west][inner sep=0.75pt]  [font=\tiny]  {$k$};
				\draw (65.58,151.86) node [anchor=north west][inner sep=0.75pt]  [font=\tiny]  {$S_{1}$};
				\draw (65.58,175.97) node [anchor=north west][inner sep=0.75pt]  [font=\tiny]  {$S_{2}$};
				\draw (65.58,203.53) node [anchor=north west][inner sep=0.75pt]  [font=\tiny]  {$S_{3}$};
				\draw (97.62,151.86) node [anchor=north west][inner sep=0.75pt]  [font=\tiny]  {$S_{4}$};
				\draw (24.75,142.02) node [anchor=north west][inner sep=0.75pt]  [font=\tiny]  {$p$};
				\draw (24.75,169.58) node [anchor=north west][inner sep=0.75pt]  [font=\tiny]  {$q$};
				\draw (24.75,210.92) node [anchor=north west][inner sep=0.75pt]  [font=\tiny]  {$r$};
				\draw (152,114) node [anchor=north west][inner sep=0.75pt]  [font=\tiny]  {$\Lambda $};
	\end{tikzpicture}}}
\end{equation}
where we have used a pair of graphs to label the equivariant map.
We will now see how central elements are used to determine the graphs $G_{\vec{R}}^{\Lambda}, G_{\vec{R}}^{\Lambda}$ and subsequently the entire invariant endomorphism tensor $Q^{\Lambda, \alpha \beta}_{ijk, pqr}$.

\subsection{Using central elements in $\mathbb{C}(S_D)$ to detect graphs} \label{subsec: T2}

The center $\mathcal{Z}(\mathbb{C}(S_D))$ of $\mathbb{C}(S_D)$, the symmetric group algebra, consists of elements
\begin{equation}
	\mathcal{Z}(\mathbb{C}(S_D)) = \{z \in \mathbb{C}(S_D) : z\sigma = \sigma z, \quad \forall \, \sigma \in \mathbb{C}(S_D)\} \, .
\end{equation}
Elements in the center are called central elements. Central elements play a special role in representation theory because Schur's lemma implies that an irreducible matrix representation of a central element is proportional to the identity matrix. The proportionality constant is a normalized character. In particular, we have
\begin{equation}
	D^{\Lambda_1}_{ab}(z) =  \frac{\chi^{\Lambda_1}(z)}{\DimSD{\Lambda_1}} \equiv \hat{\chi}^{\Lambda_1}(z)\delta_{ab} \, , \label{eq: schur lemma T2}
\end{equation}
where $\chi^{\Lambda_1}(z)$ is the character of $z$ in the irreducible representation $\Lambda_1$, and we have defined the hatted short-hand for normalised characters. Central elements act by constants on irreducible subspaces, and the constants can be used to determine the particular representation.

The element of $\mathbb{C}(S_D)$ formed by summing over all elements in a distinct conjugacy class of $S_D$ is central.
For example, we define the element $T_2 \in \mathcal{Z}(\mathbb{C}(S_D))$ as
\begin{equation}
	T_2 = \sum_{1 \leq i < j \leq D}  (ij) \, .
\end{equation} 
$T_2$ is the sum is over all transpositions. 
Normalized characters of $T_2$ can be expressed in terms of combinatorial quantities (known as the contents) of boxes of Young diagrams (see example 7 in section I.7 of \cite{Macdonald1998}).
Let $Y_{\Lambda_1}$ be the Young diagram corresponding to the integer partition $\Lambda_1 \in \cYS ( k )$, the set of valid Young diagrams with $k$ boxes. Then
\begin{equation}
	\hat{\chi}^{\Lambda_1}(T_2) = \sum_{(i,j) \in Y_{\Lambda_1}} (j-i) \, , \label{eq: norm characters of T2s}
\end{equation}
where $(i,j)$ corresponds to the cell in the $i$th row and $j$th column of the Young diagram (the top left box has coordinate $(1,1)$). 

All of the irreducible representations we need to identify appear on the right hand side of \eqref{eq: VD x VD x VD Decomposition}. The normalised characters of $T_2$ distinguish these representations, they are
\begin{align}
	&\hat{\chi}^{V_0}(T_2) = \frac{D(D-1)}{2} \, , \\
	&\hat{\chi}^{V_H}(T_2) = \frac{D(D-3)}{2} \, , \\
	&\hat{\chi}^{V_2}(T_2) = \frac{(D-1)(D-4)}{2} \, , \\
	&\hat{\chi}^{V_3}(T_2) = \frac{D(D-5)}{2} \, , \\
	&\hat{\chi}^{V_4}(T_2) = \frac{(D-4)(D-3)}{2} \, , \\
	&\hat{\chi}^{V_5}(T_2) = \frac{(D-6)(D-1)}{2} \, , \\
	&\hat{\chi}^{V_6}(T_2) = \frac{D(D-7)}{2} \, .
\end{align}
We define the following set of elements in $\mathbb{C}(S_D)^{\otimes 3}$
\begin{align} \label{eq: T_2 1}
T^{(111)}_2 &=  \sum_{i\less j} (ij) \otimes (ij) \otimes (ij) \, , \\
T^{(110)}_2 &=  \sum_{i\less j} (ij) \otimes (ij) \otimes \idn \,  , \\
T^{(100)}_2 &=  \sum_{i\less j} (ij) \otimes \idn \otimes \idn \, , \\
T^{(010)}_2 &=  \sum_{i\less j} \idn \otimes (ij) \otimes \idn \, , \\ \label{eq: T_2 5}
T^{(001)}_2 &=  \sum_{i\less j} \idn \otimes \idn \otimes (ij) \, , 
\end{align}
and refer to them collectively as $T_2^{(b)}$ where $b$ is one of the above binary strings.
Notably, as operators on $V_D^{\otimes 3}$, they commute with the action of $S_D$, that is, they are elements of $\End_{S_D}(V_D^{\otimes 3})$. Furthermore, they commute among themselves.
As we now show, the invariant endomorphism tensors $Q^{G_{\vec{R}}^\Lambda G_{\vec{S}}^{\Lambda}}_{ijk, pqr}$ are simultaneous eigenvectors of the operators \eqref{eq: T_2 1} - \eqref{eq: T_2 5}.

Consider the composition $T^{(111)}_2  Q^{G_{\vec{R}}^\Lambda G_{\vec{S}}^{\Lambda}} $ acting on $V_D^{\otimes 3}$. As a diagram equation we have
\begin{align}
	&\sum_{\substack{\gamma = (ij)\\ i<j}} \vcenter{\hbox{
	\begin{tikzpicture}[x=0.75pt,y=0.75pt,yscale=-1,xscale=1]
		\draw [color={rgb, 255:red, 0; green, 0; blue, 0 }  ,draw opacity=1 ]   (320.6,50.47) -- (237.92,93.54) ;
		\draw [color={rgb, 255:red, 0; green, 0; blue, 0 }  ,draw opacity=1 ]   (294.76,50.47) -- (268.93,24.63) ;
		\draw [color={rgb, 255:red, 0; green, 0; blue, 0 }  ,draw opacity=1 ]   (294.76,50.47) -- (268.93,50.47) ;
		\draw [color={rgb, 255:red, 0; green, 0; blue, 0 }  ,draw opacity=1 ]   (268.93,24.63) -- (237.92,24.63) ;
		\draw [color={rgb, 255:red, 0; green, 0; blue, 0 }  ,draw opacity=1 ]   (268.93,50.47) -- (237.92,50.47) ;
		\draw    (320.6,50.47) -- (294.76,50.47) ;
		\draw [shift={(294.76,50.47)}, rotate = 180] [color={rgb, 255:red, 0; green, 0; blue, 0 }  ][fill={rgb, 255:red, 0; green, 0; blue, 0 }  ][line width=0.75]      (0, 0) circle [x radius= 1.34, y radius= 1.34]   ;
		\draw [shift={(320.6,50.47)}, rotate = 180] [color={rgb, 255:red, 0; green, 0; blue, 0 }  ][fill={rgb, 255:red, 0; green, 0; blue, 0 }  ][line width=0.75]      (0, 0) circle [x radius= 1.34, y radius= 1.34]   ;
		\draw    (330.94,50.47) -- (320.6,50.47) ;
		\draw [shift={(320.6,50.47)}, rotate = 180] [color={rgb, 255:red, 0; green, 0; blue, 0 }  ][fill={rgb, 255:red, 0; green, 0; blue, 0 }  ][line width=0.75]      (0, 0) circle [x radius= 1.34, y radius= 1.34]   ;
		\draw [color={rgb, 255:red, 0; green, 0; blue, 0 }  ,draw opacity=1 ]   (320.6,143.49) -- (237.92,186.55) ;
		\draw [color={rgb, 255:red, 0; green, 0; blue, 0 }  ,draw opacity=1 ]   (294.76,143.49) -- (268.93,117.65) ;
		\draw [color={rgb, 255:red, 0; green, 0; blue, 0 }  ,draw opacity=1 ]   (294.76,143.49) -- (268.93,143.49) ;
		\draw [color={rgb, 255:red, 0; green, 0; blue, 0 }  ,draw opacity=1 ]   (268.93,117.65) -- (237.92,117.65) ;
		\draw [color={rgb, 255:red, 0; green, 0; blue, 0 }  ,draw opacity=1 ]   (268.93,143.49) -- (237.92,143.49) ;
		\draw    (320.6,143.49) -- (294.76,143.49) ;
		\draw [shift={(294.76,143.49)}, rotate = 180] [color={rgb, 255:red, 0; green, 0; blue, 0 }  ][fill={rgb, 255:red, 0; green, 0; blue, 0 }  ][line width=0.75]      (0, 0) circle [x radius= 1.34, y radius= 1.34]   ;
		\draw [shift={(320.6,143.49)}, rotate = 180] [color={rgb, 255:red, 0; green, 0; blue, 0 }  ][fill={rgb, 255:red, 0; green, 0; blue, 0 }  ][line width=0.75]      (0, 0) circle [x radius= 1.34, y radius= 1.34]   ;
		\draw    (330.94,143.49) -- (320.6,143.49) ;
		\draw [shift={(320.6,143.49)}, rotate = 180] [color={rgb, 255:red, 0; green, 0; blue, 0 }  ][fill={rgb, 255:red, 0; green, 0; blue, 0 }  ][line width=0.75]      (0, 0) circle [x radius= 1.34, y radius= 1.34]   ;
		\draw    (330.94,50.47) .. controls (356.13,49.65) and (356.57,143.11) .. (330.94,143.49) ;
		\draw (265.37,30.66) node [anchor=north west][inner sep=0.75pt]  [font=\tiny]  {$R_{1}$};
		\draw (265.37,54.78) node [anchor=north west][inner sep=0.75pt]  [font=\tiny]  {$R_{2}$};
		\draw (265.37,82.34) node [anchor=north west][inner sep=0.75pt]  [font=\tiny]  {$R_{3}$};
		\draw (297.41,30.66) node [anchor=north west][inner sep=0.75pt]  [font=\tiny]  {$R_{4}$};
		\draw (224.75,22.08) node [anchor=north west][inner sep=0.75pt]  [font=\tiny]  {$i$};
		\draw (224.75,49.64) node [anchor=north west][inner sep=0.75pt]  [font=\tiny]  {$j$};
		\draw (224.75,90.98) node [anchor=north west][inner sep=0.75pt]  [font=\tiny]  {$k$};
		\draw (265.58,124.93) node [anchor=north west][inner sep=0.75pt]  [font=\tiny]  {$S_1$};
		\draw (265.58,149.05) node [anchor=north west][inner sep=0.75pt]  [font=\tiny]  {$S_2$};
		\draw (265.58,176.61) node [anchor=north west][inner sep=0.75pt]  [font=\tiny]  {$S_3$};
		\draw (297.62,124.93) node [anchor=north west][inner sep=0.75pt]  [font=\tiny]  {$S_4$};
		\draw (224.75,115.1) node [anchor=north west][inner sep=0.75pt]  [font=\tiny]  {$p$};
		\draw (224.75,142.66) node [anchor=north west][inner sep=0.75pt]  [font=\tiny]  {$q$};
		\draw (224.75,184) node [anchor=north west][inner sep=0.75pt]  [font=\tiny]  {$r$};
		\draw  [fill={rgb, 255:red, 255; green, 255; blue, 255 }  ,fill opacity=1 ]  (248.75,18) -- (266.75,18) -- (266.75,32) -- (248.75,32) -- cycle  ;
		\draw (249.75,19) node [anchor=north west][inner sep=0.75pt]  [font=\tiny]  {$\gamma $};
		\draw  [fill={rgb, 255:red, 255; green, 255; blue, 255 }  ,fill opacity=1 ]  (249.75,43) -- (267.75,43) -- (267.75,57) -- (249.75,57) -- cycle  ;
		\draw (250.75,44) node [anchor=north west][inner sep=0.75pt]  [font=\tiny]  {$\gamma $};
		\draw  [fill={rgb, 255:red, 255; green, 255; blue, 255 }  ,fill opacity=1 ]  (249.75,78) -- (267.75,78) -- (267.75,92) -- (249.75,92) -- cycle  ;
		\draw (250.75,79) node [anchor=north west][inner sep=0.75pt]  [font=\tiny]  {$\gamma $};
		\draw (354,89) node [anchor=north west][inner sep=0.75pt]  [font=\tiny]  {$\Lambda $};
	\end{tikzpicture}}} = \sum_{\substack{\gamma = (ij)\\ i<j}} \vcenter{\hbox{
\begin{tikzpicture}[x=0.75pt,y=0.75pt,yscale=-1,xscale=1]
\draw [color={rgb, 255:red, 0; green, 0; blue, 0 }  ,draw opacity=1 ]   (455,36.92) -- (362.17,80.46) ;
\draw [color={rgb, 255:red, 0; green, 0; blue, 0 }  ,draw opacity=1 ]   (419.01,37.39) -- (393.17,11.55) ;
\draw [color={rgb, 255:red, 0; green, 0; blue, 0 }  ,draw opacity=1 ]   (419.01,37.39) -- (393.17,37.39) ;
\draw [color={rgb, 255:red, 0; green, 0; blue, 0 }  ,draw opacity=1 ]   (393.17,11.55) -- (362.17,11.55) ;
\draw [color={rgb, 255:red, 0; green, 0; blue, 0 }  ,draw opacity=1 ]   (393.17,37.39) -- (362.17,37.39) ;
\draw    (455,36.92) -- (419.01,37.39) ;
\draw [shift={(419.01,37.39)}, rotate = 179.25] [color={rgb, 255:red, 0; green, 0; blue, 0 }  ][fill={rgb, 255:red, 0; green, 0; blue, 0 }  ][line width=0.75]      (0, 0) circle [x radius= 1.34, y radius= 1.34]   ;
\draw [shift={(455,36.92)}, rotate = 179.25] [color={rgb, 255:red, 0; green, 0; blue, 0 }  ][fill={rgb, 255:red, 0; green, 0; blue, 0 }  ][line width=0.75]      (0, 0) circle [x radius= 1.34, y radius= 1.34]   ;
\draw    (460,36.92) -- (455,36.92) ;
\draw [shift={(455,36.92)}, rotate = 180] [color={rgb, 255:red, 0; green, 0; blue, 0 }  ][fill={rgb, 255:red, 0; green, 0; blue, 0 }  ][line width=0.75]      (0, 0) circle [x radius= 1.34, y radius= 1.34]   ;
\draw [color={rgb, 255:red, 0; green, 0; blue, 0 }  ,draw opacity=1 ]   (455.19,130.41) -- (362.17,173.48) ;
\draw [color={rgb, 255:red, 0; green, 0; blue, 0 }  ,draw opacity=1 ]   (419.01,130.41) -- (393.17,104.57) ;
\draw [color={rgb, 255:red, 0; green, 0; blue, 0 }  ,draw opacity=1 ]   (419.01,130.41) -- (393.17,130.41) ;
\draw [color={rgb, 255:red, 0; green, 0; blue, 0 }  ,draw opacity=1 ]   (393.17,104.57) -- (362.17,104.57) ;
\draw [color={rgb, 255:red, 0; green, 0; blue, 0 }  ,draw opacity=1 ]   (393.17,130.41) -- (362.17,130.41) ;
\draw    (455.19,130.41) -- (419.01,130.41) ;
\draw [shift={(419.01,130.41)}, rotate = 180] [color={rgb, 255:red, 0; green, 0; blue, 0 }  ][fill={rgb, 255:red, 0; green, 0; blue, 0 }  ][line width=0.75]      (0, 0) circle [x radius= 1.34, y radius= 1.34]   ;
\draw [shift={(455.19,130.41)}, rotate = 180] [color={rgb, 255:red, 0; green, 0; blue, 0 }  ][fill={rgb, 255:red, 0; green, 0; blue, 0 }  ][line width=0.75]      (0, 0) circle [x radius= 1.34, y radius= 1.34]   ;
\draw    (460.42,130.13) -- (455.19,130.41) ;
\draw [shift={(455.19,130.41)}, rotate = 176.93] [color={rgb, 255:red, 0; green, 0; blue, 0 }  ][fill={rgb, 255:red, 0; green, 0; blue, 0 }  ][line width=0.75]      (0, 0) circle [x radius= 1.34, y radius= 1.34]   ;
\draw    (460,36.92) .. controls (485.19,36.1) and (486.06,129.75) .. (460.42,130.13) ;
\draw (389.62,17.59) node [anchor=north west][inner sep=0.75pt]  [font=\tiny]  {$R_{1}$};
\draw (389.62,41.7) node [anchor=north west][inner sep=0.75pt]  [font=\tiny]  {$R_{2}$};
\draw (389.62,69.26) node [anchor=north west][inner sep=0.75pt]  [font=\tiny]  {$R_{3}$};
\draw (421.66,15.59) node [anchor=north west][inner sep=0.75pt]  [font=\tiny]  {$R_{4}$};
\draw (349,9) node [anchor=north west][inner sep=0.75pt]  [font=\tiny]  {$i$};
\draw (349,36.56) node [anchor=north west][inner sep=0.75pt]  [font=\tiny]  {$j$};
\draw (349,77.9) node [anchor=north west][inner sep=0.75pt]  [font=\tiny]  {$k$};
\draw (389.83,111.86) node [anchor=north west][inner sep=0.75pt]  [font=\tiny]  {$S_1$};
\draw (389.83,135.97) node [anchor=north west][inner sep=0.75pt]  [font=\tiny]  {$S_2$};
\draw (389.83,163.53) node [anchor=north west][inner sep=0.75pt]  [font=\tiny]  {$S_3$};
\draw (421.87,111.86) node [anchor=north west][inner sep=0.75pt]  [font=\tiny]  {$S_4$};
\draw (349,102.02) node [anchor=north west][inner sep=0.75pt]  [font=\tiny]  {$p$};
\draw (349,129.58) node [anchor=north west][inner sep=0.75pt]  [font=\tiny]  {$q$};
\draw (349,170.92) node [anchor=north west][inner sep=0.75pt]  [font=\tiny]  {$r$};
\draw  [fill={rgb, 255:red, 255; green, 255; blue, 255 }  ,fill opacity=1 ]  (423,27.92) -- (441,27.92) -- (441,41.92) -- (423,41.92) -- cycle  ;
\draw (424,28.92) node [anchor=north west][inner sep=0.75pt]  [font=\tiny]  {$\gamma $};
\draw  [fill={rgb, 255:red, 255; green, 255; blue, 255 }  ,fill opacity=1 ]  (374,64.92) -- (392,64.92) -- (392,78.92) -- (374,78.92) -- cycle  ;
\draw (375,65.92) node [anchor=north west][inner sep=0.75pt]  [font=\tiny]  {$\gamma $};
\draw (479,84) node [anchor=north west][inner sep=0.75pt]  [font=\tiny]  {$\Lambda $};
\end{tikzpicture}}} = \\
&\sum_{\substack{\gamma = (ij)\\ i<j}} \vcenter{\hbox{
\begin{tikzpicture}[x=0.75pt,y=0.75pt,yscale=-1,xscale=1]
\draw [color={rgb, 255:red, 0; green, 0; blue, 0 }  ,draw opacity=1 ]   (593.19,63.85) -- (500.36,107.38) ;
\draw [color={rgb, 255:red, 0; green, 0; blue, 0 }  ,draw opacity=1 ]   (557.2,64.32) -- (531.37,38.48) ;
\draw [color={rgb, 255:red, 0; green, 0; blue, 0 }  ,draw opacity=1 ]   (557.2,64.32) -- (531.37,64.32) ;
\draw [color={rgb, 255:red, 0; green, 0; blue, 0 }  ,draw opacity=1 ]   (531.37,38.48) -- (500.36,38.48) ;
\draw [color={rgb, 255:red, 0; green, 0; blue, 0 }  ,draw opacity=1 ]   (531.37,64.32) -- (500.36,64.32) ;
\draw    (593.19,63.85) -- (557.2,64.32) ;
\draw [shift={(557.2,64.32)}, rotate = 179.25] [color={rgb, 255:red, 0; green, 0; blue, 0 }  ][fill={rgb, 255:red, 0; green, 0; blue, 0 }  ][line width=0.75]      (0, 0) circle [x radius= 1.34, y radius= 1.34]   ;
\draw [shift={(593.19,63.85)}, rotate = 179.25] [color={rgb, 255:red, 0; green, 0; blue, 0 }  ][fill={rgb, 255:red, 0; green, 0; blue, 0 }  ][line width=0.75]      (0, 0) circle [x radius= 1.34, y radius= 1.34]   ;
\draw    (598.19,63.85) -- (593.19,63.85) ;
\draw [shift={(593.19,63.85)}, rotate = 180] [color={rgb, 255:red, 0; green, 0; blue, 0 }  ][fill={rgb, 255:red, 0; green, 0; blue, 0 }  ][line width=0.75]      (0, 0) circle [x radius= 1.34, y radius= 1.34]   ;
\draw [color={rgb, 255:red, 0; green, 0; blue, 0 }  ,draw opacity=1 ]   (583.04,157.34) -- (500.36,200.4) ;
\draw [color={rgb, 255:red, 0; green, 0; blue, 0 }  ,draw opacity=1 ]   (557.2,157.34) -- (531.37,131.5) ;
\draw [color={rgb, 255:red, 0; green, 0; blue, 0 }  ,draw opacity=1 ]   (557.2,157.34) -- (531.37,157.34) ;
\draw [color={rgb, 255:red, 0; green, 0; blue, 0 }  ,draw opacity=1 ]   (531.37,131.5) -- (500.36,131.5) ;
\draw [color={rgb, 255:red, 0; green, 0; blue, 0 }  ,draw opacity=1 ]   (531.37,157.34) -- (500.36,157.34) ;
\draw    (583.04,157.34) -- (557.2,157.34) ;
\draw [shift={(557.2,157.34)}, rotate = 180] [color={rgb, 255:red, 0; green, 0; blue, 0 }  ][fill={rgb, 255:red, 0; green, 0; blue, 0 }  ][line width=0.75]      (0, 0) circle [x radius= 1.34, y radius= 1.34]   ;
\draw [shift={(583.04,157.34)}, rotate = 180] [color={rgb, 255:red, 0; green, 0; blue, 0 }  ][fill={rgb, 255:red, 0; green, 0; blue, 0 }  ][line width=0.75]      (0, 0) circle [x radius= 1.34, y radius= 1.34]   ;
\draw    (593.38,157.34) -- (583.04,157.34) ;
\draw [shift={(583.04,157.34)}, rotate = 180] [color={rgb, 255:red, 0; green, 0; blue, 0 }  ][fill={rgb, 255:red, 0; green, 0; blue, 0 }  ][line width=0.75]      (0, 0) circle [x radius= 1.34, y radius= 1.34]   ;
\draw    (598.19,63.85) .. controls (623.38,63.02) and (619.01,156.96) .. (593.38,157.34) ;
\draw (527.81,44.51) node [anchor=north west][inner sep=0.75pt]  [font=\tiny]  {$R_{1}$};
\draw (527.81,68.62) node [anchor=north west][inner sep=0.75pt]  [font=\tiny]  {$R_{2}$};
\draw (527.81,96.19) node [anchor=north west][inner sep=0.75pt]  [font=\tiny]  {$R_{3}$};
\draw (559.85,42.51) node [anchor=north west][inner sep=0.75pt]  [font=\tiny]  {$R_{4}$};
\draw (487.19,35.92) node [anchor=north west][inner sep=0.75pt]  [font=\tiny]  {$i$};
\draw (487.19,63.48) node [anchor=north west][inner sep=0.75pt]  [font=\tiny]  {$j$};
\draw (487.19,104.83) node [anchor=north west][inner sep=0.75pt]  [font=\tiny]  {$k$};
\draw (528.02,138.78) node [anchor=north west][inner sep=0.75pt]  [font=\tiny]  {$S_1$};
\draw (528.02,162.89) node [anchor=north west][inner sep=0.75pt]  [font=\tiny]  {$S_2$};
\draw (528.02,190.46) node [anchor=north west][inner sep=0.75pt]  [font=\tiny]  {$S_3$};
\draw (560.06,138.78) node [anchor=north west][inner sep=0.75pt]  [font=\tiny]  {$S_4$};
\draw (487.19,128.94) node [anchor=north west][inner sep=0.75pt]  [font=\tiny]  {$p$};
\draw (487.19,156.5) node [anchor=north west][inner sep=0.75pt]  [font=\tiny]  {$q$};
\draw (487.19,197.85) node [anchor=north west][inner sep=0.75pt]  [font=\tiny]  {$r$};
\draw  [fill={rgb, 255:red, 255; green, 255; blue, 255 }  ,fill opacity=1 ]  (608,99.92) -- (626,99.92) -- (626,113.92) -- (608,113.92) -- cycle  ;
\draw (609,100.92) node [anchor=north west][inner sep=0.75pt]  [font=\tiny]  {$\gamma $};
\draw (625,100.92) node [anchor=north west][inner sep=0.75pt]  [font=\tiny]  {$\Lambda $};
\end{tikzpicture}}} = \vcenter{\hbox{
\begin{tikzpicture}[x=0.75pt,y=0.75pt,yscale=-1,xscale=1]
\draw [color={rgb, 255:red, 0; green, 0; blue, 0 }  ,draw opacity=1 ]   (210,316.92) -- (117.17,360.46) ;
\draw [color={rgb, 255:red, 0; green, 0; blue, 0 }  ,draw opacity=1 ]   (174.01,317.39) -- (148.17,291.55) ;
\draw [color={rgb, 255:red, 0; green, 0; blue, 0 }  ,draw opacity=1 ]   (174.01,317.39) -- (148.17,317.39) ;
\draw [color={rgb, 255:red, 0; green, 0; blue, 0 }  ,draw opacity=1 ]   (148.17,291.55) -- (117.17,291.55) ;
\draw [color={rgb, 255:red, 0; green, 0; blue, 0 }  ,draw opacity=1 ]   (148.17,317.39) -- (117.17,317.39) ;
\draw    (210,316.92) -- (174.01,317.39) ;
\draw [shift={(174.01,317.39)}, rotate = 179.25] [color={rgb, 255:red, 0; green, 0; blue, 0 }  ][fill={rgb, 255:red, 0; green, 0; blue, 0 }  ][line width=0.75]      (0, 0) circle [x radius= 1.34, y radius= 1.34]   ;
\draw [shift={(210,316.92)}, rotate = 179.25] [color={rgb, 255:red, 0; green, 0; blue, 0 }  ][fill={rgb, 255:red, 0; green, 0; blue, 0 }  ][line width=0.75]      (0, 0) circle [x radius= 1.34, y radius= 1.34]   ;
\draw    (215,316.92) -- (210,316.92) ;
\draw [shift={(210,316.92)}, rotate = 180] [color={rgb, 255:red, 0; green, 0; blue, 0 }  ][fill={rgb, 255:red, 0; green, 0; blue, 0 }  ][line width=0.75]      (0, 0) circle [x radius= 1.34, y radius= 1.34]   ;
\draw [color={rgb, 255:red, 0; green, 0; blue, 0 }  ,draw opacity=1 ]   (199.85,410.41) -- (117.17,453.48) ;
\draw [color={rgb, 255:red, 0; green, 0; blue, 0 }  ,draw opacity=1 ]   (174.01,410.41) -- (148.17,384.57) ;
\draw [color={rgb, 255:red, 0; green, 0; blue, 0 }  ,draw opacity=1 ]   (174.01,410.41) -- (148.17,410.41) ;
\draw [color={rgb, 255:red, 0; green, 0; blue, 0 }  ,draw opacity=1 ]   (148.17,384.57) -- (117.17,384.57) ;
\draw [color={rgb, 255:red, 0; green, 0; blue, 0 }  ,draw opacity=1 ]   (148.17,410.41) -- (117.17,410.41) ;
\draw    (199.85,410.41) -- (174.01,410.41) ;
\draw [shift={(174.01,410.41)}, rotate = 180] [color={rgb, 255:red, 0; green, 0; blue, 0 }  ][fill={rgb, 255:red, 0; green, 0; blue, 0 }  ][line width=0.75]      (0, 0) circle [x radius= 1.34, y radius= 1.34]   ;
\draw [shift={(199.85,410.41)}, rotate = 180] [color={rgb, 255:red, 0; green, 0; blue, 0 }  ][fill={rgb, 255:red, 0; green, 0; blue, 0 }  ][line width=0.75]      (0, 0) circle [x radius= 1.34, y radius= 1.34]   ;
\draw    (210.19,410.41) -- (199.85,410.41) ;
\draw [shift={(199.85,410.41)}, rotate = 180] [color={rgb, 255:red, 0; green, 0; blue, 0 }  ][fill={rgb, 255:red, 0; green, 0; blue, 0 }  ][line width=0.75]      (0, 0) circle [x radius= 1.34, y radius= 1.34]   ;
\draw    (215,316.92) .. controls (240.19,316.1) and (235.82,410.03) .. (210.19,410.41) ;
\draw (144.62,297.59) node [anchor=north west][inner sep=0.75pt]  [font=\tiny]  {$R_{1}$};
\draw (144.62,321.7) node [anchor=north west][inner sep=0.75pt]  [font=\tiny]  {$R_{2}$};
\draw (144.62,349.26) node [anchor=north west][inner sep=0.75pt]  [font=\tiny]  {$R_{3}$};
\draw (176.66,295.59) node [anchor=north west][inner sep=0.75pt]  [font=\tiny]  {$R_{4}$};
\draw (104,289) node [anchor=north west][inner sep=0.75pt]  [font=\tiny]  {$i$};
\draw (104,316.56) node [anchor=north west][inner sep=0.75pt]  [font=\tiny]  {$j$};
\draw (104,357.9) node [anchor=north west][inner sep=0.75pt]  [font=\tiny]  {$k$};
\draw (144.83,391.86) node [anchor=north west][inner sep=0.75pt]  [font=\tiny]  {$S_1$};
\draw (144.83,415.97) node [anchor=north west][inner sep=0.75pt]  [font=\tiny]  {$S_2$};
\draw (144.83,443.53) node [anchor=north west][inner sep=0.75pt]  [font=\tiny]  {$S_3$};
\draw (176.87,391.86) node [anchor=north west][inner sep=0.75pt]  [font=\tiny]  {$S_4$};
\draw (104,382.02) node [anchor=north west][inner sep=0.75pt]  [font=\tiny]  {$p$};
\draw (104,409.58) node [anchor=north west][inner sep=0.75pt]  [font=\tiny]  {$q$};
\draw (104,450.92) node [anchor=north west][inner sep=0.75pt]  [font=\tiny]  {$r$};
\draw  [fill={rgb, 255:red, 255; green, 255; blue, 255 }  ,fill opacity=1 ]  (224.81,353) -- (242.81,353) -- (242.81,368) -- (224.81,368) -- cycle  ;
\draw (225.81,354) node [anchor=north west][inner sep=0.75pt]  [font=\tiny]  {$T_{2}$};
\draw (241.81,354) node [anchor=north west][inner sep=0.75pt]  [font=\tiny]  {$\Lambda $};
\end{tikzpicture}}},
\end{align}
where we have made repeated use of the equivariance property \eqref{eq:CG_gamma_pull}. Because the last equation involves an irreducible representation of a central element it evaluates to multiplication of a normalized character and we get
\begin{equation}
	(T_2^{(111)} Q^{G_{\vec{R}}^\Lambda G_{\vec{S}}^\Lambda} )_{ijk,pqr}= \hat{\chi}^{\Lambda}(T_2)Q^{G_{\vec{R}}^\Lambda G_{\vec{S}}^\Lambda}_{ijk,pqr} \, . \label{eq: eigeneq1}
\end{equation}

Similar diagrammatic manipulations show that
\begin{align}
	(T_2^{(110)} Q^{G_{\vec{R}}^\Lambda G_{\vec{S}}^\Lambda} )_{ijk,pqr} = \hat{\chi}^{R_4}(T_2)Q^{G_{\vec{R}}^\Lambda G_{\vec{S}}^\Lambda}_{ijk,pqr} \, ,\label{eq: eigeneq2} \\
	(T_2^{(100)} Q^{G_{\vec{R}}^\Lambda G_{\vec{S}}^\Lambda} )_{ijk,pqr} = \hat{\chi}^{R_1}(T_2)Q^{G_{\vec{R}}^\Lambda G_{\vec{S}}^\Lambda}_{ijk,pqr} \, , \label{eq: eigeneq3}\\
	(T_2^{(010)} Q^{G_{\vec{R}}^\Lambda G_{\vec{S}}^\Lambda} )_{ijk,pqr} = \hat{\chi}^{R_2}(T_2)Q^{G_{\vec{R}}^\Lambda G_{\vec{S}}^\Lambda}_{ijk,pqr} \, , \label{eq: eigeneq4}\\
	(T_2^{(001)} Q^{G_{\vec{R}}^\Lambda G_{\vec{S}}^\Lambda} )_{ijk,pqr} = \hat{\chi}^{R_3}(T_2)Q^{G_{\vec{R}}^\Lambda G_{\vec{S}}^\Lambda}_{ijk,pqr} \, . \label{eq: eigeneq5}
\end{align}
The five eigenvalue equations \eqref{eq: eigeneq1} - \eqref{eq: eigeneq5} determine the right eigenspace in $\End(V_D^{\otimes 3})$ labelled by the graph $G_{\vec{R}}^{\Lambda}$. The left eigenspace labelled by graph $G_{\vec{S}}^{\Lambda}$ is determined through right action,
\begin{align}
	(Q^{G_{\vec{R}}^\Lambda G_{\vec{S}}^\Lambda} T_2^{(110)})_{ijk,pqr}  = \hat{\chi}^{S_4}(T_2)Q^{G_{\vec{R}}^\Lambda G_{\vec{S}}^\Lambda}_{ijk,pqr} \, , \label{eq: eigeneq6} \\
	 (Q^{G_{\vec{R}}^\Lambda G_{\vec{S}}^\Lambda} T_2^{(100)})_{ijk,pqr} = \hat{\chi}^{S_1}(T_2)Q^{G_{\vec{R}}^\Lambda G_{\vec{S}}^\Lambda}_{ijk,pqr} \, , \label{eq: eigeneq7}\\
	(Q^{G_{\vec{R}}^\Lambda G_{\vec{S}}^\Lambda} T_2^{(010)}  )_{ijk,pqr} = \hat{\chi}^{S_2}(T_2)Q^{G_{\vec{R}}^\Lambda G_{\vec{S}}^\Lambda}_{ijk,pqr} \, , \label{eq: eigeneq8}\\
	(Q^{G_{\vec{R}}^\Lambda G_{\vec{S}}^\Lambda} T_2^{(001)} )_{ijk,pqr}  = \hat{\chi}^{S_3}(T_2)Q^{G_{\vec{R}}^\Lambda G_{\vec{S}}^\Lambda}_{ijk,pqr} \, . \label{eq: eigeneq9}
\end{align}
These are eigenvector equations in a $D^6$-dimensional vector space $\End(V_D^{\otimes 3})$, and solving them directly for large $D$ would be hopeless. Luckily, we know that $Q^{G_{\vec{R}}^\Lambda G_{\vec{S}}^\Lambda}$, as well as the $T_2^{(b)}$ operators (eq  \eqref{eq: T_2 1} - \eqref{eq: T_2 5}) lie in a much smaller subspace $\End_{S_D}(V_D^{\otimes 3})$. Further, its dimension is independent of $D$ for $D \geq 6$. $\End_{S_D}(V_D^{\otimes 3})$ is a subalgebra. The composition of operators is closed in $\End_{S_D}(V_D^{\otimes 3})$. This greatly reduces the dimensionality of the problem, what remains is to give an explicit basis for $\End_{S_D}(V_D^{\otimes 3})$ in which the above eigenvector problem can be solved. This basis is given by the partition algebra, which we shall now briefly introduce. For a more detailed exposition the reader should consult \cite{PartitionAlgebras}.

\subsection{Review: partition algebra}
\label{sec: partition algebra}
The partition algebras $P_m(D)$ are a family of finite-dimensional diagram algebras, meaning they have a basis labelled by diagrams, where multiplication is defined combinatorially in terms of diagram concatenation. In particular, $\Dim \, P_m(D) = B(2m)$ the Bell numbers, which count the number of set partitions of the set $\{1, \dots, 2m\}$. For example
\begin{equation}
	\PAdiagrambig[3/2]{3}{1/-1,-2/2} \, , \quad \PAdiagrambig[3/2,-2/-3]{3}{-2/2} \, , \quad \PAdiagrambig[]{3}{-2/1,-1/3,-3/2} \,, 
\end{equation}
form a subset of the $B(6) ={203}$ basis elements in $P_3(D)$.

It is known \cite{Martin, Jones} that
\begin{equation}
	\End_{S_D}( V_D^{{\otimes 3}}) \cong P_3(D) \, ,
\end{equation}
for $D \geq 2k$, such that every diagram corresponds to an element in $\End_{S_D}(V_D^{\otimes 3})$. The correspondence is simple, an edge connecting a vertex $i$ to a vertex $j$ is mapped to $\delta_{ij}$. For example,
\begin{align}
	&\PAdiagrambig[3/2]{3}{1/-1,-2/2}(\Phi_{ijk}) = \sum_{p,q,r} \delta_{ip} \delta_{jq} \delta_{qr} \Phi_{pqr} = \Phi_{ijj} \, , \\
	&\PAdiagrambig[3/2,-2/-3]{3}{-2/2}(\Phi_{ijk}) = \sum_{p,q,r} \delta_{jq}\delta_{qr}\delta_{jk} \Phi_{pqr} = D \delta_{jk} \Phi_{ijj} \, , \\
	&\PAdiagrambig[]{3}{-2/1,-1/3,-3/2}(\Phi_{ijk}) = \sum_{p,q,r} \delta_{ir} \delta_{jp} \delta_{kq} \Phi_{pqr} = \Phi_{jki} \, .
\end{align}
This isomorphism implies that the maps $Q^{G_{\vec{R}}^\Lambda G_{\vec{S}}^\Lambda}$ can be thought of as linear combinations of diagrams. In fact, the set of all $Q$'s form a basis of $P_3(D)$ as 
\begin{align}
\text{Span}(Q^{\Lambda, \alpha \beta}) = \text{End}_{S_D}(V_D^{\otimes 3}) \cong P_3(D) \, .
\end{align}
Indeed, given the Schur-Weyl result \cite{Martin1994, Martin1996, PartitionAlgebras}
\begin{align}
V_D^{\otimes 3} \cong \bigoplus_{\Lambda} V^{\Lambda}_{S_D} \otimes V^{\Lambda}_{P_3(D)} \, , 
\end{align}
and the $S_D$ irreducible decomposition of $V_D^{\otimes 3}$ in \eqref{eq: VD x VD x VD Decomposition}, there are
\begin{align}
5^2 + 10^2 + 6^2 + 6^2 + 1^2 + 2^2 + 1^2 = 203 = B(6) = \text{dim}(P_3(D))
\end{align}
$Q$'s i.e. 203 choices of the set of labels $\Lambda, \alpha, \beta$.

As we will now see, the operators $T_2^{(b)}$ also correspond to elements in $P_3(D)$. Therefore, the eigenvector equations \eqref{eq: eigeneq1}-\eqref{eq: eigeneq9} can be understood completely in this greatly reduced space $P_3(D)$.

\subsection{$\mathbb{C}(S_D)$ central elements as partition algebra elements} \label{subsec: T2 dual}
Since the operators $T_2^{(b)}$ commute with the action of $S_D$ on $V_D^{\otimes 3}$ there exist elements $\bar{T}_2^{(b)} \in P_3(D)$ such that
\begin{equation}
	\bar{T}_2^{(b)}(\Phi_{ijk}) = T_2^{(b)}(\Phi_{ijk}) \, .
\end{equation}
The expansion of $\bar{T}_2^{(b)}$ in terms of diagrams can found using Jucys-Murphy elements \cite{PartitionAlgebras, Enyang} as we now explain.
For every $P_m(D)$, there exists a family of Jucys-Murphy elements
\begin{equation}
	L_{i} \in P_m(D) \, , \quad i = 0 \, , \, \frac{1}{2} \, , \, 1 \, , \,  \frac{3}{2} \, , \, 2 \, , \, \dots
\end{equation}
that is recursively defined in section 3 of \cite{Enyang}. Theorem 3.10 defines the following central element in $P_m(D)$
\begin{equation}
	z_{m} = L_{\tfrac{1}{2}} + L_1 + L_{\tfrac{3}{2}} + \dots + L_{m} \, .
\end{equation}
According to Proposition 5.4
\begin{equation} \label{eq: central PA element z}
	z_m = \sum_{i <j} (ij)^{\otimes m} - \qty(\binom{D}{2}-mD)\idn \, ,
\end{equation}
as operators on $V_D^{\otimes m}$.
In particular, rearranging \eqref{eq: central PA element z} and taking $m = 1$ we have
\begin{equation}
	\sum_{i<j} (ij) = L_{\tfrac{1}{2}} + L_1 + \binom{D}{2}\idn-D\idn \equiv \bar{T}_2^{(1)} \in P_{1}(D) \,,
\end{equation}
as operators on $V_D$,
\begin{equation}
	 \sum_{i<j} (ij)^{\otimes 2} = L_{\tfrac{1}{2}} + L_1 + L_{\tfrac{3}{2}} + L_{2} + \binom{D}{2}\idn-2D\idn \equiv \bar{T}_2^{(11)} \in P_2(D) \, ,
\end{equation}
as operators on $V_D^{\otimes 2}$ and
\begin{equation}
	\sum_{i<j} (ij)^{\otimes 3}= L_{\tfrac{1}{2}} +L_1 + L_{\tfrac{3}{2}} + L_{2} + L_{\tfrac{5}{2}} + L_{3}  + \binom{D}{2}\idn-3D\idn \equiv \bar{T}_2^{(111)} \in P_3(D) \,,
\end{equation}
as operators on $V_D^{\otimes 3}$. The partition algebras $P_1(D), \, P_2(D)$ are embeddable into $P_3(D)$ by adding identity strands i.e. they act trivially on all but one and two factors of $V^{\otimes 3}_D$ respectively. With the definitions
\begin{align}
	&\bar{T}_2^{(100)} = \bar{T}_2^{(1)} \otimes \idn \otimes \idn \, ,  \\
	&\bar{T}_2^{(010)} = \idn \otimes \bar{T}_2^{(1)} \otimes \idn \, , \\
	&\bar{T}_2^{(001)} = \idn \otimes \idn \otimes \bar{T}_2^{(1)} \, , \\
	&\bar{T}_2^{(110)} = \bar{T}_2^{(11)} \otimes \idn \, .
\end{align}
we find that
\begin{align}
	&\bar{T}_2^{(100)} = {T}_2^{(100)} \, , \\
	&\bar{T}_2^{(010)} = {T}_2^{(010)} \, , \\
	&\bar{T}_2^{(001)} = {T}_2^{(001)} \, , \\
	&\bar{T}_2^{(110)} = {T}_2^{(110)} \, , \\
	&\bar{T}_2^{(111)} = {T}_2^{(111)} \, .
\end{align}
as operators on $V_D^{\otimes 3}$.
For fixed $D$, it is straight-forward to solve the eigenvector equations in \eqref{eq: eigeneq1}-\eqref{eq: eigeneq8} using partition algebras. But we are interested in all $D$ constructions and this will be the subject of the next section.

\subsection{All $D$ construction of invariant endomorphism tensors} \label{sec: Q construction}
In this section, we describe an all $D \geq 6$ construction of the $B(6)=203$ Q's. In other words, we give functions
\begin{equation}
	Q^{G_{\vec{R}}^\Lambda G_{\vec{S}}^\Lambda}(D) \in P_3(D)
\end{equation}
which correspond to the invariant endomorphism tensors appearing in the two-point function \eqref{eq: 2pt schematic} for all $D$.

It will be useful to define the following set of idempotents in $P_3(D)$
\begin{equation}
	\begin{aligned}
  P_{R_1}	&= \prod_{R \neq R_1} \frac{(\bar{T}_{2}^{(100)}-\hat{\chi}^{R}(T_{2}^{(100)}))}{(\hat{\chi}^{R_1}(T_{2}^{(100)})-\hat{\chi}^{R}(T_{2}^{(100)}))} \, , \\
	P_{R_2}	&= \prod_{R \neq R_2} \frac{(\bar{T}_{2}^{(010)}-\hat{\chi}^{R}(T_{2}^{(010)}))}{(\hat{\chi}^{R_2}(T_{2}^{(010)})-\hat{\chi}^{R}(T_{2}^{(010)}))} \, , \\
	P_{R_3}	&= \prod_{R \neq R_3}\frac{(\bar{T}_{2}^{(001)}-\hat{\chi}^{R}(T_{2}^{(001)}))}{(\hat{\chi}^{R_3}(T_{2}^{(001)})-\hat{\chi}^{R}(T_{2}^{(001)}))} \, , \\
	P_{R_4}	&= \prod_{R \neq R_4} \frac{(\bar{T}_{2}^{(110)}-\hat{\chi}^{R}(T_{2}^{(110)}))}{(\hat{\chi}^{R_4}(T_{2}^{(110)})-\hat{\chi}^{R}(T_{2}^{(110)}))} \, , \\
	P_{\Lambda}	&= \prod_{R \neq \Lambda} \frac{(\bar{T}_{2}^{(111)}-\hat{\chi}^{R}(T_{2}^{(111)}))}{(\hat{\chi}^{\Lambda}(T_{2}^{(111)})-\hat{\chi}^{R}(T_{2}^{(111)}))} \, ,
\end{aligned} \label{eq: projectors}
\end{equation}
where
\begin{align} \nonumber
	&R_1, R_2, R_3 \in \{[D], [D-1,1]\} \, ,\\ \nonumber
	&R_4 \in \{[D], [D-1,1], [D-2,2], [D-2,1,1]\} \, ,\\
	&\Lambda \in \{[D], [D-1,1], [D-2,2], [D-2,1,1], [D-3,3], [D-3,2,1], [D-3,1,1,1]\} \, .
\end{align}
Given a graph $G_{\vec{R}}^\Lambda$ we construct the element
\begin{equation}
	P_{G_{\vec{R}}^{\Lambda}} = P_{\Lambda}P_{R_4}P_{R_3}P_{R_2}P_{R_1} \in P_3(D) \, , \label{eq: graph projector}
\end{equation}
and for a pair of graphs $(G_{\vec{R}}^\Lambda, G_{\vec{S}}^\Lambda)$ define the following projector on $P_3(D)$
\begin{equation}
	P_{G_{\vec{R}}^{\Lambda}, G_{\vec{S}}^{{\Lambda}}}(d) = P_{G_{\vec{S}}^{\Lambda}}dP_{G_{\vec{S}}^{\Lambda}}, \quad \forall \, d \in P_3(D) \, .
\end{equation}
It projects onto the simultaneous left and right eigenspaces of $\bar{T}_2^{(b)}$ in $P_3(D)$, determined by the pair of graphs.
Since the eigenspace is one-dimensional we have
\begin{equation}
	\mathrm{Im} \, P_{G_{\vec{R}}^{\Lambda}, G_{\vec{S}}^{{\Lambda}}} = \Span(Q^{G_{\vec{R}}^{\Lambda}, G_{\vec{S}}^{{\Lambda}}}) \, .
\end{equation}


In general, a basis for the image of a matrix can be determined from its row echelon form \cite[\textbf{2O} in Section 2.4]{strang2006linear}. This is reviewed in appendix \ref{apx: pivots} for the convenience of the reader. We will use this to find $Q^{G_{\vec{R}}^{\Lambda}, G_{\vec{S}}^{{\Lambda}}}$.
Let $\mathcal{B}$ be a basis for $P_3(D)$ (e.g. the diagram basis).
We define the matrix $(P_{G_{\vec{R}}^{\Lambda}, G_{\vec{S}}^{{\Lambda}}})_{d'd}$ in this basis by
\begin{equation}
	P_{G_{\vec{R}}^{\Lambda}, G_{\vec{S}}^{{\Lambda}}}(d) = \sum_{d' \in \mathcal{B}} (P_{G_{\vec{R}}^{\Lambda}, G_{\vec{S}}^{{\Lambda}}})_{d'd} d' \, .
\end{equation}
Since $\mathrm{Im} \, P_{G_{\vec{R}}^{\Lambda}, G_{\vec{S}}^{{\Lambda}}}$ is one-dimensional, and a basis for the image is determined by the pivot column of the row echelon form, we just have to find the first non-zero column in the first row of the echelon form.
That is, suppose
\begin{equation}
	P_{G_{\vec{R}}^{\Lambda}, G_{\vec{S}}^{{\Lambda}}} = E \widetilde{P}_{G_{\vec{R}}^{\Lambda}, G_{\vec{S}}^{{\Lambda}}} \, ,
\end{equation}
where $\widetilde{P}_{G_{\vec{R}}^{\Lambda}, G_{\vec{S}}^{{\Lambda}} }$ is in row echelon form and $E$ is an invertible matrix encoding the Gauss elimination steps. Let $(\widetilde{P}_{G_{\vec{R}}^{\Lambda}, G_{\vec{S}}^{{\Lambda}} })_{1d}$ be the pivot element then
\begin{equation}
	\sum_{d' \in \mathcal{B}} ({P}_{G_{\vec{R}}^{\Lambda}, G_{\vec{S}}^{{\Lambda}}})_{d'd} d' \propto Q^{G_{\vec{R}}^{\Lambda} \, , G_{\vec{S}}^{{\Lambda}}}.
\end{equation}

Because the matrix $P_{G_{\vec{R}}^{\Lambda}, G_{\vec{S}}^{{\Lambda}}}$ has rational functions as entries, it is non-trivial to find the row echelon form $\widetilde{P}_{G_{\vec{R}}^{\Lambda},  G_{\vec{S}}^{{\Lambda}}}$ and subsequent pivot column. Instead, to find a basis for general $D$, we will use the following method. Fix $D$ to be any integer $n \geq 6$ and find the pivot column of
\begin{equation}
	\left.P_{G_{\vec{R}}^{\Lambda}, G_{\vec{S}}^{{\Lambda}}}\right\vert_{D=n} \, .
\end{equation}
Let $d$ be the pivot column of the echelon form for $D=n$, then
\begin{equation}
	\sum_{d' \in \mathcal{B}} ({P}_{G_{\vec{R}}^{\Lambda}, G_{\vec{S}}^{{\Lambda}}})_{d'd} d' = Q^{G_{\vec{R}}^{\Lambda}, G_{\vec{S}}^{{\Lambda}}} \, .
\end{equation}
The relevant Sage code for implementing this algorithm can be found together with the arXiv version of this paper.

In practice we worked with $D = 7$ and verified that the resulting $Q^{G_{\vec{R}}^{\Lambda}, G_{\vec{S}}^{{\Lambda}}}$ satisfies the correct eigenvalue equations (e.g. \eqref{eq: T_2 1}-\eqref{eq: T_2 5}) for all $D$. Theoretically, its validity can be argued for as follows. Suppose we were able to find the row echelon form
\begin{equation}
	\widetilde{P}_{G_{\vec{R}}^{\Lambda},  G_{\vec{S}}^{{\Lambda}}} \, ,
\end{equation}
for general $D$. This matrix will have a single non-zero row, and we call the pivot column $d$. The pivot entry is a rational function
\begin{equation}
	f(D) = (\widetilde{P}_{G_{\vec{R}}^{\Lambda}, G_{\vec{S}}^{{\Lambda}}})_{1d} \, .
\end{equation}
Because it is a rational function, it has a finite number of zeros and poles. Away from these points we can consider the reduced row echelon form
\begin{equation}
	\frac{1}{f(D)}\widetilde{P}_{G_{\vec{R}}^{\Lambda},  G_{\vec{S}}^{{\Lambda}}} \, ,
\end{equation}
whose pivot element is $1$, and in particular independent of $D$. Therefore, away from the poles and zeros of $f(D)$, we can argue that the pivot column is independent of $D$.


\subsection{Examples of invariant endomorphism tensors}
In this subsection we give some examples of invariant tensors, represented by linear combinations of partition diagrams. More examples can be computed using the SageMath code accompanying the arXiv version of this paper.

For example, with 
\begin{equation}
	\vec{R}=\vec{S}= (V_{[D-1,1]},V_{[D-1,1]},V_{[D-1,1]},V_{[D-2,1,1]}) \, , \quad \Lambda = V_{[D-1,1]} \, , \label{eq: graph ex sec5}
\end{equation}
the tensor $(Q^{G_{\vec{R}}^{\Lambda} G_{\vec{S}}^{\Lambda}})_{i_1 i_2 i_3, i_4i_5 i_6}$ has the following expansion in terms of diagrams
\begin{equation}
\begin{aligned}
  Q^{G_{\vec{R}}^{\Lambda} G_{\vec{S}}^{\Lambda}} = \frac{1}{D^{2}} &\begin{tikzpicture}[scale = 0.25,thick, baseline={(0,-1ex/2)}] 
\tikzstyle{vertex} = [shape = circle, minimum size = 1pt, inner sep = 0pt] 
\node[vertex] (G--3) at (3.0, -1) [shape = circle, draw] {}; 
\node[vertex] (G--2) at (1.5, -1) [shape = circle, draw] {}; 
\node[vertex] (G--1) at (0.0, -1) [shape = circle, draw] {}; 
\node[vertex] (G-1) at (0.0, 1) [shape = circle, draw] {}; 
\node[vertex] (G-2) at (1.5, 1) [shape = circle, draw] {}; 
\node[vertex] (G-3) at (3.0, 1) [shape = circle, draw] {}; 
\draw[] (G-1) .. controls +(0, -1) and +(0, 1) .. (G--1); 
\end{tikzpicture} -\frac{1}{D} \begin{tikzpicture}[scale = 0.25,thick, baseline={(0,-1ex/2)}] 
\tikzstyle{vertex} = [shape = circle, minimum size = 1pt, inner sep = 0pt] 
\node[vertex] (G--3) at (3.0, -1) [shape = circle, draw] {}; 
\node[vertex] (G--2) at (1.5, -1) [shape = circle, draw] {}; 
\node[vertex] (G--1) at (0.0, -1) [shape = circle, draw] {}; 
\node[vertex] (G-1) at (0.0, 1) [shape = circle, draw] {}; 
\node[vertex] (G-2) at (1.5, 1) [shape = circle, draw] {}; 
\node[vertex] (G-3) at (3.0, 1) [shape = circle, draw] {}; 
\draw[] (G-1) .. controls +(0, -1) and +(0, 1) .. (G--1); 
\draw[] (G-2) .. controls +(0.5, -0.5) and +(-0.5, -0.5) .. (G-3); 
\end{tikzpicture} - \frac{1}{D^{2}} \begin{tikzpicture}[scale = 0.25,thick, baseline={(0,-1ex/2)}] 
\tikzstyle{vertex} = [shape = circle, minimum size = 1pt, inner sep = 0pt] 
\node[vertex] (G--3) at (3.0, -1) [shape = circle, draw] {}; 
\node[vertex] (G--2) at (1.5, -1) [shape = circle, draw] {}; 
\node[vertex] (G--1) at (0.0, -1) [shape = circle, draw] {}; 
\node[vertex] (G-2) at (1.5, 1) [shape = circle, draw] {}; 
\node[vertex] (G-1) at (0.0, 1) [shape = circle, draw] {}; 
\node[vertex] (G-3) at (3.0, 1) [shape = circle, draw] {}; 
\draw[] (G-2) .. controls +(-0.75, -1) and +(0.75, 1) .. (G--1); 
\end{tikzpicture} + \frac{1}{D} \begin{tikzpicture}[scale = 0.25,thick, baseline={(0,-1ex/2)}] 
\tikzstyle{vertex} = [shape = circle, minimum size = 1pt, inner sep = 0pt] 
\node[vertex] (G--3) at (3.0, -1) [shape = circle, draw] {}; 
\node[vertex] (G--2) at (1.5, -1) [shape = circle, draw] {}; 
\node[vertex] (G--1) at (0.0, -1) [shape = circle, draw] {}; 
\node[vertex] (G-2) at (1.5, 1) [shape = circle, draw] {}; 
\node[vertex] (G-1) at (0.0, 1) [shape = circle, draw] {}; 
\node[vertex] (G-3) at (3.0, 1) [shape = circle, draw] {}; 
\draw[] (G-2) .. controls +(-0.75, -1) and +(0.75, 1) .. (G--1); 
\draw[] (G-1) .. controls +(0.6, -0.6) and +(-0.6, -0.6) .. (G-3); 
\end{tikzpicture} - \frac{1}{D^{2}} \begin{tikzpicture}[scale = 0.25,thick, baseline={(0,-1ex/2)}] 
\tikzstyle{vertex} = [shape = circle, minimum size = 1pt, inner sep = 0pt] 
\node[vertex] (G--3) at (3.0, -1) [shape = circle, draw] {}; 
\node[vertex] (G--2) at (1.5, -1) [shape = circle, draw] {}; 
\node[vertex] (G-1) at (0.0, 1) [shape = circle, draw] {}; 
\node[vertex] (G--1) at (0.0, -1) [shape = circle, draw] {}; 
\node[vertex] (G-2) at (1.5, 1) [shape = circle, draw] {}; 
\node[vertex] (G-3) at (3.0, 1) [shape = circle, draw] {}; 
\draw[] (G-1) .. controls +(0.75, -1) and +(-0.75, 1) .. (G--2); 
\end{tikzpicture} + \frac{1}{D} \begin{tikzpicture}[scale = 0.25,thick, baseline={(0,-1ex/2)}] 
\tikzstyle{vertex} = [shape = circle, minimum size = 1pt, inner sep = 0pt] 
\node[vertex] (G--3) at (3.0, -1) [shape = circle, draw] {}; 
\node[vertex] (G--2) at (1.5, -1) [shape = circle, draw] {}; 
\node[vertex] (G-1) at (0.0, 1) [shape = circle, draw] {}; 
\node[vertex] (G--1) at (0.0, -1) [shape = circle, draw] {}; 
\node[vertex] (G-2) at (1.5, 1) [shape = circle, draw] {}; 
\node[vertex] (G-3) at (3.0, 1) [shape = circle, draw] {}; 
\draw[] (G-1) .. controls +(0.75, -1) and +(-0.75, 1) .. (G--2); 
\draw[] (G-2) .. controls +(0.5, -0.5) and +(-0.5, -0.5) .. (G-3); 
\end{tikzpicture} +\\
\frac{1}{D^{2}} &\begin{tikzpicture}[scale = 0.25,thick, baseline={(0,-1ex/2)}] 
\tikzstyle{vertex} = [shape = circle, minimum size = 1pt, inner sep = 0pt] 
\node[vertex] (G--3) at (3.0, -1) [shape = circle, draw] {}; 
\node[vertex] (G--2) at (1.5, -1) [shape = circle, draw] {}; 
\node[vertex] (G-2) at (1.5, 1) [shape = circle, draw] {}; 
\node[vertex] (G--1) at (0.0, -1) [shape = circle, draw] {}; 
\node[vertex] (G-1) at (0.0, 1) [shape = circle, draw] {}; 
\node[vertex] (G-3) at (3.0, 1) [shape = circle, draw] {}; 
\draw[] (G-2) .. controls +(0, -1) and +(0, 1) .. (G--2); 
\end{tikzpicture} -\frac{1}{D} \begin{tikzpicture}[scale = 0.25,thick, baseline={(0,-1ex/2)}] 
\tikzstyle{vertex} = [shape = circle, minimum size = 1pt, inner sep = 0pt] 
\node[vertex] (G--3) at (3.0, -1) [shape = circle, draw] {}; 
\node[vertex] (G--2) at (1.5, -1) [shape = circle, draw] {}; 
\node[vertex] (G-2) at (1.5, 1) [shape = circle, draw] {}; 
\node[vertex] (G--1) at (0.0, -1) [shape = circle, draw] {}; 
\node[vertex] (G-1) at (0.0, 1) [shape = circle, draw] {}; 
\node[vertex] (G-3) at (3.0, 1) [shape = circle, draw] {}; 
\draw[] (G-2) .. controls +(0, -1) and +(0, 1) .. (G--2); 
\draw[] (G-1) .. controls +(0.6, -0.6) and +(-0.6, -0.6) .. (G-3); 
\end{tikzpicture} -\frac{1}{D}\begin{tikzpicture}[scale = 0.25,thick, baseline={(0,-1ex/2)}] 
\tikzstyle{vertex} = [shape = circle, minimum size = 1pt, inner sep = 0pt] 
\node[vertex] (G--3) at (3.0, -1) [shape = circle, draw] {}; 
\node[vertex] (G--2) at (1.5, -1) [shape = circle, draw] {}; 
\node[vertex] (G--1) at (0.0, -1) [shape = circle, draw] {}; 
\node[vertex] (G-1) at (0.0, 1) [shape = circle, draw] {}; 
\node[vertex] (G-2) at (1.5, 1) [shape = circle, draw] {}; 
\node[vertex] (G-3) at (3.0, 1) [shape = circle, draw] {}; 
\draw[] (G--3) .. controls +(-0.5, 0.5) and +(0.5, 0.5) .. (G--2); 
\draw[] (G-2) .. controls +(0.5, -0.5) and +(-0.5, -0.5) .. (G-3); 
\end{tikzpicture} + 
\frac{1}{D} \begin{tikzpicture}[scale = 0.25,thick, baseline={(0,-1ex/2)}] 
\tikzstyle{vertex} = [shape = circle, minimum size = 1pt, inner sep = 0pt] 
\node[vertex] (G--3) at (3.0, -1) [shape = circle, draw] {}; 
\node[vertex] (G--2) at (1.5, -1) [shape = circle, draw] {}; 
\node[vertex] (G--1) at (0.0, -1) [shape = circle, draw] {}; 
\node[vertex] (G-1) at (0.0, 1) [shape = circle, draw] {}; 
\node[vertex] (G-3) at (3.0, 1) [shape = circle, draw] {}; 
\node[vertex] (G-2) at (1.5, 1) [shape = circle, draw] {}; 
\draw[] (G--3) .. controls +(-0.5, 0.5) and +(0.5, 0.5) .. (G--2); 
\draw[] (G-1) .. controls +(0.6, -0.6) and +(-0.6, -0.6) .. (G-3); 
\end{tikzpicture} -\frac{1}{D}\begin{tikzpicture}[scale = 0.25,thick, baseline={(0,-1ex/2)}] 
\tikzstyle{vertex} = [shape = circle, minimum size = 1pt, inner sep = 0pt] 
\node[vertex] (G--3) at (3.0, -1) [shape = circle, draw] {}; 
\node[vertex] (G--2) at (1.5, -1) [shape = circle, draw] {}; 
\node[vertex] (G--1) at (0.0, -1) [shape = circle, draw] {}; 
\node[vertex] (G-1) at (0.0, 1) [shape = circle, draw] {}; 
\node[vertex] (G-2) at (1.5, 1) [shape = circle, draw] {}; 
\node[vertex] (G-3) at (3.0, 1) [shape = circle, draw] {}; 
\draw[] (G--3) .. controls +(-0.5, 0.5) and +(0.5, 0.5) .. (G--2); 
\draw[] (G-1) .. controls +(0, -1) and +(0, 1) .. (G--1); 
\end{tikzpicture} + \begin{tikzpicture}[scale = 0.25,thick, baseline={(0,-1ex/2)}] 
\tikzstyle{vertex} = [shape = circle, minimum size = 1pt, inner sep = 0pt] 
\node[vertex] (G--3) at (3.0, -1) [shape = circle, draw] {}; 
\node[vertex] (G--2) at (1.5, -1) [shape = circle, draw] {}; 
\node[vertex] (G--1) at (0.0, -1) [shape = circle, draw] {}; 
\node[vertex] (G-1) at (0.0, 1) [shape = circle, draw] {}; 
\node[vertex] (G-2) at (1.5, 1) [shape = circle, draw] {}; 
\node[vertex] (G-3) at (3.0, 1) [shape = circle, draw] {}; 
\draw[] (G--3) .. controls +(-0.5, 0.5) and +(0.5, 0.5) .. (G--2); 
\draw[] (G-1) .. controls +(0, -1) and +(0, 1) .. (G--1); 
\draw[] (G-2) .. controls +(0.5, -0.5) and +(-0.5, -0.5) .. (G-3); 
\end{tikzpicture} + \\ \frac{1}{D} &\begin{tikzpicture}[scale = 0.25,thick, baseline={(0,-1ex/2)}] 
\tikzstyle{vertex} = [shape = circle, minimum size = 1pt, inner sep = 0pt] 
\node[vertex] (G--3) at (3.0, -1) [shape = circle, draw] {}; 
\node[vertex] (G--2) at (1.5, -1) [shape = circle, draw] {}; 
\node[vertex] (G--1) at (0.0, -1) [shape = circle, draw] {}; 
\node[vertex] (G-2) at (1.5, 1) [shape = circle, draw] {}; 
\node[vertex] (G-1) at (0.0, 1) [shape = circle, draw] {}; 
\node[vertex] (G-3) at (3.0, 1) [shape = circle, draw] {}; 
\draw[] (G--3) .. controls +(-0.5, 0.5) and +(0.5, 0.5) .. (G--2); 
\draw[] (G-2) .. controls +(-0.75, -1) and +(0.75, 1) .. (G--1); 
\end{tikzpicture} - \begin{tikzpicture}[scale = 0.25,thick, baseline={(0,-1ex/2)}] 
\tikzstyle{vertex} = [shape = circle, minimum size = 1pt, inner sep = 0pt] 
\node[vertex] (G--3) at (3.0, -1) [shape = circle, draw] {}; 
\node[vertex] (G--2) at (1.5, -1) [shape = circle, draw] {}; 
\node[vertex] (G--1) at (0.0, -1) [shape = circle, draw] {}; 
\node[vertex] (G-2) at (1.5, 1) [shape = circle, draw] {}; 
\node[vertex] (G-1) at (0.0, 1) [shape = circle, draw] {}; 
\node[vertex] (G-3) at (3.0, 1) [shape = circle, draw] {}; 
\draw[] (G--3) .. controls +(-0.5, 0.5) and +(0.5, 0.5) .. (G--2); 
\draw[] (G-2) .. controls +(-0.75, -1) and +(0.75, 1) .. (G--1); 
\draw[] (G-1) .. controls +(0.6, -0.6) and +(-0.6, -0.6) .. (G-3); 
\end{tikzpicture} + \frac{1}{D} \begin{tikzpicture}[scale = 0.25,thick, baseline={(0,-1ex/2)}] 
\tikzstyle{vertex} = [shape = circle, minimum size = 1pt, inner sep = 0pt] 
\node[vertex] (G--3) at (3.0, -1) [shape = circle, draw] {}; 
\node[vertex] (G--1) at (0.0, -1) [shape = circle, draw] {}; 
\node[vertex] (G--2) at (1.5, -1) [shape = circle, draw] {}; 
\node[vertex] (G-1) at (0.0, 1) [shape = circle, draw] {}; 
\node[vertex] (G-2) at (1.5, 1) [shape = circle, draw] {}; 
\node[vertex] (G-3) at (3.0, 1) [shape = circle, draw] {}; 
\draw[] (G--3) .. controls +(-0.6, 0.6) and +(0.6, 0.6) .. (G--1); 
\draw[] (G-2) .. controls +(0.5, -0.5) and +(-0.5, -0.5) .. (G-3); 
\end{tikzpicture} -\frac{1}{D} \begin{tikzpicture}[scale = 0.25,thick, baseline={(0,-1ex/2)}] 
\tikzstyle{vertex} = [shape = circle, minimum size = 1pt, inner sep = 0pt] 
\node[vertex] (G--3) at (3.0, -1) [shape = circle, draw] {}; 
\node[vertex] (G--1) at (0.0, -1) [shape = circle, draw] {}; 
\node[vertex] (G--2) at (1.5, -1) [shape = circle, draw] {}; 
\node[vertex] (G-1) at (0.0, 1) [shape = circle, draw] {}; 
\node[vertex] (G-3) at (3.0, 1) [shape = circle, draw] {}; 
\node[vertex] (G-2) at (1.5, 1) [shape = circle, draw] {}; 
\draw[] (G--3) .. controls +(-0.6, 0.6) and +(0.6, 0.6) .. (G--1); 
\draw[] (G-1) .. controls +(0.6, -0.6) and +(-0.6, -0.6) .. (G-3); 
\end{tikzpicture} + \frac{1}{D} \begin{tikzpicture}[scale = 0.25,thick, baseline={(0,-1ex/2)}] 
\tikzstyle{vertex} = [shape = circle, minimum size = 1pt, inner sep = 0pt] 
\node[vertex] (G--3) at (3.0, -1) [shape = circle, draw] {}; 
\node[vertex] (G--1) at (0.0, -1) [shape = circle, draw] {}; 
\node[vertex] (G--2) at (1.5, -1) [shape = circle, draw] {}; 
\node[vertex] (G-1) at (0.0, 1) [shape = circle, draw] {}; 
\node[vertex] (G-2) at (1.5, 1) [shape = circle, draw] {}; 
\node[vertex] (G-3) at (3.0, 1) [shape = circle, draw] {}; 
\draw[] (G--3) .. controls +(-0.6, 0.6) and +(0.6, 0.6) .. (G--1); 
\draw[] (G-1) .. controls +(0.75, -1) and +(-0.75, 1) .. (G--2); 
\end{tikzpicture} - \begin{tikzpicture}[scale = 0.25,thick, baseline={(0,-1ex/2)}] 
\tikzstyle{vertex} = [shape = circle, minimum size = 1pt, inner sep = 0pt] 
\node[vertex] (G--3) at (3.0, -1) [shape = circle, draw] {}; 
\node[vertex] (G--1) at (0.0, -1) [shape = circle, draw] {}; 
\node[vertex] (G--2) at (1.5, -1) [shape = circle, draw] {}; 
\node[vertex] (G-1) at (0.0, 1) [shape = circle, draw] {}; 
\node[vertex] (G-2) at (1.5, 1) [shape = circle, draw] {}; 
\node[vertex] (G-3) at (3.0, 1) [shape = circle, draw] {}; 
\draw[] (G--3) .. controls +(-0.6, 0.6) and +(0.6, 0.6) .. (G--1); 
\draw[] (G-1) .. controls +(0.75, -1) and +(-0.75, 1) .. (G--2); 
\draw[] (G-2) .. controls +(0.5, -0.5) and +(-0.5, -0.5) .. (G-3); 
\end{tikzpicture} -\frac{1}{D} \begin{tikzpicture}[scale = 0.25,thick, baseline={(0,-1ex/2)}] 
\tikzstyle{vertex} = [shape = circle, minimum size = 1pt, inner sep = 0pt] 
\node[vertex] (G--3) at (3.0, -1) [shape = circle, draw] {}; 
\node[vertex] (G--1) at (0.0, -1) [shape = circle, draw] {}; 
\node[vertex] (G--2) at (1.5, -1) [shape = circle, draw] {}; 
\node[vertex] (G-2) at (1.5, 1) [shape = circle, draw] {}; 
\node[vertex] (G-1) at (0.0, 1) [shape = circle, draw] {}; 
\node[vertex] (G-3) at (3.0, 1) [shape = circle, draw] {}; 
\draw[] (G--3) .. controls +(-0.6, 0.6) and +(0.6, 0.6) .. (G--1); 
\draw[] (G-2) .. controls +(0, -1) and +(0, 1) .. (G--2); 
\end{tikzpicture} + \begin{tikzpicture}[scale = 0.25,thick, baseline={(0,-1ex/2)}] 
\tikzstyle{vertex} = [shape = circle, minimum size = 1pt, inner sep = 0pt] 
\node[vertex] (G--3) at (3.0, -1) [shape = circle, draw] {}; 
\node[vertex] (G--1) at (0.0, -1) [shape = circle, draw] {}; 
\node[vertex] (G--2) at (1.5, -1) [shape = circle, draw] {}; 
\node[vertex] (G-2) at (1.5, 1) [shape = circle, draw] {}; 
\node[vertex] (G-1) at (0.0, 1) [shape = circle, draw] {}; 
\node[vertex] (G-3) at (3.0, 1) [shape = circle, draw] {}; 
\draw[] (G--3) .. controls +(-0.6, 0.6) and +(0.6, 0.6) .. (G--1); 
\draw[] (G-2) .. controls +(0, -1) and +(0, 1) .. (G--2); 
\draw[] (G-1) .. controls +(0.6, -0.6) and +(-0.6, -0.6) .. (G-3); 
\end{tikzpicture}
\end{aligned}
\end{equation}
For the remaining examples, we will only give the dominant diagrams in the limit $D \rightarrow \infty$. For instance, the dominant diagrams in the above equation are
\begin{equation}
  \begin{tikzpicture}[scale = 0.25,thick, baseline={(0,-1ex/2)}] 
\tikzstyle{vertex} = [shape = circle, minimum size = 1pt, inner sep = 0pt] 
\node[vertex] (G--3) at (3.0, -1) [shape = circle, draw] {}; 
\node[vertex] (G--2) at (1.5, -1) [shape = circle, draw] {}; 
\node[vertex] (G--1) at (0.0, -1) [shape = circle, draw] {}; 
\node[vertex] (G-1) at (0.0, 1) [shape = circle, draw] {}; 
\node[vertex] (G-2) at (1.5, 1) [shape = circle, draw] {}; 
\node[vertex] (G-3) at (3.0, 1) [shape = circle, draw] {}; 
\draw[] (G--3) .. controls +(-0.5, 0.5) and +(0.5, 0.5) .. (G--2); 
\draw[] (G-1) .. controls +(0, -1) and +(0, 1) .. (G--1); 
\draw[] (G-2) .. controls +(0.5, -0.5) and +(-0.5, -0.5) .. (G-3); 
\end{tikzpicture} - \begin{tikzpicture}[scale = 0.25,thick, baseline={(0,-1ex/2)}] 
\tikzstyle{vertex} = [shape = circle, minimum size = 1pt, inner sep = 0pt] 
\node[vertex] (G--3) at (3.0, -1) [shape = circle, draw] {}; 
\node[vertex] (G--2) at (1.5, -1) [shape = circle, draw] {}; 
\node[vertex] (G--1) at (0.0, -1) [shape = circle, draw] {}; 
\node[vertex] (G-2) at (1.5, 1) [shape = circle, draw] {}; 
\node[vertex] (G-1) at (0.0, 1) [shape = circle, draw] {}; 
\node[vertex] (G-3) at (3.0, 1) [shape = circle, draw] {}; 
\draw[] (G--3) .. controls +(-0.5, 0.5) and +(0.5, 0.5) .. (G--2); 
\draw[] (G-2) .. controls +(-0.75, -1) and +(0.75, 1) .. (G--1); 
\draw[] (G-1) .. controls +(0.6, -0.6) and +(-0.6, -0.6) .. (G-3); 
\end{tikzpicture} - \begin{tikzpicture}[scale = 0.25,thick, baseline={(0,-1ex/2)}] 
\tikzstyle{vertex} = [shape = circle, minimum size = 1pt, inner sep = 0pt] 
\node[vertex] (G--3) at (3.0, -1) [shape = circle, draw] {}; 
\node[vertex] (G--1) at (0.0, -1) [shape = circle, draw] {}; 
\node[vertex] (G--2) at (1.5, -1) [shape = circle, draw] {}; 
\node[vertex] (G-1) at (0.0, 1) [shape = circle, draw] {}; 
\node[vertex] (G-2) at (1.5, 1) [shape = circle, draw] {}; 
\node[vertex] (G-3) at (3.0, 1) [shape = circle, draw] {}; 
\draw[] (G--3) .. controls +(-0.6, 0.6) and +(0.6, 0.6) .. (G--1); 
\draw[] (G-1) .. controls +(0.75, -1) and +(-0.75, 1) .. (G--2); 
\draw[] (G-2) .. controls +(0.5, -0.5) and +(-0.5, -0.5) .. (G-3); 
\end{tikzpicture} + \begin{tikzpicture}[scale = 0.25,thick, baseline={(0,-1ex/2)}] 
\tikzstyle{vertex} = [shape = circle, minimum size = 1pt, inner sep = 0pt] 
\node[vertex] (G--3) at (3.0, -1) [shape = circle, draw] {}; 
\node[vertex] (G--1) at (0.0, -1) [shape = circle, draw] {}; 
\node[vertex] (G--2) at (1.5, -1) [shape = circle, draw] {}; 
\node[vertex] (G-2) at (1.5, 1) [shape = circle, draw] {}; 
\node[vertex] (G-1) at (0.0, 1) [shape = circle, draw] {}; 
\node[vertex] (G-3) at (3.0, 1) [shape = circle, draw] {}; 
\draw[] (G--3) .. controls +(-0.6, 0.6) and +(0.6, 0.6) .. (G--1); 
\draw[] (G-2) .. controls +(0, -1) and +(0, 1) .. (G--2); 
\draw[] (G-1) .. controls +(0.6, -0.6) and +(-0.6, -0.6) .. (G-3); 
\end{tikzpicture}
\end{equation}
Note that we simply extracted the diagrams whose coefficients dominate at large $D$ compared to the rest of the coefficients. This does not imply that they correspond to the diagrams which give dominant contributions to expectation values. We leave investigations of that question for future work.

For
\begin{equation}
	\vec{R}=\vec{S}= (V_{[D-1,1]},V_{[D-1,1]},V_{[D-1,1]},V_{[D-2,2]}) \, , \quad \Lambda = V_{[D-3,3]} \, ,
\end{equation}
the dominant diagram is
\begin{equation}
  \begin{aligned}
  	Q^{G_{\vec{R}}^{\Lambda} G_{\vec{S}}^{\Lambda}} = \begin{tikzpicture}[scale = 0.25,thick, baseline={(0,-1ex/2)}] 
\tikzstyle{vertex} = [shape = circle, minimum size = 1pt, inner sep = 0pt] 
\node[vertex] (G--3) at (3.0, -1) [shape = circle, draw] {}; 
\node[vertex] (G--2) at (1.5, -1) [shape = circle, draw] {}; 
\node[vertex] (G--1) at (0.0, -1) [shape = circle, draw] {}; 
\node[vertex] (G-1) at (0.0, 1) [shape = circle, draw] {}; 
\node[vertex] (G-2) at (1.5, 1) [shape = circle, draw] {}; 
\node[vertex] (G-3) at (3.0, 1) [shape = circle, draw] {}; 
\draw[] (G-1) .. controls +(0.5, -0.5) and +(-0.5, -0.5) .. (G-2); 
\draw[] (G-2) .. controls +(0.5, -0.5) and +(-0.5, -0.5) .. (G-3); 
\draw[] (G-3) .. controls +(0, -1) and +(0, 1) .. (G--3); 
\draw[] (G--3) .. controls +(-0.5, 0.5) and +(0.5, 0.5) .. (G--2); 
\draw[] (G--2) .. controls +(-0.5, 0.5) and +(0.5, 0.5) .. (G--1); 
\draw[] (G--1) .. controls +(0, 1) and +(0, -1) .. (G-1); 
\end{tikzpicture}
  \end{aligned}
\end{equation}
For
\begin{equation}
	\vec{R}=\vec{S}= (V_{[D-1,1]},V_{[D-1,1]},V_{[D-1,1]},V_{[D-2,2]}) \, , \quad \Lambda = V_{[D-3,2,1]} \, ,
\end{equation}
the dominant diagrams are
\begin{equation}
  \begin{aligned}
  	\begin{tikzpicture}[scale = 0.25,thick, baseline={(0,-1ex/2)}] 
\tikzstyle{vertex} = [shape = circle, minimum size = 1pt, inner sep = 0pt] 
\node[vertex] (G--3) at (3.0, -1) [shape = circle, draw] {}; 
\node[vertex] (G--2) at (1.5, -1) [shape = circle, draw] {}; 
\node[vertex] (G-1) at (0.0, 1) [shape = circle, draw] {}; 
\node[vertex] (G-2) at (1.5, 1) [shape = circle, draw] {}; 
\node[vertex] (G--1) at (0.0, -1) [shape = circle, draw] {}; 
\node[vertex] (G-3) at (3.0, 1) [shape = circle, draw] {}; 
\draw[] (G-1) .. controls +(0.5, -0.5) and +(-0.5, -0.5) .. (G-2); 
\draw[] (G-2) .. controls +(0.75, -1) and +(-0.75, 1) .. (G--3); 
\draw[] (G--3) .. controls +(-0.5, 0.5) and +(0.5, 0.5) .. (G--2); 
\draw[] (G--2) .. controls +(-0.75, 1) and +(0.75, -1) .. (G-1); 
\draw[] (G-3) .. controls +(-1, -1) and +(1, 1) .. (G--1); 
\end{tikzpicture} - \frac{1}{2} \begin{tikzpicture}[scale = 0.25,thick, baseline={(0,-1ex/2)}] 
\tikzstyle{vertex} = [shape = circle, minimum size = 1pt, inner sep = 0pt] 
\node[vertex] (G--3) at (3.0, -1) [shape = circle, draw] {}; 
\node[vertex] (G--2) at (1.5, -1) [shape = circle, draw] {}; 
\node[vertex] (G-1) at (0.0, 1) [shape = circle, draw] {}; 
\node[vertex] (G-3) at (3.0, 1) [shape = circle, draw] {}; 
\node[vertex] (G--1) at (0.0, -1) [shape = circle, draw] {}; 
\node[vertex] (G-2) at (1.5, 1) [shape = circle, draw] {}; 
\draw[] (G-1) .. controls +(0.6, -0.6) and +(-0.6, -0.6) .. (G-3); 
\draw[] (G-3) .. controls +(0, -1) and +(0, 1) .. (G--3); 
\draw[] (G--3) .. controls +(-0.5, 0.5) and +(0.5, 0.5) .. (G--2); 
\draw[] (G--2) .. controls +(-0.75, 1) and +(0.75, -1) .. (G-1); 
\draw[] (G-2) .. controls +(-0.75, -1) and +(0.75, 1) .. (G--1); 
\end{tikzpicture} - \frac{1}{2} \begin{tikzpicture}[scale = 0.25,thick, baseline={(0,-1ex/2)}] 
\tikzstyle{vertex} = [shape = circle, minimum size = 1pt, inner sep = 0pt] 
\node[vertex] (G--3) at (3.0, -1) [shape = circle, draw] {}; 
\node[vertex] (G--2) at (1.5, -1) [shape = circle, draw] {}; 
\node[vertex] (G-2) at (1.5, 1) [shape = circle, draw] {}; 
\node[vertex] (G-3) at (3.0, 1) [shape = circle, draw] {}; 
\node[vertex] (G--1) at (0.0, -1) [shape = circle, draw] {}; 
\node[vertex] (G-1) at (0.0, 1) [shape = circle, draw] {}; 
\draw[] (G-2) .. controls +(0.5, -0.5) and +(-0.5, -0.5) .. (G-3); 
\draw[] (G-3) .. controls +(0, -1) and +(0, 1) .. (G--3); 
\draw[] (G--3) .. controls +(-0.5, 0.5) and +(0.5, 0.5) .. (G--2); 
\draw[] (G--2) .. controls +(0, 1) and +(0, -1) .. (G-2); 
\draw[] (G-1) .. controls +(0, -1) and +(0, 1) .. (G--1); 
\end{tikzpicture} + \begin{tikzpicture}[scale = 0.25,thick, baseline={(0,-1ex/2)}] 
\tikzstyle{vertex} = [shape = circle, minimum size = 1pt, inner sep = 0pt] 
\node[vertex] (G--3) at (3.0, -1) [shape = circle, draw] {}; 
\node[vertex] (G--1) at (0.0, -1) [shape = circle, draw] {}; 
\node[vertex] (G-1) at (0.0, 1) [shape = circle, draw] {}; 
\node[vertex] (G-2) at (1.5, 1) [shape = circle, draw] {}; 
\node[vertex] (G--2) at (1.5, -1) [shape = circle, draw] {}; 
\node[vertex] (G-3) at (3.0, 1) [shape = circle, draw] {}; 
\draw[] (G-1) .. controls +(0.5, -0.5) and +(-0.5, -0.5) .. (G-2); 
\draw[] (G-2) .. controls +(0.75, -1) and +(-0.75, 1) .. (G--3); 
\draw[] (G--3) .. controls +(-0.6, 0.6) and +(0.6, 0.6) .. (G--1); 
\draw[] (G--1) .. controls +(0, 1) and +(0, -1) .. (G-1); 
\draw[] (G-3) .. controls +(-0.75, -1) and +(0.75, 1) .. (G--2); 
\end{tikzpicture} - \frac{1}{2} \begin{tikzpicture}[scale = 0.25,thick, baseline={(0,-1ex/2)}] 
\tikzstyle{vertex} = [shape = circle, minimum size = 1pt, inner sep = 0pt] 
\node[vertex] (G--3) at (3.0, -1) [shape = circle, draw] {}; 
\node[vertex] (G--1) at (0.0, -1) [shape = circle, draw] {}; 
\node[vertex] (G-1) at (0.0, 1) [shape = circle, draw] {}; 
\node[vertex] (G-3) at (3.0, 1) [shape = circle, draw] {}; 
\node[vertex] (G--2) at (1.5, -1) [shape = circle, draw] {}; 
\node[vertex] (G-2) at (1.5, 1) [shape = circle, draw] {}; 
\draw[] (G-1) .. controls +(0.6, -0.6) and +(-0.6, -0.6) .. (G-3); 
\draw[] (G-3) .. controls +(0, -1) and +(0, 1) .. (G--3); 
\draw[] (G--3) .. controls +(-0.6, 0.6) and +(0.6, 0.6) .. (G--1); 
\draw[] (G--1) .. controls +(0, 1) and +(0, -1) .. (G-1); 
\draw[] (G-2) .. controls +(0, -1) and +(0, 1) .. (G--2); 
\end{tikzpicture} - \frac{1}{2} \begin{tikzpicture}[scale = 0.25,thick, baseline={(0,-1ex/2)}] 
\tikzstyle{vertex} = [shape = circle, minimum size = 1pt, inner sep = 0pt] 
\node[vertex] (G--3) at (3.0, -1) [shape = circle, draw] {}; 
\node[vertex] (G--1) at (0.0, -1) [shape = circle, draw] {}; 
\node[vertex] (G-2) at (1.5, 1) [shape = circle, draw] {}; 
\node[vertex] (G-3) at (3.0, 1) [shape = circle, draw] {}; 
\node[vertex] (G--2) at (1.5, -1) [shape = circle, draw] {}; 
\node[vertex] (G-1) at (0.0, 1) [shape = circle, draw] {}; 
\draw[] (G-2) .. controls +(0.5, -0.5) and +(-0.5, -0.5) .. (G-3); 
\draw[] (G-3) .. controls +(0, -1) and +(0, 1) .. (G--3); 
\draw[] (G--3) .. controls +(-0.6, 0.6) and +(0.6, 0.6) .. (G--1); 
\draw[] (G--1) .. controls +(0.75, 1) and +(-0.75, -1) .. (G-2); 
\draw[] (G-1) .. controls +(0.75, -1) and +(-0.75, 1) .. (G--2); 
\end{tikzpicture} + \begin{tikzpicture}[scale = 0.25,thick, baseline={(0,-1ex/2)}] 
\tikzstyle{vertex} = [shape = circle, minimum size = 1pt, inner sep = 0pt] 
\node[vertex] (G--3) at (3.0, -1) [shape = circle, draw] {}; 
\node[vertex] (G-1) at (0.0, 1) [shape = circle, draw] {}; 
\node[vertex] (G--2) at (1.5, -1) [shape = circle, draw] {}; 
\node[vertex] (G--1) at (0.0, -1) [shape = circle, draw] {}; 
\node[vertex] (G-2) at (1.5, 1) [shape = circle, draw] {}; 
\node[vertex] (G-3) at (3.0, 1) [shape = circle, draw] {}; 
\draw[] (G-1) .. controls +(1, -1) and +(-1, 1) .. (G--3); 
\draw[] (G-2) .. controls +(0.5, -0.5) and +(-0.5, -0.5) .. (G-3); 
\draw[] (G-3) .. controls +(-0.75, -1) and +(0.75, 1) .. (G--2); 
\draw[] (G--2) .. controls +(-0.5, 0.5) and +(0.5, 0.5) .. (G--1); 
\draw[] (G--1) .. controls +(0.75, 1) and +(-0.75, -1) .. (G-2); 
\end{tikzpicture} + \begin{tikzpicture}[scale = 0.25,thick, baseline={(0,-1ex/2)}] 
\tikzstyle{vertex} = [shape = circle, minimum size = 1pt, inner sep = 0pt] 
\node[vertex] (G--3) at (3.0, -1) [shape = circle, draw] {}; 
\node[vertex] (G-2) at (1.5, 1) [shape = circle, draw] {}; 
\node[vertex] (G--2) at (1.5, -1) [shape = circle, draw] {}; 
\node[vertex] (G--1) at (0.0, -1) [shape = circle, draw] {}; 
\node[vertex] (G-1) at (0.0, 1) [shape = circle, draw] {}; 
\node[vertex] (G-3) at (3.0, 1) [shape = circle, draw] {}; 
\draw[] (G-2) .. controls +(0.75, -1) and +(-0.75, 1) .. (G--3); 
\draw[] (G-1) .. controls +(0.6, -0.6) and +(-0.6, -0.6) .. (G-3); 
\draw[] (G-3) .. controls +(-0.75, -1) and +(0.75, 1) .. (G--2); 
\draw[] (G--2) .. controls +(-0.5, 0.5) and +(0.5, 0.5) .. (G--1); 
\draw[] (G--1) .. controls +(0, 1) and +(0, -1) .. (G-1); 
\end{tikzpicture} - 2 \begin{tikzpicture}[scale = 0.25,thick, baseline={(0,-1ex/2)}] 
\tikzstyle{vertex} = [shape = circle, minimum size = 1pt, inner sep = 0pt] 
\node[vertex] (G--3) at (3.0, -1) [shape = circle, draw] {}; 
\node[vertex] (G-3) at (3.0, 1) [shape = circle, draw] {}; 
\node[vertex] (G--2) at (1.5, -1) [shape = circle, draw] {}; 
\node[vertex] (G--1) at (0.0, -1) [shape = circle, draw] {}; 
\node[vertex] (G-1) at (0.0, 1) [shape = circle, draw] {}; 
\node[vertex] (G-2) at (1.5, 1) [shape = circle, draw] {}; 
\draw[] (G-3) .. controls +(0, -1) and +(0, 1) .. (G--3); 
\draw[] (G-1) .. controls +(0.5, -0.5) and +(-0.5, -0.5) .. (G-2); 
\draw[] (G-2) .. controls +(0, -1) and +(0, 1) .. (G--2); 
\draw[] (G--2) .. controls +(-0.5, 0.5) and +(0.5, 0.5) .. (G--1); 
\draw[] (G--1) .. controls +(0, 1) and +(0, -1) .. (G-1); 
\end{tikzpicture}
  \end{aligned}
\end{equation}
For
\begin{equation}
	\vec{R}=\vec{S}= (V_{[D-1,1]},V_{[D-1,1]},V_{[D-1,1]},V_{[D-2,1,1]}) \, , \quad \Lambda = V_{[D-3,1,1,1]} \, ,
\end{equation}
the dominant diagrams are
\begin{equation}
  \begin{aligned}
  	\begin{tikzpicture}[scale = 0.25,thick, baseline={(0,-1ex/2)}] 
\tikzstyle{vertex} = [shape = circle, minimum size = 1pt, inner sep = 0pt] 
\node[vertex] (G--3) at (3.0, -1) [shape = circle, draw] {}; 
\node[vertex] (G-1) at (0.0, 1) [shape = circle, draw] {}; 
\node[vertex] (G--2) at (1.5, -1) [shape = circle, draw] {}; 
\node[vertex] (G-2) at (1.5, 1) [shape = circle, draw] {}; 
\node[vertex] (G--1) at (0.0, -1) [shape = circle, draw] {}; 
\node[vertex] (G-3) at (3.0, 1) [shape = circle, draw] {}; 
\draw[] (G-1) .. controls +(1, -1) and +(-1, 1) .. (G--3); 
\draw[] (G-2) .. controls +(0, -1) and +(0, 1) .. (G--2); 
\draw[] (G-3) .. controls +(-1, -1) and +(1, 1) .. (G--1); 
\end{tikzpicture} - \begin{tikzpicture}[scale = 0.25,thick, baseline={(0,-1ex/2)}] 
\tikzstyle{vertex} = [shape = circle, minimum size = 1pt, inner sep = 0pt] 
\node[vertex] (G--3) at (3.0, -1) [shape = circle, draw] {}; 
\node[vertex] (G-1) at (0.0, 1) [shape = circle, draw] {}; 
\node[vertex] (G--2) at (1.5, -1) [shape = circle, draw] {}; 
\node[vertex] (G-3) at (3.0, 1) [shape = circle, draw] {}; 
\node[vertex] (G--1) at (0.0, -1) [shape = circle, draw] {}; 
\node[vertex] (G-2) at (1.5, 1) [shape = circle, draw] {}; 
\draw[] (G-1) .. controls +(1, -1) and +(-1, 1) .. (G--3); 
\draw[] (G-3) .. controls +(-0.75, -1) and +(0.75, 1) .. (G--2); 
\draw[] (G-2) .. controls +(-0.75, -1) and +(0.75, 1) .. (G--1); 
\end{tikzpicture} - \begin{tikzpicture}[scale = 0.25,thick, baseline={(0,-1ex/2)}] 
\tikzstyle{vertex} = [shape = circle, minimum size = 1pt, inner sep = 0pt] 
\node[vertex] (G--3) at (3.0, -1) [shape = circle, draw] {}; 
\node[vertex] (G-2) at (1.5, 1) [shape = circle, draw] {}; 
\node[vertex] (G--2) at (1.5, -1) [shape = circle, draw] {}; 
\node[vertex] (G-1) at (0.0, 1) [shape = circle, draw] {}; 
\node[vertex] (G--1) at (0.0, -1) [shape = circle, draw] {}; 
\node[vertex] (G-3) at (3.0, 1) [shape = circle, draw] {}; 
\draw[] (G-2) .. controls +(0.75, -1) and +(-0.75, 1) .. (G--3); 
\draw[] (G-1) .. controls +(0.75, -1) and +(-0.75, 1) .. (G--2); 
\draw[] (G-3) .. controls +(-1, -1) and +(1, 1) .. (G--1); 
\end{tikzpicture} + \begin{tikzpicture}[scale = 0.25,thick, baseline={(0,-1ex/2)}] 
\tikzstyle{vertex} = [shape = circle, minimum size = 1pt, inner sep = 0pt] 
\node[vertex] (G--3) at (3.0, -1) [shape = circle, draw] {}; 
\node[vertex] (G-2) at (1.5, 1) [shape = circle, draw] {}; 
\node[vertex] (G--2) at (1.5, -1) [shape = circle, draw] {}; 
\node[vertex] (G-3) at (3.0, 1) [shape = circle, draw] {}; 
\node[vertex] (G--1) at (0.0, -1) [shape = circle, draw] {}; 
\node[vertex] (G-1) at (0.0, 1) [shape = circle, draw] {}; 
\draw[] (G-2) .. controls +(0.75, -1) and +(-0.75, 1) .. (G--3); 
\draw[] (G-3) .. controls +(-0.75, -1) and +(0.75, 1) .. (G--2); 
\draw[] (G-1) .. controls +(0, -1) and +(0, 1) .. (G--1); 
\end{tikzpicture} + \begin{tikzpicture}[scale = 0.25,thick, baseline={(0,-1ex/2)}] 
\tikzstyle{vertex} = [shape = circle, minimum size = 1pt, inner sep = 0pt] 
\node[vertex] (G--3) at (3.0, -1) [shape = circle, draw] {}; 
\node[vertex] (G-3) at (3.0, 1) [shape = circle, draw] {}; 
\node[vertex] (G--2) at (1.5, -1) [shape = circle, draw] {}; 
\node[vertex] (G-1) at (0.0, 1) [shape = circle, draw] {}; 
\node[vertex] (G--1) at (0.0, -1) [shape = circle, draw] {}; 
\node[vertex] (G-2) at (1.5, 1) [shape = circle, draw] {}; 
\draw[] (G-3) .. controls +(0, -1) and +(0, 1) .. (G--3); 
\draw[] (G-1) .. controls +(0.75, -1) and +(-0.75, 1) .. (G--2); 
\draw[] (G-2) .. controls +(-0.75, -1) and +(0.75, 1) .. (G--1); 
\end{tikzpicture} - \begin{tikzpicture}[scale = 0.25,thick, baseline={(0,-1ex/2)}] 
\tikzstyle{vertex} = [shape = circle, minimum size = 1pt, inner sep = 0pt] 
\node[vertex] (G--3) at (3.0, -1) [shape = circle, draw] {}; 
\node[vertex] (G-3) at (3.0, 1) [shape = circle, draw] {}; 
\node[vertex] (G--2) at (1.5, -1) [shape = circle, draw] {}; 
\node[vertex] (G-2) at (1.5, 1) [shape = circle, draw] {}; 
\node[vertex] (G--1) at (0.0, -1) [shape = circle, draw] {}; 
\node[vertex] (G-1) at (0.0, 1) [shape = circle, draw] {}; 
\draw[] (G-3) .. controls +(0, -1) and +(0, 1) .. (G--3); 
\draw[] (G-2) .. controls +(0, -1) and +(0, 1) .. (G--2); 
\draw[] (G-1) .. controls +(0, -1) and +(0, 1) .. (G--1); 
\end{tikzpicture}
  \end{aligned}
\end{equation}

\section{Counting $S_D$ invariant tensor observables}
\label{sec: counting}
In this section we derive a formula for the counting of invariant tensor observables using characters of permutations in the natural representation $V_D$. We then prove that a basis for the space of degree $m$ observables of matrices of size $D$ is in bijection with bipartite 3-coloured graphs with $m$ white vertices and up to $D$ black vertices.

\subsection{Representation theoretic counting}

The dimension of the space of degree $m$ observables is the same as the multiplicity of the trivial representation of $S_D$ in the decomposition into irreducibles of
\begin{align}
\text{Sym}^m(V_D \otimes V_D \otimes V_D) \, .
\end{align}
We call this
\begin{equation}
  Dim(D,m) = \text{Multiplicity of $V_0$ in $\text{Sym}^m(V_D \otimes V_D \otimes V_D)$.}
\end{equation}

Writing the linear operator for $\sigma$ in $V_D$ as $\rho(\sigma)$, the linear operator in $V^{\otimes 3}_{D}$ we write as
\begin{align}
\rho_{V^{\otimes 3}_D}(\sigma) = \rho(\sigma) \otimes \rho(\sigma) \otimes \rho(\sigma).
\end{align}
The tensor product $(V^{\otimes 3}_D)^{\otimes m}$ has an action of
\begin{align}
\rho_{V^{\otimes 3m}_D}(\sigma) = \underbrace{\rho_{V^{\otimes 3}_D}(\sigma) \otimes \dots \otimes \rho_{V^{\otimes 3}_D}(\sigma)}_{m \text{ tensor factors}}.
\end{align}
The symmetric group $S_m$ acts on $(V_D^{\otimes 3})^{\otimes m}$ by permuting tensor factors. For $\tau \in S_m$ the action is given by
\begin{equation}
  \tau (\Phi_{i_1 j_1 k_1} \otimes \dots \otimes \Phi_{i_m j_m k_m}) = \Phi_{ i_{\tau(1)} j_{\tau(1)} k_{\tau(1)} } \otimes \dots \otimes \Phi_{ i_{\tau(m)} j_{\tau(m)} k_{\tau(m)} }
\end{equation}

We are interested in the symmetric subspace of $V_{D}^{\otimes 3m}$, which corresponds to the trivial representation of $S_m$. Define the projector to the trivial representation of $S_m$
\begin{equation}
	P_0^{S_m} = \frac{1}{m!}\sum_{\tau \in S_m} \tau
\end{equation}
and the corresponding projector for $S_D$
\begin{equation}
  P_0^{S_D} = \frac{1}{D!}\sum_{\sigma \in S_D} \rho_{V^{\otimes 3m}_D}(\sigma).
\end{equation}
The dimension of the space of degree $m$ observables is
\begin{align} \label{eq: Dim(D, m) 1} \nonumber
Dim(D,m) &= tr_{V^{\otimes 3m}_{D}} \big( P_0^{S_D} P_0^{S_k}\big) =\frac{1}{D! m!} \sum_{\sigma \in S_D} \sum_{\tau \in S_m} tr_{V^{\otimes 3m}_{D}} \big( \rho_{V^{\otimes 3m}_D}(\sigma) \tau \big) \\
&= \frac{1}{D! m!} \sum_{\sigma \in S_D} \sum_{\tau \in S_m} \prod_{i=1}^m tr_{V^{\otimes 3}_{D}} \big( \rho_{V^{\otimes 3}_D}(\sigma^i) \big)^{C_i(\tau)} .
\end{align}
Using
\begin{align}
tr_{V^{\otimes 3}_{D}} \big( \rho_{V^{\otimes 3}_D}(\sigma) \big) = \Big( tr_{V_{D}} \big( \rho(\sigma) \big) \Big)^3,
\end{align}
and
\begin{align}
tr_{V_D} \big( \rho(\sigma^i) \big) = \sum_{l | i} l C_l (\sigma),
\end{align}
where the sum is over the divisors of $i$, we can rewrite \eqref{eq: Dim(D, m) 1} 
\begin{align}
Dim(D,m) =  \frac{1}{D! m!} \sum_{\sigma \in S_D} \sum_{\tau \in S_m} \prod_{i=1}^m \Big( \sum_{l | i} l C_l (\sigma) \Big)^{3 C_i (\tau)}. \label{eq: dim D m}
\end{align}
We collapse the sums over permutations into sums over conjugacy classes to get
\begin{align} \label{eqn: rep theory counting}
Dim(D, m) = \sum_{p \vdash D} \sum_{q \vdash m} \frac{1}{\prod_{i=1} i^{p_i + q_i} p_i! q_i!} \prod_{i=1}^{m} \Big( \sum_{l | i} l p_l \Big)^{3q_i},
\end{align}
with $p$ and $q$ partitions obeying $\sum_i i p_i = D$ and $\sum_i i q_i = m$ respectively. For $m= 1,2,3,4$ and $D=3m$ this gives $Dim(D, m) = 5, 117, 3813, 187584$.

\subsection{Graph counting}
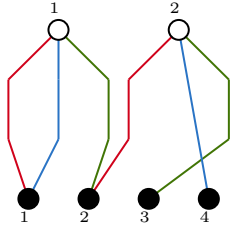
\begin{figure}[hbt]
\centering
{
\tikzset{every picture/.style={line width=0.75pt}} 
\begin{tikzpicture}[x=0.75pt,y=0.75pt,yscale=-1,xscale=1]
\draw [color={rgb, 255:red, 65; green, 117; blue, 5 }  ,draw opacity=1 ]   (115,70) -- (75,100) ;
\draw [color={rgb, 255:red, 208; green, 2; blue, 27 }  ,draw opacity=1 ]   (65,70) -- (45,100) ;
\draw [color={rgb, 255:red, 65; green, 117; blue, 5 }  ,draw opacity=1 ]   (90,15) -- (115,40) ;
\draw [color={rgb, 255:red, 65; green, 117; blue, 5 }  ,draw opacity=1 ]   (115,40) -- (115,70) ;
\draw [color={rgb, 255:red, 65; green, 117; blue, 5 }  ,draw opacity=1 ]   (55,70) -- (45,100) ;
\draw [color={rgb, 255:red, 65; green, 117; blue, 5 }  ,draw opacity=1 ]   (30,15) -- (55,40) ;
\draw [color={rgb, 255:red, 208; green, 2; blue, 27 }  ,draw opacity=1 ]   (30,15) -- (5,40) ;
\draw [color={rgb, 255:red, 42; green, 116; blue, 197 }  ,draw opacity=1 ]   (30,15) -- (30,40) ;
\draw  [fill={rgb, 255:red, 255; green, 255; blue, 255 }  ,fill opacity=1 ] (35,15) .. controls (35,12.24) and (32.76,10) .. (30,10) .. controls (27.24,10) and (25,12.24) .. (25,15) .. controls (25,17.76) and (27.24,20) .. (30,20) .. controls (32.76,20) and (35,17.76) .. (35,15) -- cycle ;
\draw [color={rgb, 255:red, 208; green, 2; blue, 27 }  ,draw opacity=1 ]   (90,15) -- (65,40) ;
\draw [color={rgb, 255:red, 42; green, 116; blue, 197 }  ,draw opacity=1 ]   (90,15) -- (105,100) ;
\draw  [fill={rgb, 255:red, 255; green, 255; blue, 255 }  ,fill opacity=1 ] (95,15) .. controls (95,12.24) and (92.76,10) .. (90,10) .. controls (87.24,10) and (85,12.24) .. (85,15) .. controls (85,17.76) and (87.24,20) .. (90,20) .. controls (92.76,20) and (95,17.76) .. (95,15) -- cycle ;
\draw [color={rgb, 255:red, 208; green, 2; blue, 27 }  ,draw opacity=1 ]   (5,40) -- (5,70) ;
\draw [color={rgb, 255:red, 208; green, 2; blue, 27 }  ,draw opacity=1 ]   (65,40) -- (65,70) ;
\draw [color={rgb, 255:red, 208; green, 2; blue, 27 }  ,draw opacity=1 ]   (15,100) -- (5,70) ;
\draw [color={rgb, 255:red, 42; green, 116; blue, 197 }  ,draw opacity=1 ]   (30,40) -- (30,70) ;
\draw [color={rgb, 255:red, 42; green, 116; blue, 197 }  ,draw opacity=1 ]   (30,70) -- (15,100) ;
\draw  [fill={rgb, 255:red, 0; green, 0; blue, 0 }  ,fill opacity=1 ] (20,100) .. controls (20,97.24) and (17.76,95) .. (15,95) .. controls (12.24,95) and (10,97.24) .. (10,100) .. controls (10,102.76) and (12.24,105) .. (15,105) .. controls (17.76,105) and (20,102.76) .. (20,100) -- cycle ;
\draw  [fill={rgb, 255:red, 0; green, 0; blue, 0 }  ,fill opacity=1 ] (50,100) .. controls (50,97.24) and (47.76,95) .. (45,95) .. controls (42.24,95) and (40,97.24) .. (40,100) .. controls (40,102.76) and (42.24,105) .. (45,105) .. controls (47.76,105) and (50,102.76) .. (50,100) -- cycle ;
\draw  [fill={rgb, 255:red, 0; green, 0; blue, 0 }  ,fill opacity=1 ] (80,100) .. controls (80,97.24) and (77.76,95) .. (75,95) .. controls (72.24,95) and (70,97.24) .. (70,100) .. controls (70,102.76) and (72.24,105) .. (75,105) .. controls (77.76,105) and (80,102.76) .. (80,100) -- cycle ;
\draw [color={rgb, 255:red, 65; green, 117; blue, 5 }  ,draw opacity=1 ]   (55,40) -- (55,70) ;
\draw  [fill={rgb, 255:red, 0; green, 0; blue, 0 }  ,fill opacity=1 ] (110,100) .. controls (110,97.24) and (107.76,95) .. (105,95) .. controls (102.24,95) and (100,97.24) .. (100,100) .. controls (100,102.76) and (102.24,105) .. (105,105) .. controls (107.76,105) and (110,102.76) .. (110,100) -- cycle ;
\draw (9,105) node [anchor=north west][inner sep=0.75pt]  [font=\tiny] [align=left] {1};
\draw (39,105) node [anchor=north west][inner sep=0.75pt]  [font=\tiny] [align=left] {2};
\draw (69,105) node [anchor=north west][inner sep=0.75pt]  [font=\tiny] [align=left] {3};
\draw (99,105) node [anchor=north west][inner sep=0.75pt]  [font=\tiny] [align=left] {4};
\draw (24,0) node [anchor=north west][inner sep=0.75pt]  [font=\tiny] [align=left] {1};
\draw (84,0) node [anchor=north west][inner sep=0.75pt]  [font=\tiny] [align=left] {2};
\end{tikzpicture}
}
\caption{An example of a bipartite 3-coloured labelled graph with two white vertices and three black vertices. It corresponds to the pair of triplets $((1,1,2), (2,4,3))$, which in turn corresponds to the tensor invariant $\sum_{i,j,k,l} \Phi_{iij}\Phi_{jkl}$.}
\label{fig: graph ex}
\end{figure}

Consider a bi-partite 3-coloured graph with $m$ (labelled) white vertices and $k$ (labelled) black vertices and demand that the white vertices have exactly one red, one blue and one green edge.
Because the graph is bi-partite, all edges coming out of a white vertex end on a black vertex. We encode the incidence of edges coming out of the $i$th white vertex using a triplet $(r_i,b_i,g_i)$, where $r_i,b_i,g_i \in \{1, \dots, k\}$. Given this data -- a list of $m$ elements in $\{1,\dots, k\}^{\times 3}$ -- we have a labeled graph. For an example of this correspondence, see Figure \ref{fig: graph ex}.

Let $N(k,m)$ be the number of unlabelled bi-partite 3-coloured graphs with $m$ white vertices (with one red, one blue and one green edge) and up to $k$ black vertices. As we will now show,
\begin{equation}
  N(k,m) = Dim(k,m).
\end{equation}

To count unlabelled graphs, we have to forget (form orbits under relabelling) the labels of white as well as black vertices. We start with the white vertex labels. An element $\tau \in S_m$ acts on lists of the above type as follows
\begin{equation}
	(r_1, b_1, g_1), \dots, (r_m, b_m, g_m) \mapsto (r_{\tau(1)}, b_{\tau(1)}, g_{\tau(1)}), \dots, (r_{\tau(m)}, b_{\tau(m)}, g_{\tau(m)}). \label{eq: sk action}
\end{equation}
Orbits under this action corresponds to graphs with unlabelled white vertices but labelled black vertices. For the example given in Figure \ref{fig: graph ex}, the orbit under $S_2$ acting on the white vertices is given in terms of triplets by
\begin{equation}
  \{((1,1,2), (2,4,3)), ((2,4,3), (1,1,2))\}.
\end{equation}
The corresponding set of graphs is
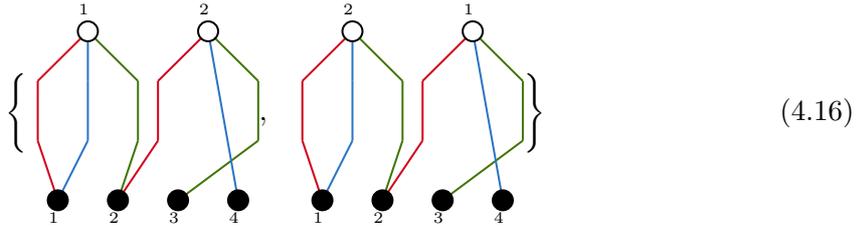
\begin{equation}
\tikzset{every picture/.style={line width=0.75pt}} 
\Bigg\{\vcenter{\hbox{
\begin{tikzpicture}[x=0.75pt,y=0.75pt,yscale=-1,xscale=1]
\draw [color={rgb, 255:red, 65; green, 117; blue, 5 }  ,draw opacity=1 ]   (115,70) -- (75,100) ;
\draw [color={rgb, 255:red, 208; green, 2; blue, 27 }  ,draw opacity=1 ]   (65,70) -- (45,100) ;
\draw [color={rgb, 255:red, 65; green, 117; blue, 5 }  ,draw opacity=1 ]   (90,15) -- (115,40) ;
\draw [color={rgb, 255:red, 65; green, 117; blue, 5 }  ,draw opacity=1 ]   (115,40) -- (115,70) ;
\draw [color={rgb, 255:red, 65; green, 117; blue, 5 }  ,draw opacity=1 ]   (55,70) -- (45,100) ;
\draw [color={rgb, 255:red, 65; green, 117; blue, 5 }  ,draw opacity=1 ]   (30,15) -- (55,40) ;
\draw [color={rgb, 255:red, 208; green, 2; blue, 27 }  ,draw opacity=1 ]   (30,15) -- (5,40) ;
\draw [color={rgb, 255:red, 42; green, 116; blue, 197 }  ,draw opacity=1 ]   (30,15) -- (30,40) ;
\draw  [fill={rgb, 255:red, 255; green, 255; blue, 255 }  ,fill opacity=1 ] (35,15) .. controls (35,12.24) and (32.76,10) .. (30,10) .. controls (27.24,10) and (25,12.24) .. (25,15) .. controls (25,17.76) and (27.24,20) .. (30,20) .. controls (32.76,20) and (35,17.76) .. (35,15) -- cycle ;
\draw [color={rgb, 255:red, 208; green, 2; blue, 27 }  ,draw opacity=1 ]   (90,15) -- (65,40) ;
\draw [color={rgb, 255:red, 42; green, 116; blue, 197 }  ,draw opacity=1 ]   (90,15) -- (105,100) ;
\draw  [fill={rgb, 255:red, 255; green, 255; blue, 255 }  ,fill opacity=1 ] (95,15) .. controls (95,12.24) and (92.76,10) .. (90,10) .. controls (87.24,10) and (85,12.24) .. (85,15) .. controls (85,17.76) and (87.24,20) .. (90,20) .. controls (92.76,20) and (95,17.76) .. (95,15) -- cycle ;
\draw [color={rgb, 255:red, 208; green, 2; blue, 27 }  ,draw opacity=1 ]   (5,40) -- (5,70) ;
\draw [color={rgb, 255:red, 208; green, 2; blue, 27 }  ,draw opacity=1 ]   (65,40) -- (65,70) ;
\draw [color={rgb, 255:red, 208; green, 2; blue, 27 }  ,draw opacity=1 ]   (15,100) -- (5,70) ;
\draw [color={rgb, 255:red, 42; green, 116; blue, 197 }  ,draw opacity=1 ]   (30,40) -- (30,70) ;
\draw [color={rgb, 255:red, 42; green, 116; blue, 197 }  ,draw opacity=1 ]   (30,70) -- (15,100) ;
\draw  [fill={rgb, 255:red, 0; green, 0; blue, 0 }  ,fill opacity=1 ] (20,100) .. controls (20,97.24) and (17.76,95) .. (15,95) .. controls (12.24,95) and (10,97.24) .. (10,100) .. controls (10,102.76) and (12.24,105) .. (15,105) .. controls (17.76,105) and (20,102.76) .. (20,100) -- cycle ;
\draw  [fill={rgb, 255:red, 0; green, 0; blue, 0 }  ,fill opacity=1 ] (50,100) .. controls (50,97.24) and (47.76,95) .. (45,95) .. controls (42.24,95) and (40,97.24) .. (40,100) .. controls (40,102.76) and (42.24,105) .. (45,105) .. controls (47.76,105) and (50,102.76) .. (50,100) -- cycle ;
\draw  [fill={rgb, 255:red, 0; green, 0; blue, 0 }  ,fill opacity=1 ] (80,100) .. controls (80,97.24) and (77.76,95) .. (75,95) .. controls (72.24,95) and (70,97.24) .. (70,100) .. controls (70,102.76) and (72.24,105) .. (75,105) .. controls (77.76,105) and (80,102.76) .. (80,100) -- cycle ;
\draw [color={rgb, 255:red, 65; green, 117; blue, 5 }  ,draw opacity=1 ]   (55,40) -- (55,70) ;
\draw  [fill={rgb, 255:red, 0; green, 0; blue, 0 }  ,fill opacity=1 ] (110,100) .. controls (110,97.24) and (107.76,95) .. (105,95) .. controls (102.24,95) and (100,97.24) .. (100,100) .. controls (100,102.76) and (102.24,105) .. (105,105) .. controls (107.76,105) and (110,102.76) .. (110,100) -- cycle ;

\draw (9,105) node [anchor=north west][inner sep=0.75pt]  [font=\tiny] [align=left] {1};
\draw (39,105) node [anchor=north west][inner sep=0.75pt]  [font=\tiny] [align=left] {2};
\draw (69,105) node [anchor=north west][inner sep=0.75pt]  [font=\tiny] [align=left] {3};
\draw (99,105) node [anchor=north west][inner sep=0.75pt]  [font=\tiny] [align=left] {4};
\draw (24,0) node [anchor=north west][inner sep=0.75pt]  [font=\tiny] [align=left] {1};
\draw (84,0) node [anchor=north west][inner sep=0.75pt]  [font=\tiny] [align=left] {2};
\end{tikzpicture}}}
,\quad      
\vcenter{\hbox{\begin{tikzpicture}[x=0.75pt,y=0.75pt,yscale=-1,xscale=1]
\draw [color={rgb, 255:red, 65; green, 117; blue, 5 }  ,draw opacity=1 ]   (115,70) -- (75,100) ;
\draw [color={rgb, 255:red, 208; green, 2; blue, 27 }  ,draw opacity=1 ]   (65,70) -- (45,100) ;
\draw [color={rgb, 255:red, 65; green, 117; blue, 5 }  ,draw opacity=1 ]   (90,15) -- (115,40) ;
\draw [color={rgb, 255:red, 65; green, 117; blue, 5 }  ,draw opacity=1 ]   (115,40) -- (115,70) ;
\draw [color={rgb, 255:red, 65; green, 117; blue, 5 }  ,draw opacity=1 ]   (55,70) -- (45,100) ;
\draw [color={rgb, 255:red, 65; green, 117; blue, 5 }  ,draw opacity=1 ]   (30,15) -- (55,40) ;
\draw [color={rgb, 255:red, 208; green, 2; blue, 27 }  ,draw opacity=1 ]   (30,15) -- (5,40) ;
\draw [color={rgb, 255:red, 42; green, 116; blue, 197 }  ,draw opacity=1 ]   (30,15) -- (30,40) ;
\draw  [fill={rgb, 255:red, 255; green, 255; blue, 255 }  ,fill opacity=1 ] (35,15) .. controls (35,12.24) and (32.76,10) .. (30,10) .. controls (27.24,10) and (25,12.24) .. (25,15) .. controls (25,17.76) and (27.24,20) .. (30,20) .. controls (32.76,20) and (35,17.76) .. (35,15) -- cycle ;
\draw [color={rgb, 255:red, 208; green, 2; blue, 27 }  ,draw opacity=1 ]   (90,15) -- (65,40) ;
\draw [color={rgb, 255:red, 42; green, 116; blue, 197 }  ,draw opacity=1 ]   (90,15) -- (105,100) ;
\draw  [fill={rgb, 255:red, 255; green, 255; blue, 255 }  ,fill opacity=1 ] (95,15) .. controls (95,12.24) and (92.76,10) .. (90,10) .. controls (87.24,10) and (85,12.24) .. (85,15) .. controls (85,17.76) and (87.24,20) .. (90,20) .. controls (92.76,20) and (95,17.76) .. (95,15) -- cycle ;
\draw [color={rgb, 255:red, 208; green, 2; blue, 27 }  ,draw opacity=1 ]   (5,40) -- (5,70) ;
\draw [color={rgb, 255:red, 208; green, 2; blue, 27 }  ,draw opacity=1 ]   (65,40) -- (65,70) ;
\draw [color={rgb, 255:red, 208; green, 2; blue, 27 }  ,draw opacity=1 ]   (15,100) -- (5,70) ;
\draw [color={rgb, 255:red, 42; green, 116; blue, 197 }  ,draw opacity=1 ]   (30,40) -- (30,70) ;
\draw [color={rgb, 255:red, 42; green, 116; blue, 197 }  ,draw opacity=1 ]   (30,70) -- (15,100) ;
\draw  [fill={rgb, 255:red, 0; green, 0; blue, 0 }  ,fill opacity=1 ] (20,100) .. controls (20,97.24) and (17.76,95) .. (15,95) .. controls (12.24,95) and (10,97.24) .. (10,100) .. controls (10,102.76) and (12.24,105) .. (15,105) .. controls (17.76,105) and (20,102.76) .. (20,100) -- cycle ;
\draw  [fill={rgb, 255:red, 0; green, 0; blue, 0 }  ,fill opacity=1 ] (50,100) .. controls (50,97.24) and (47.76,95) .. (45,95) .. controls (42.24,95) and (40,97.24) .. (40,100) .. controls (40,102.76) and (42.24,105) .. (45,105) .. controls (47.76,105) and (50,102.76) .. (50,100) -- cycle ;
\draw  [fill={rgb, 255:red, 0; green, 0; blue, 0 }  ,fill opacity=1 ] (80,100) .. controls (80,97.24) and (77.76,95) .. (75,95) .. controls (72.24,95) and (70,97.24) .. (70,100) .. controls (70,102.76) and (72.24,105) .. (75,105) .. controls (77.76,105) and (80,102.76) .. (80,100) -- cycle ;
\draw [color={rgb, 255:red, 65; green, 117; blue, 5 }  ,draw opacity=1 ]   (55,40) -- (55,70) ;
\draw  [fill={rgb, 255:red, 0; green, 0; blue, 0 }  ,fill opacity=1 ] (110,100) .. controls (110,97.24) and (107.76,95) .. (105,95) .. controls (102.24,95) and (100,97.24) .. (100,100) .. controls (100,102.76) and (102.24,105) .. (105,105) .. controls (107.76,105) and (110,102.76) .. (110,100) -- cycle ;

\draw (9,105) node [anchor=north west][inner sep=0.75pt]  [font=\tiny] [align=left] {1};
\draw (39,105) node [anchor=north west][inner sep=0.75pt]  [font=\tiny] [align=left] {2};
\draw (69,105) node [anchor=north west][inner sep=0.75pt]  [font=\tiny] [align=left] {3};
\draw (99,105) node [anchor=north west][inner sep=0.75pt]  [font=\tiny] [align=left] {4};
\draw (24,0) node [anchor=north west][inner sep=0.75pt]  [font=\tiny] [align=left] {2};
\draw (84,0) node [anchor=north west][inner sep=0.75pt]  [font=\tiny] [align=left] {1};
\end{tikzpicture}}}\Bigg\}
\end{equation}
To forget the labels on black vertices we take orbits under $S_k$, where $\sigma \in S_k$ acts on the list as
\begin{equation}
	(r_1, b_1, g_1), \dots, (r_m, b_m, g_m) \mapsto (\sigma(r_1), \sigma(b_1), \sigma(g_1)), \dots, (\sigma(r_m), \sigma(b_m), \sigma(g_m)). \label{eq: sl action}
\end{equation}
Combining the two actions gives an orbit that we identify with the unlabelled graph,
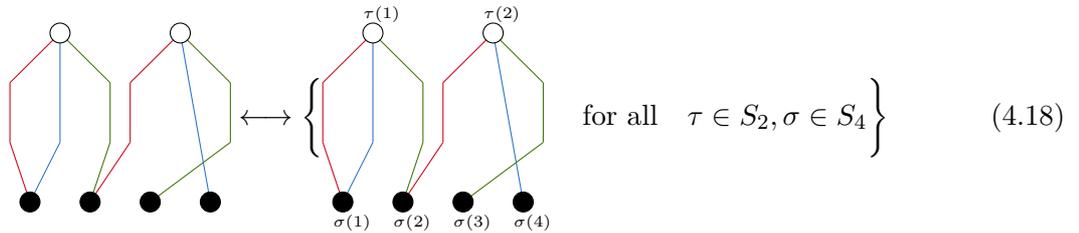
\begin{equation}
\vcenter{\hbox{\begin{tikzpicture}[x=0.75pt,y=0.75pt,yscale=-1,xscale=1]
\draw [color={rgb, 255:red, 65; green, 117; blue, 5 }  ,draw opacity=1 ]   (115,70) -- (75,100) ;
\draw [color={rgb, 255:red, 208; green, 2; blue, 27 }  ,draw opacity=1 ]   (65,70) -- (45,100) ;
\draw [color={rgb, 255:red, 65; green, 117; blue, 5 }  ,draw opacity=1 ]   (90,15) -- (115,40) ;
\draw [color={rgb, 255:red, 65; green, 117; blue, 5 }  ,draw opacity=1 ]   (115,40) -- (115,70) ;
\draw [color={rgb, 255:red, 65; green, 117; blue, 5 }  ,draw opacity=1 ]   (55,70) -- (45,100) ;
\draw [color={rgb, 255:red, 65; green, 117; blue, 5 }  ,draw opacity=1 ]   (30,15) -- (55,40) ;
\draw [color={rgb, 255:red, 208; green, 2; blue, 27 }  ,draw opacity=1 ]   (30,15) -- (5,40) ;
\draw [color={rgb, 255:red, 42; green, 116; blue, 197 }  ,draw opacity=1 ]   (30,15) -- (30,40) ;
\draw  [fill={rgb, 255:red, 255; green, 255; blue, 255 }  ,fill opacity=1 ] (35,15) .. controls (35,12.24) and (32.76,10) .. (30,10) .. controls (27.24,10) and (25,12.24) .. (25,15) .. controls (25,17.76) and (27.24,20) .. (30,20) .. controls (32.76,20) and (35,17.76) .. (35,15) -- cycle ;
\draw [color={rgb, 255:red, 208; green, 2; blue, 27 }  ,draw opacity=1 ]   (90,15) -- (65,40) ;
\draw [color={rgb, 255:red, 42; green, 116; blue, 197 }  ,draw opacity=1 ]   (90,15) -- (105,100) ;
\draw  [fill={rgb, 255:red, 255; green, 255; blue, 255 }  ,fill opacity=1 ] (95,15) .. controls (95,12.24) and (92.76,10) .. (90,10) .. controls (87.24,10) and (85,12.24) .. (85,15) .. controls (85,17.76) and (87.24,20) .. (90,20) .. controls (92.76,20) and (95,17.76) .. (95,15) -- cycle ;
\draw [color={rgb, 255:red, 208; green, 2; blue, 27 }  ,draw opacity=1 ]   (5,40) -- (5,70) ;
\draw [color={rgb, 255:red, 208; green, 2; blue, 27 }  ,draw opacity=1 ]   (65,40) -- (65,70) ;
\draw [color={rgb, 255:red, 208; green, 2; blue, 27 }  ,draw opacity=1 ]   (15,100) -- (5,70) ;
\draw [color={rgb, 255:red, 42; green, 116; blue, 197 }  ,draw opacity=1 ]   (30,40) -- (30,70) ;
\draw [color={rgb, 255:red, 42; green, 116; blue, 197 }  ,draw opacity=1 ]   (30,70) -- (15,100) ;
\draw  [fill={rgb, 255:red, 0; green, 0; blue, 0 }  ,fill opacity=1 ] (20,100) .. controls (20,97.24) and (17.76,95) .. (15,95) .. controls (12.24,95) and (10,97.24) .. (10,100) .. controls (10,102.76) and (12.24,105) .. (15,105) .. controls (17.76,105) and (20,102.76) .. (20,100) -- cycle ;
\draw  [fill={rgb, 255:red, 0; green, 0; blue, 0 }  ,fill opacity=1 ] (50,100) .. controls (50,97.24) and (47.76,95) .. (45,95) .. controls (42.24,95) and (40,97.24) .. (40,100) .. controls (40,102.76) and (42.24,105) .. (45,105) .. controls (47.76,105) and (50,102.76) .. (50,100) -- cycle ;
\draw  [fill={rgb, 255:red, 0; green, 0; blue, 0 }  ,fill opacity=1 ] (80,100) .. controls (80,97.24) and (77.76,95) .. (75,95) .. controls (72.24,95) and (70,97.24) .. (70,100) .. controls (70,102.76) and (72.24,105) .. (75,105) .. controls (77.76,105) and (80,102.76) .. (80,100) -- cycle ;
\draw [color={rgb, 255:red, 65; green, 117; blue, 5 }  ,draw opacity=1 ]   (55,40) -- (55,70) ;
\draw  [fill={rgb, 255:red, 0; green, 0; blue, 0 }  ,fill opacity=1 ] (110,100) .. controls (110,97.24) and (107.76,95) .. (105,95) .. controls (102.24,95) and (100,97.24) .. (100,100) .. controls (100,102.76) and (102.24,105) .. (105,105) .. controls (107.76,105) and (110,102.76) .. (110,100) -- cycle ;
\end{tikzpicture}%
}} \longleftrightarrow \Bigg\{\vcenter{\hbox{%
\begin{tikzpicture}[x=0.75pt,y=0.75pt,yscale=-1,xscale=1]
\draw [color={rgb, 255:red, 65; green, 117; blue, 5 }  ,draw opacity=1 ]   (115,70) -- (75,100) ;
\draw [color={rgb, 255:red, 208; green, 2; blue, 27 }  ,draw opacity=1 ]   (65,70) -- (45,100) ;
\draw [color={rgb, 255:red, 65; green, 117; blue, 5 }  ,draw opacity=1 ]   (90,15) -- (115,40) ;
\draw [color={rgb, 255:red, 65; green, 117; blue, 5 }  ,draw opacity=1 ]   (115,40) -- (115,70) ;
\draw [color={rgb, 255:red, 65; green, 117; blue, 5 }  ,draw opacity=1 ]   (55,70) -- (45,100) ;
\draw [color={rgb, 255:red, 65; green, 117; blue, 5 }  ,draw opacity=1 ]   (30,15) -- (55,40) ;
\draw [color={rgb, 255:red, 208; green, 2; blue, 27 }  ,draw opacity=1 ]   (30,15) -- (5,40) ;
\draw [color={rgb, 255:red, 42; green, 116; blue, 197 }  ,draw opacity=1 ]   (30,15) -- (30,40) ;
\draw  [fill={rgb, 255:red, 255; green, 255; blue, 255 }  ,fill opacity=1 ] (35,15) .. controls (35,12.24) and (32.76,10) .. (30,10) .. controls (27.24,10) and (25,12.24) .. (25,15) .. controls (25,17.76) and (27.24,20) .. (30,20) .. controls (32.76,20) and (35,17.76) .. (35,15) -- cycle ;
\draw [color={rgb, 255:red, 208; green, 2; blue, 27 }  ,draw opacity=1 ]   (90,15) -- (65,40) ;
\draw [color={rgb, 255:red, 42; green, 116; blue, 197 }  ,draw opacity=1 ]   (90,15) -- (105,100) ;
\draw  [fill={rgb, 255:red, 255; green, 255; blue, 255 }  ,fill opacity=1 ] (95,15) .. controls (95,12.24) and (92.76,10) .. (90,10) .. controls (87.24,10) and (85,12.24) .. (85,15) .. controls (85,17.76) and (87.24,20) .. (90,20) .. controls (92.76,20) and (95,17.76) .. (95,15) -- cycle ;
\draw [color={rgb, 255:red, 208; green, 2; blue, 27 }  ,draw opacity=1 ]   (5,40) -- (5,70) ;
\draw [color={rgb, 255:red, 208; green, 2; blue, 27 }  ,draw opacity=1 ]   (65,40) -- (65,70) ;
\draw [color={rgb, 255:red, 208; green, 2; blue, 27 }  ,draw opacity=1 ]   (15,100) -- (5,70) ;
\draw [color={rgb, 255:red, 42; green, 116; blue, 197 }  ,draw opacity=1 ]   (30,40) -- (30,70) ;
\draw [color={rgb, 255:red, 42; green, 116; blue, 197 }  ,draw opacity=1 ]   (30,70) -- (15,100) ;
\draw  [fill={rgb, 255:red, 0; green, 0; blue, 0 }  ,fill opacity=1 ] (20,100) .. controls (20,97.24) and (17.76,95) .. (15,95) .. controls (12.24,95) and (10,97.24) .. (10,100) .. controls (10,102.76) and (12.24,105) .. (15,105) .. controls (17.76,105) and (20,102.76) .. (20,100) -- cycle ;
\draw  [fill={rgb, 255:red, 0; green, 0; blue, 0 }  ,fill opacity=1 ] (50,100) .. controls (50,97.24) and (47.76,95) .. (45,95) .. controls (42.24,95) and (40,97.24) .. (40,100) .. controls (40,102.76) and (42.24,105) .. (45,105) .. controls (47.76,105) and (50,102.76) .. (50,100) -- cycle ;
\draw  [fill={rgb, 255:red, 0; green, 0; blue, 0 }  ,fill opacity=1 ] (80,100) .. controls (80,97.24) and (77.76,95) .. (75,95) .. controls (72.24,95) and (70,97.24) .. (70,100) .. controls (70,102.76) and (72.24,105) .. (75,105) .. controls (77.76,105) and (80,102.76) .. (80,100) -- cycle ;
\draw [color={rgb, 255:red, 65; green, 117; blue, 5 }  ,draw opacity=1 ]   (55,40) -- (55,70) ;
\draw  [fill={rgb, 255:red, 0; green, 0; blue, 0 }  ,fill opacity=1 ] (110,100) .. controls (110,97.24) and (107.76,95) .. (105,95) .. controls (102.24,95) and (100,97.24) .. (100,100) .. controls (100,102.76) and (102.24,105) .. (105,105) .. controls (107.76,105) and (110,102.76) .. (110,100) -- cycle ;
\draw (9,105) node [anchor=north west][inner sep=0.75pt]  [font=\tiny] [align=left] {$\sigma(1)$};
\draw (39,105) node [anchor=north west][inner sep=0.75pt]  [font=\tiny] [align=left] {$\sigma(2)$};
\draw (69,105) node [anchor=north west][inner sep=0.75pt]  [font=\tiny] [align=left] {$\sigma(3)$};
\draw (99,105) node [anchor=north west][inner sep=0.75pt]  [font=\tiny] [align=left] {$\sigma(4)$};
\draw (24,0) node [anchor=north west][inner sep=0.75pt]  [font=\tiny] [align=left] {$\tau(1)$};
\draw (84,0) node [anchor=north west][inner sep=0.75pt]  [font=\tiny] [align=left] {$\tau(2)$};
\end{tikzpicture}}}  \quad \text{for all} \quad \tau \in S_2, \sigma \in S_4 \Bigg\}
\end{equation}

Burnside's lemma allows us to count the orbits of the above actions on the set of a labelled graphs with $m$ white and $k$ black vertices,
\begin{align} \nonumber
	N(k,m) &= \text{Average number of elements fixed by the above actions} \\
	&= \frac{1}{k! m!} \sum_{\sigma \in S_k, \tau \in S_m} \text{Number of elements fixed by $\sigma$ and $\tau$}.
\end{align}
This has an interpretation in terms of permutation representations. Let
\begin{equation}
	U_{k,m} \cong (V_k \otimes V_k \otimes V_k)^{\otimes m},
\end{equation}
be the vector space with basis
\begin{equation}
	\ket{r_1, b_1, g_1, \dots, r_m, b_m, g_m}
\end{equation}
where $\tau \in S_m$ and $\sigma \in S_k$ act as in equation \eqref{eq: sk action} and \eqref{eq: sl action} respectively.
It follows that
\begin{equation}
	\text{Number of elements fixed by $\sigma$ and $\tau$} = \Tr_{U_{k,m}}(\sigma \tau)
\end{equation}
and therefore
\begin{equation}
	N(k,m) = \frac{1}{k! m!} \sum_{\sigma \in S_k, \tau \in S_m} \Tr_{U_{k,m}}(\sigma \tau) = \Tr_{U_{k,m}}(P_0^{S_k} P_0^{S_m}).
\end{equation}
We have
\begin{equation}
	\Tr_{U_{k,m}}(P_0^{S_k} P_0^{S_m}) =  \sum_{p \vdash k, q \vdash m} \frac{1}{\prod_{i=1} i^{p_i + q_i} p_i! q_i!} \prod_{i=1}^{m} (\sum_{r | i} r p_r)^{3q_i},
\end{equation}
and from equation \eqref{eqn: rep theory counting}
\begin{equation} \label{eq: counting equality}
	N(k, m) = \Dim(k,m),
\end{equation}
which proves the correspondence between unlabelled bi-partite 3-coloured graphs and observables.

We emphasise that \eqref{eq: counting equality} holds for all $k$. Consequently, bipartite 3-coloured graphs with up to $D$ vertices count tensor observables in the unstable limit $D < 3m$ and the stable limit $D \geq 3m$. This is analogous to the correspondence between directed graphs and matrix observables given in \cite{PIG2MM}.
\section{Summary and Outlook} \label{sec: summary and outlook}
In this paper we developed a permutation invariant statistical model of $D$-dimensional $3$-index tensors $\Phi_{ijk}$. The most general Gaussian model contains $5+117$ parameters. The basic structure of the model was given in section \ref{sec: the model}, where it was solved using representation theoretic variables. These variables give the most efficient descriptions of the one-point and two-point functions of $\Phi_{ijk}$, in terms of invariant tensors $C^{G_{\vec{R}}^{[D]}}_{ijk}$ and $Q^{G_{\vec{R}}^{\Lambda} G_{\vec{S}}^{\Lambda}}_{ijk; pqr}$ which in general depend on $D$. Section \ref{sec: the Qs} was devoted to developing techniques for determining the invariant tensors $Q^{G_{\vec{R}}^{\Lambda} G_{\vec{S}}^{\Lambda}}_{ijk; pqr}$. These techniques used a combination of representation theory of $S_D$ and partition algebras to set up a system of eigenvalue equations whose solutions determine the invariant tensors $Q^{G_{\vec{R}}^{\Lambda} G_{\vec{S}}^{\Lambda}}_{ijk; pqr}$. Importantly, the resulting algorithm is able to find the invariant tensors as exact functions of $D$. This is important for large $D$ studies of the model. Section \ref{sec: counting} contains a study of observables in this model. Observables are permutation invariant polynomials in the tensor. Observables were counted using representation theory, and by counting bi-partite 3-colored graphs we proved a one-to-one correspondence between permutation invariant tensor observables and 3-colored bi-partite graphs.

This work opens up several avenues of future research. A combinatorial algorithm for computing expectation values of permutation invariant 2-matrix observables, including computer code implementing the algorithm was developed in \cite{PIG2MM}. The algorithm is based on Wick's theorem which holds for Gaussian tensor models as well. Developing analogous computer code for the tensor model studied here  will be a useful project. 
Applications to computational linguistics along the lines of \cite{RSS, HCRS} should be possible given such an algorithm. In these approaches the meanings of words are modelled by vector and tensor objects in a vector space, the changes of meaning under composition of words are given by the composition of these objects. The type of structure used to model the meaning of each word is dictated by the word's grammatical role in the sentence. For example, nouns correspond to vectors, adjectives to matrices, and transitive verbs to three index tensors. The statistics of the objects in this final grammatical category could be fruitfully studied with the model presented in this paper, and in fact the tensorial data has already been produced \cite{wijnholds2020} for such a study.
Two point functions of permutation invariant matrix observables were shown to exhibit large $N$ factorisation in \cite{PIMO_Factor}. The proof relied on the close connection between permutation invariant matrix observables and set partitions, which naturally form a partially ordered set. The ordering and corresponding Hasse diagram was used to determine the powers of $1/N$ in the expansion of two-point functions of permutation invariant matrix observables and was therefore a crucial ingredient in the proof. Since tensor model observables also correspond to set partitions, we expect similar factorisation results to hold for permutation invariant tensor observables.
	In \cite{PIMQM}, a representation basis for the space of permutation invariant observables was developed in the context of quantum mechanical models of matrices. The representation basis is an eigenbasis for a set of commuting operators with known eigenvalues. The operators were used to define algebraic Hamiltonians with specified degeneracy patterns, including Hamiltonians with permutation invariant ground states with large degeneracy. It would be interesting to explore quantum mechanical models of tensors with permutation symmetry using similar techniques.

\vskip1cm 

\begin{center} 
{\bf Acknowledgments} 
\end{center} 

SR is supported by the Science and Technology Facilities Council (STFC) Consolidated
Grants ST/P000754/1 “String theory, gauge theory and  duality” and ST/T000686/1
“Amplitudes, strings and  duality”. AP is funded by the Deutsche Forschungs- gemeinschaft (DFG) grant SFB 1283/2 2021 – 317210226. We thank Joseph Ben Geloun for useful discussions. 

\vskip1cm

\begin{appendices}
\section{Clebsch-Gordan coefficients for the trivial representation}
\label{apx: CGC}
In this appendix we derive explicit expressions for the five Clebsch-Gordan coefficients for the trivial representation that appear in the linear part of the action \eqref{eq: action graph basis} and subsequently in the formulae for one-point functions of tensors \eqref{eq: tensor 1pt}. 

To compute expectation values we need formulas for the five Clebsch-Gordan coefficients
\begin{equation}
	C_{i_1 i_2 i_3}^{V_D^{\otimes 3} \rightarrow V_0; M} = \mytikz{	
		\node (t1) at (0,0) [circle,fill,inner sep=0.5mm,label=above:] {};	
		\node (t2) at (1.5,0) [circle,fill,inner sep=0.5mm,label=above:] {};	
		\node (i1) at (-1.5,0.9) {$i_1$};
		\node (i2) at (-1.5,0) {$i_2$};
		\node (i3) at (-1.5,-0.9) {$i_3$};
		\node (m) at (3,0) {};
		\draw [postaction={decorate}] (t2) to node[above]{$V_0$} (m);
		\draw [postaction={decorate}] (t1) to node[above]{$R_4$} (t2);
		\draw [postaction={decorate}] (i1) to node[above]{$R_1$} (t1);
		\draw [postaction={decorate}] (i2) to node[below]{$R_2$} (t1);
		\draw [postaction={decorate}] (i3) to node[below]{$R_3$} (t2);
	},
\end{equation}
where the multiplicity label $M=(R_1, R_2, R_3, R_4, V_0)$.
The non-zero Clebsch-Gordan coefficients are associated with the graphs
\begin{align}
	&\mytikz{	
		\node (t1) at (0,0) [circle,fill,inner sep=0.5mm,label=above:] {};	
		\node (t2) at (1.5,0) [circle,fill,inner sep=0.5mm,label=above:] {};	
		\node (i1) at (-1.5,0.9) {$i_1$};
		\node (i2) at (-1.5,0) {$i_2$};
		\node (i3) at (-1.5,-0.9) {$i_3$};
		\node (m) at (3,0) {};
		\draw [postaction={decorate}] (t2) to node[above]{$V_0$} (m);
		\draw [postaction={decorate}] (t1) to node[above]{$V_0$} (t2);
		\draw [postaction={decorate}] (i1) to node[above]{$V_0$} (t1);
		\draw [postaction={decorate}] (i2) to node[below]{$V_0$} (t1);
		\draw [postaction={decorate}] (i3) to node[below]{$V_0$} (t2);
	}, \\
	&\mytikz{	
		\node (t1) at (0,0) [circle,fill,inner sep=0.5mm,label=above:] {};	
		\node (t2) at (1.5,0) [circle,fill,inner sep=0.5mm,label=above:] {};	
		\node (i1) at (-1.5,0.9) {$i_1$};
		\node (i2) at (-1.5,0) {$i_2$};
		\node (i3) at (-1.5,-0.9) {$i_3$};
		\node (m) at (3,0) {};
		\draw [postaction={decorate}] (t2) to node[above]{$V_0$} (m);
		\draw [postaction={decorate}] (t1) to node[above]{$V_H$} (t2);
		\draw [postaction={decorate}] (i1) to node[above]{$V_0$} (t1);
		\draw [postaction={decorate}] (i2) to node[below]{$V_H$} (t1);
		\draw [postaction={decorate}] (i3) to node[below]{$V_H$} (t2);
	}, \\
	&\mytikz{	
		\node (t1) at (0,0) [circle,fill,inner sep=0.5mm,label=above:] {};	
		\node (t2) at (1.5,0) [circle,fill,inner sep=0.5mm,label=above:] {};	
		\node (i1) at (-1.5,0.9) {$i_1$};
		\node (i2) at (-1.5,0) {$i_2$};
		\node (i3) at (-1.5,-0.9) {$i_3$};
		\node (m) at (3,0) {};
		\draw [postaction={decorate}] (t2) to node[above]{$V_0$} (m);
		\draw [postaction={decorate}] (t1) to node[above]{$V_H$} (t2);
		\draw [postaction={decorate}] (i1) to node[above]{$V_H$} (t1);
		\draw [postaction={decorate}] (i2) to node[below]{$V_0$} (t1);
		\draw [postaction={decorate}] (i3) to node[below]{$V_H$} (t2);
	}, \\
	&\mytikz{	
		\node (t1) at (0,0) [circle,fill,inner sep=0.5mm,label=above:] {};	
		\node (t2) at (1.5,0) [circle,fill,inner sep=0.5mm,label=above:] {};	
		\node (i1) at (-1.5,0.9) {$i_1$};
		\node (i2) at (-1.5,0) {$i_2$};
		\node (i3) at (-1.5,-0.9) {$i_3$};
		\node (m) at (3,0) {};
		\draw [postaction={decorate}] (t2) to node[above]{$V_0$} (m);
		\draw [postaction={decorate}] (t1) to node[above]{$V_H$} (t2);
		\draw [postaction={decorate}] (i1) to node[above]{$V_H$} (t1);
		\draw [postaction={decorate}] (i2) to node[below]{$V_H$} (t1);
		\draw [postaction={decorate}] (i3) to node[below]{$V_0$} (t2);
	}, \\
	&\mytikz{	
		\node (t1) at (0,0) [circle,fill,inner sep=0.5mm,label=above:] {};	
		\node (t2) at (1.5,0) [circle,fill,inner sep=0.5mm,label=above:] {};	
		\node (i1) at (-1.5,0.9) {$i_1$};
		\node (i2) at (-1.5,0) {$i_2$};
		\node (i3) at (-1.5,-0.9) {$i_3$};
		\node (m) at (3,0) {};
		\draw [postaction={decorate}] (t2) to node[above]{$V_0$} (m);
		\draw [postaction={decorate}] (t1) to node[above]{$V_H$} (t2);
		\draw [postaction={decorate}] (i1) to node[above]{$V_H$} (t1);
		\draw [postaction={decorate}] (i2) to node[below]{$V_H$} (t1);
		\draw [postaction={decorate}] (i3) to node[below]{$V_H$} (t2);
	}.
\end{align}
From equation \eqref{eq:CG_CG_diag} these only involve known (see \cite{PIGMM}) Clebsch-Gordan coefficients.

We now describe these in details.
The matrix elements of the maps $V_D \rightarrow V_0$ and $V_D \rightarrow V_H$ are
\begin{equation}
	C_{0,i} = \frac{1}{\sqrt{D}}, \quad C_{m,i} = \frac{1}{\sqrt{m(m+1)}}\qty(-m\delta_{i,m+1} + \sum_{j=1}^m \delta_{ji}),
\end{equation}
respectively.
It will be useful to define
\begin{equation}
	F_{ij} = \sum_{a=1}^{D-1} C_{a,i} C_{a,j} = \delta_{ij} - \frac{1}{D}.
\end{equation}
The Clebsch-Gordan coefficient for $V_H \otimes V_H \rightarrow V_0$ is given by
\begin{equation}
	C_{mm'}^{V_H \otimes V_H \rightarrow V_0} = \frac{\delta_{mm'}}{\sqrt{D-1}},
\end{equation}
and for $V_H \otimes V_H \rightarrow V_H$
\begin{equation}
	C_{mm',m''}^{V_H \otimes V_H \rightarrow V_H} = \sqrt{\frac{D}{D-2}}\sum_{i=1}^D C_{m,i} C_{m',i} C_{m'',i}.
\end{equation}
We will also use
\begin{equation}
	C^{V_0 \otimes V_0 \rightarrow V_0} = 1, \quad C_{m,m'}^{V_0 \otimes V_H \rightarrow V_H} = \delta_{mm'}.
\end{equation}

The simplest Clebsch-Gordan coefficient to evaluate is
\begin{equation}
	\mytikz{	
		\node (t1) at (0,0) [circle,fill,inner sep=0.5mm,label=above:] {};	
		\node (t2) at (1.5,0) [circle,fill,inner sep=0.5mm,label=above:] {};	
		\node (i1) at (-1.5,0.9) {$i_1$};
		\node (i2) at (-1.5,0) {$i_2$};
		\node (i3) at (-1.5,-0.9) {$i_3$};
		\node (m) at (3,0) {};
		\draw [postaction={decorate}] (t2) to node[above]{$V_0$} (m);
		\draw [postaction={decorate}] (t1) to node[above]{$V_0$} (t2);
		\draw [postaction={decorate}] (i1) to node[above]{$V_0$} (t1);
		\draw [postaction={decorate}] (i2) to node[below]{$V_0$} (t1);
		\draw [postaction={decorate}] (i3) to node[below]{$V_0$} (t2);
	} = C_{0,i_1} C_{0,i_2} C_{0,i_3}= \frac{1}{\sqrt{D^3}}.
\end{equation}
followed by
\begin{equation}
	\begin{aligned}
		\mytikz{	
			\node (t1) at (0,0) [circle,fill,inner sep=0.5mm,label=above:] {};	
			\node (t2) at (1.5,0) [circle,fill,inner sep=0.5mm,label=above:] {};	
			\node (i1) at (-1.5,0.9) {$i_1$};
			\node (i2) at (-1.5,0) {$i_2$};
			\node (i3) at (-1.5,-0.9) {$i_3$};
			\node (m) at (3,0) {};
			\draw [postaction={decorate}] (t2) to node[above]{$V_0$} (m);
			\draw [postaction={decorate}] (t1) to node[above]{$V_H$} (t2);
			\draw [postaction={decorate}] (i1) to node[above]{$V_0$} (t1);
			\draw [postaction={decorate}] (i2) to node[below]{$V_H$} (t1);
			\draw [postaction={decorate}] (i3) to node[below]{$V_H$} (t2);
		} &= C_{0,i_1} C_{m,i_2} C_{m',i_3} C_{m,m''}^{V_0 \otimes V_H \rightarrow V_H} C_{m'm''}^{V_H \otimes V_H \rightarrow V_0} \\
		&= \frac{1}{\sqrt{D}} C_{m,i_2} C_{m',i_3} \delta_{mm''} \frac{\delta_{m'm''}}{\sqrt{D-1}} \\
		&= \frac{1}{\sqrt{D(D-1)}} F(i_2,i_3) \\
		&= \frac{\delta_{i_2 i_3}}{\sqrt{D(D-1)}} -  \frac{1}{\sqrt{D^3(D-1)}}
	\end{aligned}
\end{equation}
and its variations given by cyclic permutations of $i_1, i_2, i_3$. The last coefficient is
\begin{equation}
	\begin{aligned}
		\mytikz{	
			\node (t1) at (0,0) [circle,fill,inner sep=0.5mm,label=above:] {};	
			\node (t2) at (1.5,0) [circle,fill,inner sep=0.5mm,label=above:] {};	
			\node (i1) at (-1.5,0.9) {$i_1$};
			\node (i2) at (-1.5,0) {$i_2$};
			\node (i3) at (-1.5,-0.9) {$i_3$};
			\node (m) at (3,0) {};
			\draw [postaction={decorate}] (t2) to node[above]{$V_0$} (m);
			\draw [postaction={decorate}] (t1) to node[above]{$V_H$} (t2);
			\draw [postaction={decorate}] (i1) to node[above]{$V_H$} (t1);
			\draw [postaction={decorate}] (i2) to node[below]{$V_H$} (t1);
			\draw [postaction={decorate}] (i3) to node[below]{$V_H$} (t2);
		} &= \sum_{m,m',m'',n}C_{m,i_1} C_{m',i_2} C_{m'',i_3} C_{mm',n}^{V_H \otimes V_H \rightarrow V_H} C_{nm''}^{V_H \otimes V_H \rightarrow V_0} \\
		&= \sum_{m,m',m'',n=1}^{D-1} C_{m,i_1} C_{m',i_2} C_{m'',i_3} \sqrt{\frac{D}{D-2}} \sum_{l=1}^{D} C_{m,l} C_{m',l} C_{n,l}  \frac{\delta_{nm''}}{\sqrt{D-1}} \\
		&= \sqrt{\frac{D}{(D-1)(D-2)}} \sum_{l=1}^D F_{i_1l}F_{i_2l} F_{i_3l}.
	\end{aligned}
\end{equation}
We compute
\begin{equation}
	\sum_{l=1}^D F_{i_1l}F_{i_2l} F_{i_3l} = \delta_{i_1i_2}\delta_{i_2i_3} - \frac{1}{D}\qty(\delta_{i_1i_2} + \delta_{i_1i_3}+ \delta_{i_2i_3}) + \frac{2}{D^2}.
\end{equation}

\section{Irreducible decomposition of $V_D \otimes V_D \otimes V_D$}\label{apx: V_D3 decomp}
The aim of this appendix is to explain the origin of each of the terms appearing in the irreducible decomposition of $V_D \otimes V_D \otimes V_D$:
\begin{align} \label{eq: VD x VD x VD Decomposition 2} \nonumber
\Span(\, \Phi_{ijk}\, ) &\cong  V_D \otimes V_D \otimes V_D \\
&\cong 5 V_0 \oplus 10 V_H \oplus 6 V_2 \oplus 6 V_3 \oplus V_4 \oplus 2 V_5 \oplus V_6.
\end{align}
Recall the notation defined earlier for Young diagrams in equation \eqref{eq: Young diagram map}
\begin{align} \nonumber
&V_{[D]} \equiv V_0 \, , \quad V_{[D-1,1]} \equiv V_H \, , \quad V_{[D-2,2]} \equiv V_2 \, , \quad V_{[D-2,1,1]} \equiv V_3 \, , \quad  \\ &V_{[D-3,3]} \equiv V_4 \, , \quad  V_{[D-3,2,1]} \equiv V_5 \, , \quad  V_{[D-3,1,1,1]} \equiv V_6 \, . 
\end{align}
As mentioned in the main text this can be found with the rule \cite[Section 7.13]{Hamermesh} for decomposing tensor products of the form $V_{R} \otimes V_{[D-1,1]}$ (also see \cite{Quasi-partition}). This rule give the following multiplicity free decompositions
\begin{align}
V_H \otimes V_H &\cong V_0 \oplus  V_H \oplus  V_2 \oplus  V_3 \, , \\ 
V_2 \otimes V_H &\cong V_H \oplus  V_2 \oplus  V_3 \oplus  V_4 \oplus V_5 \, , \\ 
V_3 \otimes V_H &\cong V_H \oplus  V_2 \oplus  V_3 \oplus  V_5 \oplus V_6 \, ,
\end{align} 
which allows for a graphical description of the resulting multiplicities appearing on the RHS of \eqref{eq: VD x VD x VD Decomposition 2} which we give here. Each of the following graphs corresponds to a single irreducible representation appearing in the decomposition

\noindent
$\Lambda = V_0$:
\begin{align} \nonumber
	&\mytikz{	
		\node (t1) at (0,0) [circle,fill,inner sep=0.5mm,label=above:] {};	
		\node (t2) at (1.5,0) [circle,fill,inner sep=0.5mm,label=above:] {};	
		\node (i1) at (-1.5,0.9) {$i_1$};
		\node (i2) at (-1.5,0) {$i_2$};
		\node (i3) at (-1.5,-0.9) {$i_3$};
		\node (m) at (3,0) {};
		\draw [postaction={decorate}] (t2) to node[above]{$V_0$} (m);
		\draw [postaction={decorate}] (t1) to node[above]{$V_0$} (t2);
		\draw [postaction={decorate}] (i1) to node[above]{$V_0$} (t1);
		\draw [postaction={decorate}] (i2) to node[below]{$V_0$} (t1);
		\draw [postaction={decorate}] (i3) to node[below]{$V_0$} (t2);
	},
	&&\mytikz{	
		\node (t1) at (0,0) [circle,fill,inner sep=0.5mm,label=above:] {};	
		\node (t2) at (1.5,0) [circle,fill,inner sep=0.5mm,label=above:] {};	
		\node (i1) at (-1.5,0.9) {$i_1$};
		\node (i2) at (-1.5,0) {$i_2$};
		\node (i3) at (-1.5,-0.9) {$i_3$};
		\node (m) at (3,0) {};
		\draw [postaction={decorate}] (t2) to node[above]{$V_0$} (m);
		\draw [postaction={decorate}] (t1) to node[above]{$V_H$} (t2);
		\draw [postaction={decorate}] (i1) to node[above]{$V_0$} (t1);
		\draw [postaction={decorate}] (i2) to node[below]{$V_H$} (t1);
		\draw [postaction={decorate}] (i3) to node[below]{$V_H$} (t2);
	}, \\ \nonumber
	&\mytikz{	
		\node (t1) at (0,0) [circle,fill,inner sep=0.5mm,label=above:] {};	
		\node (t2) at (1.5,0) [circle,fill,inner sep=0.5mm,label=above:] {};	
		\node (i1) at (-1.5,0.9) {$i_1$};
		\node (i2) at (-1.5,0) {$i_2$};
		\node (i3) at (-1.5,-0.9) {$i_3$};
		\node (m) at (3,0) {};
		\draw [postaction={decorate}] (t2) to node[above]{$V_0$} (m);
		\draw [postaction={decorate}] (t1) to node[above]{$V_H$} (t2);
		\draw [postaction={decorate}] (i1) to node[above]{$V_H$} (t1);
		\draw [postaction={decorate}] (i2) to node[below]{$V_0$} (t1);
		\draw [postaction={decorate}] (i3) to node[below]{$V_H$} (t2);
	}, 
	&&\mytikz{	
		\node (t1) at (0,0) [circle,fill,inner sep=0.5mm,label=above:] {};	
		\node (t2) at (1.5,0) [circle,fill,inner sep=0.5mm,label=above:] {};	
		\node (i1) at (-1.5,0.9) {$i_1$};
		\node (i2) at (-1.5,0) {$i_2$};
		\node (i3) at (-1.5,-0.9) {$i_3$};
		\node (m) at (3,0) {};
		\draw [postaction={decorate}] (t2) to node[above]{$V_0$} (m);
		\draw [postaction={decorate}] (t1) to node[above]{$V_H$} (t2);
		\draw [postaction={decorate}] (i1) to node[above]{$V_H$} (t1);
		\draw [postaction={decorate}] (i2) to node[below]{$V_H$} (t1);
		\draw [postaction={decorate}] (i3) to node[below]{$V_0$} (t2);
	}, \\
	&\mytikz{	
		\node (t1) at (0,0) [circle,fill,inner sep=0.5mm,label=above:] {};	
		\node (t2) at (1.5,0) [circle,fill,inner sep=0.5mm,label=above:] {};	
		\node (i1) at (-1.5,0.9) {$i_1$};
		\node (i2) at (-1.5,0) {$i_2$};
		\node (i3) at (-1.5,-0.9) {$i_3$};
		\node (m) at (3,0) {};
		\draw [postaction={decorate}] (t2) to node[above]{$V_0$} (m);
		\draw [postaction={decorate}] (t1) to node[above]{$V_H$} (t2);
		\draw [postaction={decorate}] (i1) to node[above]{$V_H$} (t1);
		\draw [postaction={decorate}] (i2) to node[below]{$V_H$} (t1);
		\draw [postaction={decorate}] (i3) to node[below]{$V_H$} (t2);
	}.
\end{align}

\noindent
$\Lambda = V_H$:
\begin{align} \nonumber
	&\mytikz{	
		\node (t1) at (0,0) [circle,fill,inner sep=0.5mm,label=above:] {};	
		\node (t2) at (1.5,0) [circle,fill,inner sep=0.5mm,label=above:] {};	
		\node (i1) at (-1.5,0.9) {$i_1$};
		\node (i2) at (-1.5,0) {$i_2$};
		\node (i3) at (-1.5,-0.9) {$i_3$};
		\node (m) at (3,0) {};
		\draw [postaction={decorate}] (i1) to node[above]{$V_H$} (t1);
		\draw [postaction={decorate}] (i2) to node[below]{$V_0$} (t1);
		\draw [postaction={decorate}] (i3) to node[below]{$V_0$} (t2);
		\draw [postaction={decorate}] (t1) to node[above]{$V_H$} (t2);
		\draw [postaction={decorate}] (t2) to node[above]{$V_H$} (m);
	},
	&&\mytikz{	
		\node (t1) at (0,0) [circle,fill,inner sep=0.5mm,label=above:] {};	
		\node (t2) at (1.5,0) [circle,fill,inner sep=0.5mm,label=above:] {};	
		\node (i1) at (-1.5,0.9) {$i_1$};
		\node (i2) at (-1.5,0) {$i_2$};
		\node (i3) at (-1.5,-0.9) {$i_3$};
		\node (m) at (3,0) {};
		\draw [postaction={decorate}] (i1) to node[above]{$V_0$} (t1);
		\draw [postaction={decorate}] (i2) to node[below]{$V_H$} (t1);
		\draw [postaction={decorate}] (i3) to node[below]{$V_0$} (t2);
		\draw [postaction={decorate}] (t1) to node[above]{$V_H$} (t2);
		\draw [postaction={decorate}] (t2) to node[above]{$V_H$} (m);
	}, \\ \nonumber
	&\mytikz{	
		\node (t1) at (0,0) [circle,fill,inner sep=0.5mm,label=above:] {};	
		\node (t2) at (1.5,0) [circle,fill,inner sep=0.5mm,label=above:] {};	
		\node (i1) at (-1.5,0.9) {$i_1$};
		\node (i2) at (-1.5,0) {$i_2$};
		\node (i3) at (-1.5,-0.9) {$i_3$};
		\node (m) at (3,0) {};
		\draw [postaction={decorate}] (i1) to node[above]{$V_0$} (t1);
		\draw [postaction={decorate}] (i2) to node[below]{$V_0$} (t1);
		\draw [postaction={decorate}] (i3) to node[below]{$V_H$} (t2);
		\draw [postaction={decorate}] (t1) to node[above]{$V_0$} (t2);
		\draw [postaction={decorate}] (t2) to node[above]{$V_H$} (m);
	}, 
	&&\mytikz{	
		\node (t1) at (0,0) [circle,fill,inner sep=0.5mm,label=above:] {};	
		\node (t2) at (1.5,0) [circle,fill,inner sep=0.5mm,label=above:] {};	
		\node (i1) at (-1.5,0.9) {$i_1$};
		\node (i2) at (-1.5,0) {$i_2$};
		\node (i3) at (-1.5,-0.9) {$i_3$};
		\node (m) at (3,0) {};
		\draw [postaction={decorate}] (i1) to node[above]{$V_H$} (t1);
		\draw [postaction={decorate}] (i2) to node[below]{$V_H$} (t1);
		\draw [postaction={decorate}] (i3) to node[below]{$V_0$} (t2);
		\draw [postaction={decorate}] (t1) to node[above]{$V_H$} (t2);
		\draw [postaction={decorate}] (t2) to node[above]{$V_H$} (m);
	}, \\ \nonumber
	&\mytikz{	
		\node (t1) at (0,0) [circle,fill,inner sep=0.5mm,label=above:] {};	
		\node (t2) at (1.5,0) [circle,fill,inner sep=0.5mm,label=above:] {};	
		\node (i1) at (-1.5,0.9) {$i_1$};
		\node (i2) at (-1.5,0) {$i_2$};
		\node (i3) at (-1.5,-0.9) {$i_3$};
		\node (m) at (3,0) {};
		\draw [postaction={decorate}] (i1) to node[above]{$V_H$} (t1);
		\draw [postaction={decorate}] (i2) to node[below]{$V_0$} (t1);
		\draw [postaction={decorate}] (i3) to node[below]{$V_H$} (t2);
		\draw [postaction={decorate}] (t1) to node[above]{$V_H$} (t2);
		\draw [postaction={decorate}] (t2) to node[above]{$V_H$} (m);
	}, 
		&&\mytikz{	
		\node (t1) at (0,0) [circle,fill,inner sep=0.5mm,label=above:] {};	
		\node (t2) at (1.5,0) [circle,fill,inner sep=0.5mm,label=above:] {};	
		\node (i1) at (-1.5,0.9) {$i_1$};
		\node (i2) at (-1.5,0) {$i_2$};
		\node (i3) at (-1.5,-0.9) {$i_3$};
		\node (m) at (3,0) {};
		\draw [postaction={decorate}] (i1) to node[above]{$V_0$} (t1);
		\draw [postaction={decorate}] (i2) to node[below]{$V_H$} (t1);
		\draw [postaction={decorate}] (i3) to node[below]{$V_H$} (t2);
		\draw [postaction={decorate}] (t1) to node[above]{$V_H$} (t2);
		\draw [postaction={decorate}] (t2) to node[above]{$V_H$} (m);
	}, \\ \nonumber
		&\mytikz{	
		\node (t1) at (0,0) [circle,fill,inner sep=0.5mm,label=above:] {};	
		\node (t2) at (1.5,0) [circle,fill,inner sep=0.5mm,label=above:] {};	
		\node (i1) at (-1.5,0.9) {$i_1$};
		\node (i2) at (-1.5,0) {$i_2$};
		\node (i3) at (-1.5,-0.9) {$i_3$};
		\node (m) at (3,0) {};
		\draw [postaction={decorate}] (i1) to node[above]{$V_H$} (t1);
		\draw [postaction={decorate}] (i2) to node[below]{$V_H$} (t1);
		\draw [postaction={decorate}] (i3) to node[below]{$V_H$} (t2);
		\draw [postaction={decorate}] (t1) to node[above]{$V_0$} (t2);
		\draw [postaction={decorate}] (t2) to node[above]{$V_H$} (m);
	}, 
		&&\mytikz{	
		\node (t1) at (0,0) [circle,fill,inner sep=0.5mm,label=above:] {};	
		\node (t2) at (1.5,0) [circle,fill,inner sep=0.5mm,label=above:] {};	
		\node (i1) at (-1.5,0.9) {$i_1$};
		\node (i2) at (-1.5,0) {$i_2$};
		\node (i3) at (-1.5,-0.9) {$i_3$};
		\node (m) at (3,0) {};
		\draw [postaction={decorate}] (i1) to node[above]{$V_H$} (t1);
		\draw [postaction={decorate}] (i2) to node[below]{$V_H$} (t1);
		\draw [postaction={decorate}] (i3) to node[below]{$V_H$} (t2);
		\draw [postaction={decorate}] (t1) to node[above]{$V_H$} (t2);
		\draw [postaction={decorate}] (t2) to node[above]{$V_H$} (m);
	}, \\
		&\mytikz{	
		\node (t1) at (0,0) [circle,fill,inner sep=0.5mm,label=above:] {};	
		\node (t2) at (1.5,0) [circle,fill,inner sep=0.5mm,label=above:] {};	
		\node (i1) at (-1.5,0.9) {$i_1$};
		\node (i2) at (-1.5,0) {$i_2$};
		\node (i3) at (-1.5,-0.9) {$i_3$};
		\node (m) at (3,0) {};
		\draw [postaction={decorate}] (i1) to node[above]{$V_H$} (t1);
		\draw [postaction={decorate}] (i2) to node[below]{$V_H$} (t1);
		\draw [postaction={decorate}] (i3) to node[below]{$V_H$} (t2);
		\draw [postaction={decorate}] (t1) to node[above]{$V_2$} (t2);
		\draw [postaction={decorate}] (t2) to node[above]{$V_H$} (m);
	}, 
		&&\mytikz{	
		\node (t1) at (0,0) [circle,fill,inner sep=0.5mm,label=above:] {};	
		\node (t2) at (1.5,0) [circle,fill,inner sep=0.5mm,label=above:] {};	
		\node (i1) at (-1.5,0.9) {$i_1$};
		\node (i2) at (-1.5,0) {$i_2$};
		\node (i3) at (-1.5,-0.9) {$i_3$};
		\node (m) at (3,0) {};
		\draw [postaction={decorate}] (i1) to node[above]{$V_H$} (t1);
		\draw [postaction={decorate}] (i2) to node[below]{$V_H$} (t1);
		\draw [postaction={decorate}] (i3) to node[below]{$V_H$} (t2);
		\draw [postaction={decorate}] (t1) to node[above]{$V_3$} (t2);
		\draw [postaction={decorate}] (t2) to node[above]{$V_H$} (m);
	}.
\end{align}

\noindent
$\Lambda = V_2$:
\begin{align} \nonumber
	&\mytikz{	
		\node (t1) at (0,0) [circle,fill,inner sep=0.5mm,label=above:] {};	
		\node (t2) at (1.5,0) [circle,fill,inner sep=0.5mm,label=above:] {};	
		\node (i1) at (-1.5,0.9) {$i_1$};
		\node (i2) at (-1.5,0) {$i_2$};
		\node (i3) at (-1.5,-0.9) {$i_3$};
		\node (m) at (3,0) {};
		\draw [postaction={decorate}] (i1) to node[above]{$V_H$} (t1);
		\draw [postaction={decorate}] (i2) to node[below]{$V_H$} (t1);
		\draw [postaction={decorate}] (i3) to node[below]{$V_0$} (t2);
		\draw [postaction={decorate}] (t1) to node[above]{$V_2$} (t2);
		\draw [postaction={decorate}] (t2) to node[above]{$V_2$} (m);
	},
	&&\mytikz{	
		\node (t1) at (0,0) [circle,fill,inner sep=0.5mm,label=above:] {};	
		\node (t2) at (1.5,0) [circle,fill,inner sep=0.5mm,label=above:] {};	
		\node (i1) at (-1.5,0.9) {$i_1$};
		\node (i2) at (-1.5,0) {$i_2$};
		\node (i3) at (-1.5,-0.9) {$i_3$};
		\node (m) at (3,0) {};
		\draw [postaction={decorate}] (i1) to node[above]{$V_H$} (t1);
		\draw [postaction={decorate}] (i2) to node[below]{$V_0$} (t1);
		\draw [postaction={decorate}] (i3) to node[below]{$V_H$} (t2);
		\draw [postaction={decorate}] (t1) to node[above]{$V_H$} (t2);
		\draw [postaction={decorate}] (t2) to node[above]{$V_2$} (m);
	}, \\ \nonumber
	&\mytikz{	
		\node (t1) at (0,0) [circle,fill,inner sep=0.5mm,label=above:] {};	
		\node (t2) at (1.5,0) [circle,fill,inner sep=0.5mm,label=above:] {};	
		\node (i1) at (-1.5,0.9) {$i_1$};
		\node (i2) at (-1.5,0) {$i_2$};
		\node (i3) at (-1.5,-0.9) {$i_3$};
		\node (m) at (3,0) {};
		\draw [postaction={decorate}] (i1) to node[above]{$V_0$} (t1);
		\draw [postaction={decorate}] (i2) to node[below]{$V_H$} (t1);
		\draw [postaction={decorate}] (i3) to node[below]{$V_H$} (t2);
		\draw [postaction={decorate}] (t1) to node[above]{$V_H$} (t2);
		\draw [postaction={decorate}] (t2) to node[above]{$V_2$} (m);
	}, 
	&&\mytikz{	
		\node (t1) at (0,0) [circle,fill,inner sep=0.5mm,label=above:] {};	
		\node (t2) at (1.5,0) [circle,fill,inner sep=0.5mm,label=above:] {};	
		\node (i1) at (-1.5,0.9) {$i_1$};
		\node (i2) at (-1.5,0) {$i_2$};
		\node (i3) at (-1.5,-0.9) {$i_3$};
		\node (m) at (3,0) {};
		\draw [postaction={decorate}] (i1) to node[above]{$V_H$} (t1);
		\draw [postaction={decorate}] (i2) to node[below]{$V_H$} (t1);
		\draw [postaction={decorate}] (i3) to node[below]{$V_H$} (t2);
		\draw [postaction={decorate}] (t1) to node[above]{$V_H$} (t2);
		\draw [postaction={decorate}] (t2) to node[above]{$V_2$} (m);
	}, \\
	&\mytikz{	
		\node (t1) at (0,0) [circle,fill,inner sep=0.5mm,label=above:] {};	
		\node (t2) at (1.5,0) [circle,fill,inner sep=0.5mm,label=above:] {};	
		\node (i1) at (-1.5,0.9) {$i_1$};
		\node (i2) at (-1.5,0) {$i_2$};
		\node (i3) at (-1.5,-0.9) {$i_3$};
		\node (m) at (3,0) {};
		\draw [postaction={decorate}] (i1) to node[above]{$V_H$} (t1);
		\draw [postaction={decorate}] (i2) to node[below]{$V_H$} (t1);
		\draw [postaction={decorate}] (i3) to node[below]{$V_H$} (t2);
		\draw [postaction={decorate}] (t1) to node[above]{$V_2$} (t2);
		\draw [postaction={decorate}] (t2) to node[above]{$V_2$} (m);
	}, 
		&&\mytikz{	
		\node (t1) at (0,0) [circle,fill,inner sep=0.5mm,label=above:] {};	
		\node (t2) at (1.5,0) [circle,fill,inner sep=0.5mm,label=above:] {};	
		\node (i1) at (-1.5,0.9) {$i_1$};
		\node (i2) at (-1.5,0) {$i_2$};
		\node (i3) at (-1.5,-0.9) {$i_3$};
		\node (m) at (3,0) {};
		\draw [postaction={decorate}] (i1) to node[above]{$V_H$} (t1);
		\draw [postaction={decorate}] (i2) to node[below]{$V_H$} (t1);
		\draw [postaction={decorate}] (i3) to node[below]{$V_H$} (t2);
		\draw [postaction={decorate}] (t1) to node[above]{$V_3$} (t2);
		\draw [postaction={decorate}] (t2) to node[above]{$V_2$} (m);
	}.
\end{align}

\noindent
$\Lambda = V_3$:
\begin{align} \nonumber
	&\mytikz{	
		\node (t1) at (0,0) [circle,fill,inner sep=0.5mm,label=above:] {};	
		\node (t2) at (1.5,0) [circle,fill,inner sep=0.5mm,label=above:] {};	
		\node (i1) at (-1.5,0.9) {$i_1$};
		\node (i2) at (-1.5,0) {$i_2$};
		\node (i3) at (-1.5,-0.9) {$i_3$};
		\node (m) at (3,0) {};
		\draw [postaction={decorate}] (i1) to node[above]{$V_H$} (t1);
		\draw [postaction={decorate}] (i2) to node[below]{$V_H$} (t1);
		\draw [postaction={decorate}] (i3) to node[below]{$V_0$} (t2);
		\draw [postaction={decorate}] (t1) to node[above]{$V_3$} (t2);
		\draw [postaction={decorate}] (t2) to node[above]{$V_3$} (m);
	},
	&&\mytikz{	
		\node (t1) at (0,0) [circle,fill,inner sep=0.5mm,label=above:] {};	
		\node (t2) at (1.5,0) [circle,fill,inner sep=0.5mm,label=above:] {};	
		\node (i1) at (-1.5,0.9) {$i_1$};
		\node (i2) at (-1.5,0) {$i_2$};
		\node (i3) at (-1.5,-0.9) {$i_3$};
		\node (m) at (3,0) {};
		\draw [postaction={decorate}] (i1) to node[above]{$V_H$} (t1);
		\draw [postaction={decorate}] (i2) to node[below]{$V_0$} (t1);
		\draw [postaction={decorate}] (i3) to node[below]{$V_H$} (t2);
		\draw [postaction={decorate}] (t1) to node[above]{$V_H$} (t2);
		\draw [postaction={decorate}] (t2) to node[above]{$V_3$} (m);
	}, \\ \nonumber
	&\mytikz{	
		\node (t1) at (0,0) [circle,fill,inner sep=0.5mm,label=above:] {};	
		\node (t2) at (1.5,0) [circle,fill,inner sep=0.5mm,label=above:] {};	
		\node (i1) at (-1.5,0.9) {$i_1$};
		\node (i2) at (-1.5,0) {$i_2$};
		\node (i3) at (-1.5,-0.9) {$i_3$};
		\node (m) at (3,0) {};
		\draw [postaction={decorate}] (i1) to node[above]{$V_0$} (t1);
		\draw [postaction={decorate}] (i2) to node[below]{$V_H$} (t1);
		\draw [postaction={decorate}] (i3) to node[below]{$V_H$} (t2);
		\draw [postaction={decorate}] (t1) to node[above]{$V_H$} (t2);
		\draw [postaction={decorate}] (t2) to node[above]{$V_3$} (m);
	}, 
	&&\mytikz{	
		\node (t1) at (0,0) [circle,fill,inner sep=0.5mm,label=above:] {};	
		\node (t2) at (1.5,0) [circle,fill,inner sep=0.5mm,label=above:] {};	
		\node (i1) at (-1.5,0.9) {$i_1$};
		\node (i2) at (-1.5,0) {$i_2$};
		\node (i3) at (-1.5,-0.9) {$i_3$};
		\node (m) at (3,0) {};
		\draw [postaction={decorate}] (i1) to node[above]{$V_H$} (t1);
		\draw [postaction={decorate}] (i2) to node[below]{$V_H$} (t1);
		\draw [postaction={decorate}] (i3) to node[below]{$V_H$} (t2);
		\draw [postaction={decorate}] (t1) to node[above]{$V_H$} (t2);
		\draw [postaction={decorate}] (t2) to node[above]{$V_3$} (m);
	}, \\
	&\mytikz{	
		\node (t1) at (0,0) [circle,fill,inner sep=0.5mm,label=above:] {};	
		\node (t2) at (1.5,0) [circle,fill,inner sep=0.5mm,label=above:] {};	
		\node (i1) at (-1.5,0.9) {$i_1$};
		\node (i2) at (-1.5,0) {$i_2$};
		\node (i3) at (-1.5,-0.9) {$i_3$};
		\node (m) at (3,0) {};
		\draw [postaction={decorate}] (i1) to node[above]{$V_H$} (t1);
		\draw [postaction={decorate}] (i2) to node[below]{$V_H$} (t1);
		\draw [postaction={decorate}] (i3) to node[below]{$V_H$} (t2);
		\draw [postaction={decorate}] (t1) to node[above]{$V_3$} (t2);
		\draw [postaction={decorate}] (t2) to node[above]{$V_3$} (m);
	}, 
		&&\mytikz{	
		\node (t1) at (0,0) [circle,fill,inner sep=0.5mm,label=above:] {};	
		\node (t2) at (1.5,0) [circle,fill,inner sep=0.5mm,label=above:] {};	
		\node (i1) at (-1.5,0.9) {$i_1$};
		\node (i2) at (-1.5,0) {$i_2$};
		\node (i3) at (-1.5,-0.9) {$i_3$};
		\node (m) at (3,0) {};
		\draw [postaction={decorate}] (i1) to node[above]{$V_H$} (t1);
		\draw [postaction={decorate}] (i2) to node[below]{$V_H$} (t1);
		\draw [postaction={decorate}] (i3) to node[below]{$V_H$} (t2);
		\draw [postaction={decorate}] (t1) to node[above]{$V_3$} (t2);
		\draw [postaction={decorate}] (t2) to node[above]{$V_3$} (m);
	}.
\end{align}

\noindent
$\Lambda = V_4$:
\begin{align} 
	&\mytikz{	
		\node (t1) at (0,0) [circle,fill,inner sep=0.5mm,label=above:] {};	
		\node (t2) at (1.5,0) [circle,fill,inner sep=0.5mm,label=above:] {};	
		\node (i1) at (-1.5,0.9) {$i_1$};
		\node (i2) at (-1.5,0) {$i_2$};
		\node (i3) at (-1.5,-0.9) {$i_3$};
		\node (m) at (3,0) {};
		\draw [postaction={decorate}] (i1) to node[above]{$V_H$} (t1);
		\draw [postaction={decorate}] (i2) to node[below]{$V_H$} (t1);
		\draw [postaction={decorate}] (i3) to node[below]{$V_H$} (t2);
		\draw [postaction={decorate}] (t1) to node[above]{$V_2$} (t2);
		\draw [postaction={decorate}] (t2) to node[above]{$V_4$} (m);
	}.
\end{align}

\noindent
$\Lambda = V_5$:
\begin{align} 
	&\mytikz{	
		\node (t1) at (0,0) [circle,fill,inner sep=0.5mm,label=above:] {};	
		\node (t2) at (1.5,0) [circle,fill,inner sep=0.5mm,label=above:] {};	
		\node (i1) at (-1.5,0.9) {$i_1$};
		\node (i2) at (-1.5,0) {$i_2$};
		\node (i3) at (-1.5,-0.9) {$i_3$};
		\node (m) at (3,0) {};
		\draw [postaction={decorate}] (i1) to node[above]{$V_H$} (t1);
		\draw [postaction={decorate}] (i2) to node[below]{$V_H$} (t1);
		\draw [postaction={decorate}] (i3) to node[below]{$V_H$} (t2);
		\draw [postaction={decorate}] (t1) to node[above]{$V_2$} (t2);
		\draw [postaction={decorate}] (t2) to node[above]{$V_5$} (m);
	}, 
	&&\mytikz{	
		\node (t1) at (0,0) [circle,fill,inner sep=0.5mm,label=above:] {};	
		\node (t2) at (1.5,0) [circle,fill,inner sep=0.5mm,label=above:] {};	
		\node (i1) at (-1.5,0.9) {$i_1$};
		\node (i2) at (-1.5,0) {$i_2$};
		\node (i3) at (-1.5,-0.9) {$i_3$};
		\node (m) at (3,0) {};
		\draw [postaction={decorate}] (i1) to node[above]{$V_H$} (t1);
		\draw [postaction={decorate}] (i2) to node[below]{$V_H$} (t1);
		\draw [postaction={decorate}] (i3) to node[below]{$V_H$} (t2);
		\draw [postaction={decorate}] (t1) to node[above]{$V_3$} (t2);
		\draw [postaction={decorate}] (t2) to node[above]{$V_5$} (m);
	}.
\end{align}

\noindent
$\Lambda = V_6$:
\begin{align}
	&\mytikz{	
		\node (t1) at (0,0) [circle,fill,inner sep=0.5mm,label=above:] {};	
		\node (t2) at (1.5,0) [circle,fill,inner sep=0.5mm,label=above:] {};	
		\node (i1) at (-1.5,0.9) {$i_1$};
		\node (i2) at (-1.5,0) {$i_2$};
		\node (i3) at (-1.5,-0.9) {$i_3$};
		\node (m) at (3,0) {};
		\draw [postaction={decorate}] (i1) to node[above]{$V_H$} (t1);
		\draw [postaction={decorate}] (i2) to node[below]{$V_H$} (t1);
		\draw [postaction={decorate}] (i3) to node[below]{$V_H$} (t2);
		\draw [postaction={decorate}] (t1) to node[above]{$V_3$} (t2);
		\draw [postaction={decorate}] (t2) to node[above]{$V_6$} (m);
	}.
\end{align}

\section{Pivot columns and basis of image}\label{apx: pivots}
In this appendix we review pivot columns of matrices and how they are used to find bases for the image of a matrix (see \cite[\textbf{2O} in Section 2.4]{strang2006linear}). This is used in section \ref{sec: Q construction} to give an all $D$ construction of invariant endomorphism tensors.

Given a matrix $B$, it is said to be in row echelon form if
\begin{itemize}
	\item All rows consisting of only zeros are at the bottom.
	\item Reading the rows from the left, the first non-zero element in every row is to the right of the first non-zero element in every row above it.
\end{itemize}
For a matrix in row echelon form, the first non-zero element of a row is called the pivot.
For example,
\begin{equation}
	B = \mqty(b_{11} & b_{12} & b_{13} & b_{14 }\\
					0	& 0 & b_{23}& b_{24}\\
					0 	& 0	& 0	& 0),
\end{equation}
is in row echelon form and the pivots are $b_{11}, b_{23}$. Consider $B$ as list of column vectors
\begin{equation}
	B = [b_1 \, b_2 \, b_3 \, b_4],
\end{equation}
The vectors
\begin{equation}
	b_1 = \mqty(b_{11} \\ 0 \\ 0), \quad b_3 = \mqty(b_{13} \\ b_{23} \\ 0),
\end{equation}
form a basis for the image of $B$. In general, the columns containing pivots form a basis for the image of $B$. That is, all $y$ such that
\begin{equation}
	y = Bx = \sum_i b_i x_i
\end{equation}
can be expanded in terms of pivot columns. This follows since every non-pivot column is a linear combination of the set of pivot columns before it.

Let $A,B,E$ be $n \times n$ matrices where
\begin{equation}
	A = EB, \label{eq: row echelon form}
\end{equation}
such that $B$ is in row echelon form and $E$ is invertible. Consider $A,B$ as lists of column vectors
\begin{equation}
	A = [a_1 \, \dots \, a_n], B = [b_1 \, \dots \, b_n].
\end{equation}
Equation \eqref{eq: row echelon form} reads
\begin{equation}
	a_i = Eb_i, \quad \forall i = 1,\dots,n. \label{eq: a_i = Eb_i}
\end{equation}
As we will now show, if $\{b_{i_1}, \dots, b_{i_k}\}$ is the set of pivot columns of $B$,
the set of vectors $\{a_{i_1}, \dots, a_{i_k}\}$ form a basis for the image of $A$. First we show that they constitute a set of linearly independent vectors. That is $x_p = 0$ is the only solution to
\begin{equation}
	\sum_{p=1}^k x_{p} a_{i_p} = 0.
\end{equation}
Using \eqref{eq: a_i = Eb_i} we have
\begin{equation}
	\sum_{p=1}^k x_{p} E b_{i_p} = 0,
\end{equation}
and because $E$ is invertible this is equivalent to solving
\begin{equation}
	\sum_{p=1}^k x_{p} b_{i_p} = 0.
\end{equation}
But the pivot columns of $B$ are linearly independent, therefore $x_p = 0$.
Next, we show that all $y$ such that
\begin{equation}
	y = Ax,
\end{equation}
can be expanded in terms of $\{a_{i_1}, \dots, a_{i_k}\}$. Using \eqref{eq: row echelon form} we have
\begin{equation}
	y = EBx,
\end{equation}
or
\begin{equation}
	E^{-1}y = Bx.
\end{equation}
Define $y' = E^{-1}y$, $y'$ is in the image of $B$, which has a basis of pivot colums. That is, there exists a unique set of numbers $y'_{1}, \dots, y'_{p}$ such that
\begin{equation}
	y' = \sum_{p=1}^k y'_{p} b_{i_p}.
\end{equation}
Using the inverse of \eqref{eq: a_i = Eb_i} we have
\begin{equation}
	E^{-1}y = \sum_{p=1}^k y'_{p} E^{-1} a_{i_p},
\end{equation}
or
\begin{equation}
	y = \sum_{p=1}^k y'_{p} a_{i_p}.
\end{equation}
We have proven that the set $\{a_{i_1}, \dots, a_{i_k}\}$ of columns corresponding to pivot columns of $B$ form a linearly independent set of vectors that span the image of $A$. That is, they form a basis for the image of $A$.

\section{Finding invariant endomorphism tensors (code)} \label{app: code}
This appendix explains the code implementing the algorithm in section \ref{sec: Q construction} for constructing invariant endomorphism tensors. The code is implemented in SageMath \cite{sagemath}.

The first step in computing the invariant endomorphism tensors, using partition algebras in SageMath, is to define the partition algebras $P_1(D), P_2(D), P_3(D)$ over the polynomial ring $\mathbb{Q}[D]$. This is done in the first cell of the Jupyter notebook.
    \begin{tcolorbox}[breakable, size=fbox, boxrule=1pt, pad at break*=1mm,colback=cellbackground, colframe=cellborder]
\prompt{In}{incolor}{1}{\boxspacing}
\begin{Verbatim}[commandchars=\\\{\}]
\PY{c+c1}{\PYZsh{}\PYZsh{} Define the partition algebras P\PYZus{}1(D) P\PYZus{}2(D) P\PYZus{}3(D) over the polynomial ring QQ[D]}
\PY{n}{R}\PY{o}{.}\PY{o}{\PYZlt{}}\PY{n}{D}\PY{o}{\PYZgt{}} \PY{o}{=} \PY{n}{QQ}\PY{p}{[}\PY{p}{]}
\PY{n}{PA3} \PY{o}{=} \PY{n}{PartitionAlgebra}\PY{p}{(}\PY{l+m+mi}{3}\PY{p}{,}\PY{n}{D}\PY{p}{,}\PY{n}{R}\PY{p}{)}
\PY{n}{PA2} \PY{o}{=} \PY{n}{PartitionAlgebra}\PY{p}{(}\PY{l+m+mi}{2}\PY{p}{,}\PY{n}{D}\PY{p}{,}\PY{n}{R}\PY{p}{)}
\PY{n}{PA1} \PY{o}{=} \PY{n}{PartitionAlgebra}\PY{p}{(}\PY{l+m+mi}{1}\PY{p}{,}\PY{n}{D}\PY{p}{,}\PY{n}{R}\PY{p}{)}
\end{Verbatim}
\end{tcolorbox}

The second step is to compute the projectors in \eqref{eq: projectors}. Each projector is defined through a function that takes an irreducible representation $R$ of $S_D$ as input and returns the corresponding projector. The irreducible representations are to be given as integer partitions of $D$. In particular, for an integer partition $R = (D-k, \lambda)$ of $D$, where $\lambda$ is an integer partition of $k$, only $\lambda$ should be given as input. For example, for $R = [D]$, the corresponding input partition is $Partition([])$ -- the empty partition. For $R = [D-1,1]$ one should input $Partition([1])$ and so on. Note that in the current implementation, the projectors do not include the denominator in \eqref{eq: projectors}. This allows us to do the computations using partition algebras over polynomial rings, which is faster than computations using partition algebras over symbolic rings. Excluding the denominators does not change the image of the projectors but affects the overall normalisation of the output vector in the algorithm. If desired, the normalisation can be adjusted as a final step in the algorithm.
    \begin{tcolorbox}[breakable, size=fbox, boxrule=1pt, pad at break*=1mm,colback=cellbackground, colframe=cellborder]
\prompt{In}{incolor}{2}{\boxspacing}
\begin{Verbatim}[commandchars=\\\{\}]
\PY{c+c1}{\PYZsh{}\PYZsh{} Define the \PYZdq{}projector\PYZdq{} P\PYZus{}\PYZob{}R\PYZus{}1\PYZcb{} on the first representation in G\PYZus{}A\PYZca{}\PYZbs{}Lambda, see equation (3.52) in associated paper}
\PY{k}{def} \PY{n+nf}{P\PYZus{}R1}\PY{p}{(}\PY{n}{R1}\PY{p}{)}\PY{p}{:}
    \PY{n}{R}\PY{o}{.}\PY{o}{\PYZlt{}}\PY{n}{D}\PY{o}{\PYZgt{}} \PY{o}{=} \PY{n}{QQ}\PY{p}{[}\PY{p}{]}
    \PY{c+c1}{\PYZsh{}\PYZsh{} Construct T2 \PYZbs{}otimes \PYZbs{}idn \PYZbs{}otimes \PYZbs{}idn defined in equation (3.42)}
    \PY{n}{T2\PYZus{}id\PYZus{}id} \PY{o}{=} \PY{n}{PA3}\PY{p}{(}\PY{n+nb}{sum}\PY{p}{(}\PY{n}{PA1}\PY{o}{.}\PY{n}{jucys\PYZus{}murphy\PYZus{}element}\PY{p}{(}\PY{n}{i}\PY{o}{/}\PY{l+m+mi}{2}\PY{p}{)} \PY{k}{for} \PY{n}{i} \PY{o+ow}{in} \PY{p}{[}\PY{l+m+mf}{1.}\PY{l+m+mf}{.2}\PY{o}{*}\PY{l+m+mi}{1}\PY{p}{]}\PY{p}{)}\PY{o}{+}\PY{p}{(}\PY{n}{D}\PY{o}{*}\PY{p}{(}\PY{n}{D}\PY{o}{\PYZhy{}}\PY{l+m+mi}{1}\PY{p}{)}\PY{o}{/}\PY{l+m+mi}{2}\PY{o}{\PYZhy{}}\PY{n}{D}\PY{o}{*}\PY{l+m+mi}{1}\PY{p}{)}\PY{o}{*}\PY{n}{PA1}\PY{o}{.}\PY{n}{one}\PY{p}{(}\PY{p}{)}\PY{p}{)}
   	\PY{c+c1}{\PYZsh{}\PYZsh{} Define the set of irreps and corresponding normalized characters to take a product over in equation (3.52)}
    \PY{n}{IrrepsEigenvaluesDictionary} \PY{o}{=} \PY{p}{\PYZob{}}\PY{n}{Partition}\PY{p}{(}\PY{p}{[}\PY{p}{]}\PY{p}{)}\PY{p}{:} \PY{n}{R}\PY{p}{(}\PY{l+m+mi}{1}\PY{o}{/}\PY{l+m+mi}{2}\PY{o}{*}\PY{p}{(}\PY{n}{D}\PY{o}{\PYZhy{}}\PY{l+m+mi}{1}\PY{p}{)}\PY{o}{*}\PY{n}{D}\PY{p}{)}\PY{p}{,} \PY{n}{Partition}\PY{p}{(}\PY{p}{[}\PY{l+m+mi}{1}\PY{p}{]}\PY{p}{)}\PY{p}{:} \PY{n}{R}\PY{p}{(}\PY{p}{(}\PY{n}{D}\PY{o}{\PYZhy{}}\PY{l+m+mi}{3}\PY{p}{)}\PY{o}{*}\PY{n}{D}\PY{o}{/}\PY{l+m+mi}{2}\PY{p}{)}\PY{p}{\PYZcb{}}
    \PY{n}{proj} \PY{o}{=} \PY{n}{prod}\PY{p}{(}\PY{p}{(}\PY{n}{T2\PYZus{}id\PYZus{}id}\PY{o}{\PYZhy{}}\PY{n}{ev2}\PY{o}{*}\PY{n}{PA3}\PY{o}{.}\PY{n}{one}\PY{p}{(}\PY{p}{)}\PY{p}{)} \PY{k}{for} \PY{p}{(}\PY{n}{rep}\PY{p}{,} \PY{n}{ev2}\PY{p}{)} \PY{o+ow}{in} \PY{n}{IrrepsEigenvaluesDictionary}\PY{o}{.}\PY{n}{items}\PY{p}{(}\PY{p}{)} \PY{k}{if} \PY{n}{rep} \PY{o}{!=} \PY{n}{R1}\PY{p}{)}
    \PY{k}{return} \PY{n}{proj}
\end{Verbatim}
\end{tcolorbox}
 \begin{tcolorbox}[breakable, size=fbox, boxrule=1pt, pad at break*=1mm,colback=cellbackground, colframe=cellborder]
\prompt{In}{incolor}{3}{\boxspacing}
\begin{Verbatim}[commandchars=\\\{\}]
\PY{c+c1}{\PYZsh{}\PYZsh{} Define the \PYZdq{}projector\PYZdq{} P\PYZus{}\PYZob{}R\PYZus{}2\PYZcb{} on the second representation in G\PYZus{}A\PYZca{}\PYZbs{}Lambda, see equation (3.53) in associated paper}
\PY{k}{def} \PY{n+nf}{P\PYZus{}R2}\PY{p}{(}\PY{n}{R2}\PY{p}{)}\PY{p}{:}
    \PY{n}{R}\PY{o}{.}\PY{o}{\PYZlt{}}\PY{n}{D}\PY{o}{\PYZgt{}} \PY{o}{=} \PY{n}{QQ}\PY{p}{[}\PY{p}{]}
    \PY{c+c1}{\PYZsh{}\PYZsh{} Construct \PYZbs{}idn \PYZbs{}otimes T2 \PYZbs{}otimes \PYZbs{}idn defined in equation (3.43)}
    \PY{n}{id\PYZus{}T2\PYZus{}id} \PY{o}{=} \PY{n}{PA3}\PY{p}{(}\PY{p}{[}\PY{p}{[}\PY{o}{\PYZhy{}}\PY{l+m+mi}{1}\PY{p}{,}\PY{l+m+mi}{2}\PY{p}{]}\PY{p}{,}\PY{p}{[}\PY{o}{\PYZhy{}}\PY{l+m+mi}{2}\PY{p}{,}\PY{l+m+mi}{1}\PY{p}{]}\PY{p}{,}\PY{p}{[}\PY{o}{\PYZhy{}}\PY{l+m+mi}{3}\PY{p}{,}\PY{l+m+mi}{3}\PY{p}{]}\PY{p}{]}\PY{p}{)}\PY{o}{*}\PY{n}{PA3}\PY{p}{(}\PY{n+nb}{sum}\PY{p}{(}\PY{n}{PA1}\PY{o}{.}\PY{n}{jucys\PYZus{}murphy\PYZus{}element}\PY{p}{(}\PY{n}{i}\PY{o}{/}\PY{l+m+mi}{2}\PY{p}{)} \PY{k}{for} \PY{n}{i} \PY{o+ow}{in} \PY{p}{[}\PY{l+m+mf}{1.}\PY{l+m+mf}{.2}\PY{o}{*}\PY{l+m+mi}{1}\PY{p}{]}\PY{p}{)}\PY{o}{+}\PY{p}{(}\PY{n}{D}\PY{o}{*}\PY{p}{(}\PY{n}{D}\PY{o}{\PYZhy{}}\PY{l+m+mi}{1}\PY{p}{)}\PY{o}{/}\PY{l+m+mi}{2}\PY{o}{\PYZhy{}}\PY{n}{D}\PY{o}{*}\PY{l+m+mi}{1}\PY{p}{)}\PY{o}{*}\PY{n}{PA1}\PY{o}{.}\PY{n}{one}\PY{p}{(}\PY{p}{)}\PY{p}{)}\PY{o}{*}\PY{n}{PA3}\PY{p}{(}\PY{p}{[}\PY{p}{[}\PY{o}{\PYZhy{}}\PY{l+m+mi}{1}\PY{p}{,}\PY{l+m+mi}{2}\PY{p}{]}\PY{p}{,}\PY{p}{[}\PY{o}{\PYZhy{}}\PY{l+m+mi}{2}\PY{p}{,}\PY{l+m+mi}{1}\PY{p}{]}\PY{p}{,}\PY{p}{[}\PY{o}{\PYZhy{}}\PY{l+m+mi}{3}\PY{p}{,}\PY{l+m+mi}{3}\PY{p}{]}\PY{p}{]}\PY{p}{)}
    \PY{c+c1}{\PYZsh{}\PYZsh{} Define the set of irreps and corresponding normalized characters to take a product over in equation (3.53)}
    \PY{n}{IrrepsEigenvaluesDictionary} \PY{o}{=} \PY{p}{\PYZob{}}\PY{n}{Partition}\PY{p}{(}\PY{p}{[}\PY{p}{]}\PY{p}{)}\PY{p}{:} \PY{n}{R}\PY{p}{(}\PY{l+m+mi}{1}\PY{o}{/}\PY{l+m+mi}{2}\PY{o}{*}\PY{p}{(}\PY{n}{D}\PY{o}{\PYZhy{}}\PY{l+m+mi}{1}\PY{p}{)}\PY{o}{*}\PY{n}{D}\PY{p}{)}\PY{p}{,} \PY{n}{Partition}\PY{p}{(}\PY{p}{[}\PY{l+m+mi}{1}\PY{p}{]}\PY{p}{)}\PY{p}{:} \PY{n}{R}\PY{p}{(}\PY{p}{(}\PY{n}{D}\PY{o}{\PYZhy{}}\PY{l+m+mi}{3}\PY{p}{)}\PY{o}{*}\PY{n}{D}\PY{o}{/}\PY{l+m+mi}{2}\PY{p}{)}\PY{p}{\PYZcb{}}
    \PY{n}{proj} \PY{o}{=} \PY{n}{prod}\PY{p}{(}\PY{p}{(}\PY{n}{id\PYZus{}T2\PYZus{}id}\PY{o}{\PYZhy{}}\PY{n}{ev2}\PY{o}{*}\PY{n}{PA3}\PY{o}{.}\PY{n}{one}\PY{p}{(}\PY{p}{)}\PY{p}{)} \PY{k}{for} \PY{p}{(}\PY{n}{rep}\PY{p}{,} \PY{n}{ev2}\PY{p}{)} \PY{o+ow}{in} \PY{n}{IrrepsEigenvaluesDictionary}\PY{o}{.}\PY{n}{items}\PY{p}{(}\PY{p}{)} \PY{k}{if} \PY{n}{rep} \PY{o}{!=} \PY{n}{R2}\PY{p}{)}
    \PY{k}{return} \PY{n}{proj}
 \end{Verbatim}
\end{tcolorbox}
 \begin{tcolorbox}[breakable, size=fbox, boxrule=1pt, pad at break*=1mm,colback=cellbackground, colframe=cellborder]
\prompt{In}{incolor}{4}{\boxspacing}
\begin{Verbatim}[commandchars=\\\{\}]
\PY{c+c1}{\PYZsh{}\PYZsh{} Define the \PYZdq{}projector\PYZdq{} P\PYZus{}\PYZob{}R\PYZus{}3\PYZcb{} on the third representation in G\PYZus{}A\PYZca{}\PYZbs{}Lambda, see equation (3.54) in associated paper}
\PY{k}{def} \PY{n+nf}{P\PYZus{}R3}\PY{p}{(}\PY{n}{R3}\PY{p}{)}\PY{p}{:}
    \PY{n}{R}\PY{o}{.}\PY{o}{\PYZlt{}}\PY{n}{D}\PY{o}{\PYZgt{}} \PY{o}{=} \PY{n}{QQ}\PY{p}{[}\PY{p}{]}
    \PY{c+c1}{\PYZsh{}\PYZsh{} Construct \PYZbs{}idn \PYZbs{}otimes \PYZbs{}idn \PYZbs{}otimes T2 defined in equation (3.44)}
    \PY{n}{id\PYZus{}T2\PYZus{}id} \PY{o}{=} \PY{n}{PA3}\PY{p}{(}\PY{p}{[}\PY{p}{[}\PY{o}{\PYZhy{}}\PY{l+m+mi}{1}\PY{p}{,}\PY{l+m+mi}{3}\PY{p}{]}\PY{p}{,}\PY{p}{[}\PY{o}{\PYZhy{}}\PY{l+m+mi}{3}\PY{p}{,}\PY{l+m+mi}{1}\PY{p}{]}\PY{p}{,}\PY{p}{[}\PY{o}{\PYZhy{}}\PY{l+m+mi}{2}\PY{p}{,}\PY{l+m+mi}{2}\PY{p}{]}\PY{p}{]}\PY{p}{)}\PY{o}{*}\PY{n}{PA3}\PY{p}{(}\PY{n+nb}{sum}\PY{p}{(}\PY{n}{PA1}\PY{o}{.}\PY{n}{jucys\PYZus{}murphy\PYZus{}element}\PY{p}{(}\PY{n}{i}\PY{o}{/}\PY{l+m+mi}{2}\PY{p}{)} \PY{k}{for} \PY{n}{i} \PY{o+ow}{in} \PY{p}{[}\PY{l+m+mf}{1.}\PY{l+m+mf}{.2}\PY{o}{*}\PY{l+m+mi}{1}\PY{p}{]}\PY{p}{)}\PY{o}{+}\PY{p}{(}\PY{n}{D}\PY{o}{*}\PY{p}{(}\PY{n}{D}\PY{o}{\PYZhy{}}\PY{l+m+mi}{1}\PY{p}{)}\PY{o}{/}\PY{l+m+mi}{2}\PY{o}{\PYZhy{}}\PY{n}{D}\PY{o}{*}\PY{l+m+mi}{1}\PY{p}{)}\PY{o}{*}\PY{n}{PA1}\PY{o}{.}\PY{n}{one}\PY{p}{(}\PY{p}{)}\PY{p}{)}\PY{o}{*}\PY{n}{PA3}\PY{p}{(}\PY{p}{[}\PY{p}{[}\PY{o}{\PYZhy{}}\PY{l+m+mi}{1}\PY{p}{,}\PY{l+m+mi}{3}\PY{p}{]}\PY{p}{,}\PY{p}{[}\PY{o}{\PYZhy{}}\PY{l+m+mi}{3}\PY{p}{,}\PY{l+m+mi}{1}\PY{p}{]}\PY{p}{,}\PY{p}{[}\PY{o}{\PYZhy{}}\PY{l+m+mi}{2}\PY{p}{,}\PY{l+m+mi}{2}\PY{p}{]}\PY{p}{]}\PY{p}{)}
    \PY{c+c1}{\PYZsh{}\PYZsh{} Define the set of irreps and corresponding normalized characters to take a product over in equation (3.54)}
    \PY{n}{IrrepsEigenvaluesDictionary} \PY{o}{=} \PY{p}{\PYZob{}}\PY{n}{Partition}\PY{p}{(}\PY{p}{[}\PY{p}{]}\PY{p}{)}\PY{p}{:} \PY{n}{R}\PY{p}{(}\PY{l+m+mi}{1}\PY{o}{/}\PY{l+m+mi}{2}\PY{o}{*}\PY{p}{(}\PY{n}{D}\PY{o}{\PYZhy{}}\PY{l+m+mi}{1}\PY{p}{)}\PY{o}{*}\PY{n}{D}\PY{p}{)}\PY{p}{,} \PY{n}{Partition}\PY{p}{(}\PY{p}{[}\PY{l+m+mi}{1}\PY{p}{]}\PY{p}{)}\PY{p}{:} \PY{n}{R}\PY{p}{(}\PY{p}{(}\PY{n}{D}\PY{o}{\PYZhy{}}\PY{l+m+mi}{3}\PY{p}{)}\PY{o}{*}\PY{n}{D}\PY{o}{/}\PY{l+m+mi}{2}\PY{p}{)}\PY{p}{\PYZcb{}}
    \PY{n}{proj} \PY{o}{=} \PY{n}{prod}\PY{p}{(}\PY{p}{(}\PY{n}{id\PYZus{}T2\PYZus{}id}\PY{o}{\PYZhy{}}\PY{n}{ev2}\PY{o}{*}\PY{n}{PA3}\PY{o}{.}\PY{n}{one}\PY{p}{(}\PY{p}{)}\PY{p}{)} \PY{k}{for} \PY{p}{(}\PY{n}{rep}\PY{p}{,} \PY{n}{ev2}\PY{p}{)} \PY{o+ow}{in} \PY{n}{IrrepsEigenvaluesDictionary}\PY{o}{.}\PY{n}{items}\PY{p}{(}\PY{p}{)} \PY{k}{if} \PY{n}{rep} \PY{o}{!=} \PY{n}{R3}\PY{p}{)}
    \PY{k}{return} \PY{n}{proj}
\end{Verbatim}
\end{tcolorbox}
 \begin{tcolorbox}[breakable, size=fbox, boxrule=1pt, pad at break*=1mm,colback=cellbackground, colframe=cellborder]
\prompt{In}{incolor}{5}{\boxspacing}
\begin{Verbatim}[commandchars=\\\{\}]
\PY{c+c1}{\PYZsh{}\PYZsh{} Define the \PYZdq{}projector\PYZdq{} P\PYZus{}\PYZob{}R\PYZus{}4\PYZcb{} on the fourth representation in G\PYZus{}A\PYZca{}\PYZbs{}Lambda, see equation (3.55) in associated paper}
\PY{k}{def} \PY{n+nf}{P\PYZus{}R4}\PY{p}{(}\PY{n}{R4}\PY{p}{)}\PY{p}{:}
    \PY{n}{R}\PY{o}{.}\PY{o}{\PYZlt{}}\PY{n}{D}\PY{o}{\PYZgt{}} \PY{o}{=} \PY{n}{QQ}\PY{p}{[}\PY{p}{]}
    \PY{c+c1}{\PYZsh{}\PYZsh{} Construct T2 \PYZbs{}otimes \PYZbs{}idn defined in equation (3.45)}
    \PY{n}{T2\PYZus{}id} \PY{o}{=} \PY{n}{PA3}\PY{p}{(}\PY{n+nb}{sum}\PY{p}{(}\PY{n}{PA2}\PY{o}{.}\PY{n}{jucys\PYZus{}murphy\PYZus{}element}\PY{p}{(}\PY{n}{i}\PY{o}{/}\PY{l+m+mi}{2}\PY{p}{)} \PY{k}{for} \PY{n}{i} \PY{o+ow}{in} \PY{p}{[}\PY{l+m+mf}{1.}\PY{l+m+mf}{.2}\PY{o}{*}\PY{l+m+mi}{2}\PY{p}{]}\PY{p}{)}\PY{o}{+}\PY{p}{(}\PY{n}{D}\PY{o}{*}\PY{p}{(}\PY{n}{D}\PY{o}{\PYZhy{}}\PY{l+m+mi}{1}\PY{p}{)}\PY{o}{/}\PY{l+m+mi}{2}\PY{o}{\PYZhy{}}\PY{n}{D}\PY{o}{*}\PY{l+m+mi}{2}\PY{p}{)}\PY{o}{*}\PY{n}{PA2}\PY{o}{.}\PY{n}{one}\PY{p}{(}\PY{p}{)}\PY{p}{)}
    \PY{c+c1}{\PYZsh{}\PYZsh{} Define the set of irreps and corresponding normalized characters to take a product over in equation (3.55)}
    \PY{n}{IrrepsEigenvaluesDictionary} \PY{o}{=} \PY{p}{\PYZob{}}\PY{n}{Partition}\PY{p}{(}\PY{p}{[}\PY{p}{]}\PY{p}{)}\PY{p}{:} \PY{n}{R}\PY{p}{(}\PY{l+m+mi}{1}\PY{o}{/}\PY{l+m+mi}{2}\PY{o}{*}\PY{p}{(}\PY{n}{D}\PY{o}{\PYZhy{}}\PY{l+m+mi}{1}\PY{p}{)}\PY{o}{*}\PY{n}{D}\PY{p}{)}\PY{p}{,} \PY{n}{Partition}\PY{p}{(}\PY{p}{[}\PY{l+m+mi}{1}\PY{p}{]}\PY{p}{)}\PY{p}{:} \PY{n}{R}\PY{p}{(}\PY{p}{(}\PY{n}{D}\PY{o}{\PYZhy{}}\PY{l+m+mi}{3}\PY{p}{)}\PY{o}{*}\PY{n}{D}\PY{o}{/}\PY{l+m+mi}{2}\PY{p}{)}\PY{p}{,} \PY{n}{Partition}\PY{p}{(}\PY{p}{[}\PY{l+m+mi}{2}\PY{p}{]}\PY{p}{)}\PY{p}{:} \PY{n}{R}\PY{p}{(}\PY{l+m+mi}{1}\PY{o}{/}\PY{l+m+mi}{2}\PY{o}{*}\PY{p}{(}\PY{n}{D} \PY{o}{\PYZhy{}} \PY{l+m+mi}{1}\PY{p}{)}\PY{o}{*}\PY{p}{(}\PY{n}{D} \PY{o}{\PYZhy{}} \PY{l+m+mi}{4}\PY{p}{)}\PY{p}{)}\PY{p}{,} \PY{n}{Partition}\PY{p}{(}\PY{p}{[}\PY{l+m+mi}{1}\PY{p}{,}\PY{l+m+mi}{1}\PY{p}{]}\PY{p}{)}\PY{p}{:} \PY{n}{R}\PY{p}{(}\PY{l+m+mi}{1}\PY{o}{/}\PY{l+m+mi}{2}\PY{o}{*}\PY{p}{(}\PY{n}{D} \PY{o}{\PYZhy{}} \PY{l+m+mi}{5}\PY{p}{)}\PY{o}{*}\PY{n}{D}\PY{p}{)}\PY{p}{\PYZcb{}}
    \PY{n}{proj} \PY{o}{=} \PY{n}{prod}\PY{p}{(}\PY{p}{(}\PY{n}{T2\PYZus{}id}\PY{o}{\PYZhy{}}\PY{n}{ev2}\PY{o}{*}\PY{n}{PA3}\PY{o}{.}\PY{n}{one}\PY{p}{(}\PY{p}{)}\PY{p}{)} \PY{k}{for} \PY{p}{(}\PY{n}{rep}\PY{p}{,} \PY{n}{ev2}\PY{p}{)} \PY{o+ow}{in} \PY{n}{IrrepsEigenvaluesDictionary}\PY{o}{.}\PY{n}{items}\PY{p}{(}\PY{p}{)} \PY{k}{if} \PY{n}{rep} \PY{o}{!=} \PY{n}{R4}\PY{p}{)}
    \PY{k}{return} \PY{n}{proj}
 \end{Verbatim}
\end{tcolorbox}
 \begin{tcolorbox}[breakable, size=fbox, boxrule=1pt, pad at break*=1mm,colback=cellbackground, colframe=cellborder]
\prompt{In}{incolor}{6}{\boxspacing}
\begin{Verbatim}[commandchars=\\\{\}]
\PY{c+c1}{\PYZsh{}\PYZsh{} Define the \PYZdq{}projector\PYZdq{} P\PYZus{}\PYZob{}\PYZbs{}Lambda\PYZcb{} on the last representation in G\PYZus{}A\PYZca{}\PYZbs{}Lambda, see equation (3.55) in associated paper}
\PY{k}{def} \PY{n+nf}{P\PYZus{}Lambda}\PY{p}{(}\PY{n}{Lambda}\PY{p}{)}\PY{p}{:}
    \PY{n}{R}\PY{o}{.}\PY{o}{\PYZlt{}}\PY{n}{D}\PY{o}{\PYZgt{}} \PY{o}{=} \PY{n}{QQ}\PY{p}{[}\PY{p}{]}
    \PY{c+c1}{\PYZsh{}\PYZsh{} Construct T2 defined in equation (3.41)}
    \PY{n}{T2} \PY{o}{=} \PY{n+nb}{sum}\PY{p}{(}\PY{n}{PA3}\PY{o}{.}\PY{n}{jucys\PYZus{}murphy\PYZus{}element}\PY{p}{(}\PY{n}{i}\PY{o}{/}\PY{l+m+mi}{2}\PY{p}{)} \PY{k}{for} \PY{n}{i} \PY{o+ow}{in} \PY{p}{[}\PY{l+m+mf}{1.}\PY{l+m+mf}{.2}\PY{o}{*}\PY{l+m+mi}{3}\PY{p}{]}\PY{p}{)}\PY{o}{+}\PY{p}{(}\PY{n}{D}\PY{o}{*}\PY{p}{(}\PY{n}{D}\PY{o}{\PYZhy{}}\PY{l+m+mi}{1}\PY{p}{)}\PY{o}{/}\PY{l+m+mi}{2}\PY{o}{\PYZhy{}}\PY{n}{D}\PY{o}{*}\PY{l+m+mi}{3}\PY{p}{)}\PY{o}{*}\PY{n}{PA3}\PY{o}{.}\PY{n}{one}\PY{p}{(}\PY{p}{)}
    \PY{c+c1}{\PYZsh{}\PYZsh{} Define the set of irreps and corresponding normalized characters to take a product over in equation (3.55)}
    \PY{n}{IrrepsEigenvaluesDictionary} \PY{o}{=} \PY{p}{\PYZob{}}\PY{n}{Partition}\PY{p}{(}\PY{p}{[}\PY{p}{]}\PY{p}{)}\PY{p}{:} \PY{n}{R}\PY{p}{(}\PY{l+m+mi}{1}\PY{o}{/}\PY{l+m+mi}{2}\PY{o}{*}\PY{p}{(}\PY{n}{D}\PY{o}{\PYZhy{}}\PY{l+m+mi}{1}\PY{p}{)}\PY{o}{*}\PY{n}{D}\PY{p}{)}\PY{p}{,} \PY{n}{Partition}\PY{p}{(}\PY{p}{[}\PY{l+m+mi}{1}\PY{p}{]}\PY{p}{)}\PY{p}{:} \PY{n}{R}\PY{p}{(}\PY{p}{(}\PY{n}{D}\PY{o}{\PYZhy{}}\PY{l+m+mi}{3}\PY{p}{)}\PY{o}{*}\PY{n}{D}\PY{o}{/}\PY{l+m+mi}{2}\PY{p}{)}\PY{p}{,} \PY{n}{Partition}\PY{p}{(}\PY{p}{[}\PY{l+m+mi}{2}\PY{p}{]}\PY{p}{)}\PY{p}{:} \PY{n}{R}\PY{p}{(}\PY{l+m+mi}{1}\PY{o}{/}\PY{l+m+mi}{2}\PY{o}{*}\PY{p}{(}\PY{n}{D} \PY{o}{\PYZhy{}} \PY{l+m+mi}{1}\PY{p}{)}\PY{o}{*}\PY{p}{(}\PY{n}{D} \PY{o}{\PYZhy{}} \PY{l+m+mi}{4}\PY{p}{)}\PY{p}{)}\PY{p}{,} \PY{n}{Partition}\PY{p}{(}\PY{p}{[}\PY{l+m+mi}{1}\PY{p}{,}\PY{l+m+mi}{1}\PY{p}{]}\PY{p}{)}\PY{p}{:} \PY{n}{R}\PY{p}{(}\PY{l+m+mi}{1}\PY{o}{/}\PY{l+m+mi}{2}\PY{o}{*}\PY{p}{(}\PY{n}{D} \PY{o}{\PYZhy{}} \PY{l+m+mi}{5}\PY{p}{)}\PY{o}{*}\PY{n}{D}\PY{p}{)}\PY{p}{,} \PY{n}{Partition}\PY{p}{(}\PY{p}{[}\PY{l+m+mi}{3}\PY{p}{]}\PY{p}{)}\PY{p}{:} \PY{n}{R}\PY{p}{(}\PY{l+m+mi}{1}\PY{o}{/}\PY{l+m+mi}{2}\PY{o}{*}\PY{p}{(}\PY{n}{D}\PY{o}{\PYZhy{}}\PY{l+m+mi}{3}\PY{p}{)}\PY{o}{*}\PY{p}{(}\PY{n}{D}\PY{o}{\PYZhy{}}\PY{l+m+mi}{4}\PY{p}{)}\PY{p}{)}\PY{p}{,} \PY{n}{Partition}\PY{p}{(}\PY{p}{[}\PY{l+m+mi}{2}\PY{p}{,}\PY{l+m+mi}{1}\PY{p}{]}\PY{p}{)}\PY{p}{:} \PY{n}{R}\PY{p}{(}\PY{l+m+mi}{1}\PY{o}{/}\PY{l+m+mi}{2}\PY{o}{*}\PY{p}{(}\PY{n}{D}\PY{o}{\PYZhy{}}\PY{l+m+mi}{1}\PY{p}{)}\PY{o}{*}\PY{p}{(}\PY{n}{D}\PY{o}{\PYZhy{}}\PY{l+m+mi}{6}\PY{p}{)}\PY{p}{)}\PY{p}{,} \PY{n}{Partition}\PY{p}{(}\PY{p}{[}\PY{l+m+mi}{1}\PY{p}{,}\PY{l+m+mi}{1}\PY{p}{,}\PY{l+m+mi}{1}\PY{p}{]}\PY{p}{)}\PY{p}{:} \PY{n}{R}\PY{p}{(}\PY{l+m+mi}{1}\PY{o}{/}\PY{l+m+mi}{2}\PY{o}{*}\PY{n}{D}\PY{o}{*}\PY{p}{(}\PY{n}{D}\PY{o}{\PYZhy{}}\PY{l+m+mi}{7}\PY{p}{)}\PY{p}{)}\PY{p}{\PYZcb{}}
    \PY{n}{proj} \PY{o}{=} \PY{n}{prod}\PY{p}{(}\PY{p}{(}\PY{n}{T2}\PY{o}{\PYZhy{}}\PY{n}{ev2}\PY{o}{*}\PY{n}{PA3}\PY{o}{.}\PY{n}{one}\PY{p}{(}\PY{p}{)}\PY{p}{)} \PY{k}{for} \PY{p}{(}\PY{n}{rep}\PY{p}{,} \PY{n}{ev2}\PY{p}{)} \PY{o+ow}{in} \PY{n}{IrrepsEigenvaluesDictionary}\PY{o}{.}\PY{n}{items}\PY{p}{(}\PY{p}{)} \PY{k}{if} \PY{n}{rep} \PY{o}{!=} \PY{n}{Lambda}\PY{p}{)}
    \PY{k}{return} \PY{n}{proj}
\end{Verbatim}
\end{tcolorbox}
With all the projectors defined, we can compute the elements \eqref{eq: graph projector} for a pair of multiplicity graphs by taking products of several projectors. The graphs are specified by two lists of partitions, LeftReps and RightReps, respectively. The corresponding left and right projectors are computed as LeftProj and RightProj.
    \begin{tcolorbox}[breakable, size=fbox, boxrule=1pt, pad at break*=1mm,colback=cellbackground, colframe=cellborder]
\prompt{In}{incolor}{7}{\boxspacing}
\begin{Verbatim}[commandchars=\\\{\}]
\PY{c+c1}{\PYZsh{}\PYZsh{} Define the set of irreducible representations R\PYZus{}1, R\PYZus{}2, R\PYZus{}3, R\PYZus{}4, \PYZbs{}Lambda appearing in G\PYZus{}A\PYZca{}\PYZbs{}Lambda}
\PY{n}{LeftReps} \PY{o}{=} \PY{p}{(}\PY{n}{Partition}\PY{p}{(}\PY{p}{[}\PY{l+m+mi}{1}\PY{p}{]}\PY{p}{)}\PY{p}{,} \PY{n}{Partition}\PY{p}{(}\PY{p}{[}\PY{l+m+mi}{1}\PY{p}{]}\PY{p}{)}\PY{p}{,} \PY{n}{Partition}\PY{p}{(}\PY{p}{[}\PY{l+m+mi}{1}\PY{p}{]}\PY{p}{)}\PY{p}{,} \PY{n}{Partition}\PY{p}{(}\PY{p}{[}\PY{l+m+mi}{1}\PY{p}{,}\PY{l+m+mi}{1}\PY{p}{]}\PY{p}{)}\PY{p}{,} \PY{n}{Partition}\PY{p}{(}\PY{p}{[}\PY{l+m+mi}{1}\PY{p}{]}\PY{p}{)}\PY{p}{)}
\PY{c+c1}{\PYZsh{}\PYZsh{} and the total \PYZdq{}projector\PYZdq{} P\PYZus{}\PYZob{}\PYZbs{}Lambda\PYZcb{} P\PYZus{}\PYZob{}R\PYZus{}4\PYZcb{} P\PYZus{}\PYZob{}R\PYZus{}3\PYZcb{} P\PYZus{}\PYZob{}R\PYZus{}2\PYZcb{} P\PYZus{}\PYZob{}R\PYZus{}1\PYZcb{}}
\PY{n}{LeftProj} \PY{o}{=} \PY{n}{P\PYZus{}Lambda}\PY{p}{(}\PY{n}{LeftReps}\PY{p}{[}\PY{l+m+mi}{4}\PY{p}{]}\PY{p}{)}\PY{o}{*}\PY{n}{P\PYZus{}R4}\PY{p}{(}\PY{n}{LeftReps}\PY{p}{[}\PY{l+m+mi}{3}\PY{p}{]}\PY{p}{)}\PY{o}{*}\PY{n}{P\PYZus{}R3}\PY{p}{(}\PY{n}{LeftReps}\PY{p}{[}\PY{l+m+mi}{2}\PY{p}{]}\PY{p}{)}\PY{o}{*}
	\PY{n}{P\PYZus{}R2}\PY{p}{(}\PY{n}{LeftReps}\PY{p}{[}\PY{l+m+mi}{1}\PY{p}{]}\PY{p}{)}\PY{o}{*}\PY{n}{P\PYZus{}R1}\PY{p}{(}\PY{n}{LeftReps}\PY{p}{[}\PY{l+m+mi}{0}\PY{p}{]}\PY{p}{)}

\PY{c+c1}{\PYZsh{}\PYZsh{} Define the set of irreducible representations R\PYZus{}1, R\PYZus{}2, R\PYZus{}3, R\PYZus{}4, \PYZbs{}Lambda appearing in G\PYZus{}B\PYZca{}\PYZbs{}Lambda}
\PY{n}{RightReps} \PY{o}{=} \PY{p}{(}\PY{n}{Partition}\PY{p}{(}\PY{p}{[}\PY{l+m+mi}{1}\PY{p}{]}\PY{p}{)}\PY{p}{,} \PY{n}{Partition}\PY{p}{(}\PY{p}{[}\PY{l+m+mi}{1}\PY{p}{]}\PY{p}{)}\PY{p}{,} \PY{n}{Partition}\PY{p}{(}\PY{p}{[}\PY{l+m+mi}{1}\PY{p}{]}\PY{p}{)}\PY{p}{,} \PY{n}{Partition}\PY{p}{(}\PY{p}{[}\PY{l+m+mi}{1}\PY{p}{,}\PY{l+m+mi}{1}\PY{p}{]}\PY{p}{)}\PY{p}{,} \PY{n}{Partition}\PY{p}{(}\PY{p}{[}\PY{l+m+mi}{1}\PY{p}{]}\PY{p}{)}\PY{p}{)}
\PY{c+c1}{\PYZsh{}\PYZsh{} and the total \PYZdq{}projector\PYZdq{} P\PYZus{}\PYZob{}\PYZbs{}Lambda\PYZcb{} P\PYZus{}\PYZob{}R\PYZus{}4\PYZcb{} P\PYZus{}\PYZob{}R\PYZus{}3\PYZcb{} P\PYZus{}\PYZob{}R\PYZus{}2\PYZcb{} P\PYZus{}\PYZob{}R\PYZus{}1\PYZcb{}}
\PY{n}{RightProj} \PY{o}{=} \PY{n}{P\PYZus{}Lambda}\PY{p}{(}\PY{n}{RightReps}\PY{p}{[}\PY{l+m+mi}{4}\PY{p}{]}\PY{p}{)}\PY{o}{*}\PY{n}{P\PYZus{}R4}\PY{p}{(}\PY{n}{RightReps}\PY{p}{[}\PY{l+m+mi}{3}\PY{p}{]}\PY{p}{)}\PY{o}{*}\PY{n}{P\PYZus{}R3}\PY{p}{(}\PY{n}{RightReps}\PY{p}{[}\PY{l+m+mi}{2}\PY{p}{]}\PY{p}{)}\PY{o}{*}
	\PY{n}{P\PYZus{}R2}\PY{p}{(}\PY{n}{RightReps}\PY{p}{[}\PY{l+m+mi}{1}\PY{p}{]}\PY{p}{)}\PY{o}{*}\PY{n}{P\PYZus{}R1}\PY{p}{(}\PY{n}{RightReps}\PY{p}{[}\PY{l+m+mi}{0}\PY{p}{]}\PY{p}{)}
\end{Verbatim}
\end{tcolorbox}
To find the image of the left and right multiplication of LeftProj and RightProj, we compute their corresponding matrix representations. SageMath has a built-in method called to\_matrix() for this purpose. To get the left and right action, respectively, we give this method the input "side=left" and "side=right". The matrix of the combined action is stored in ProjMatrix. The last step is to find the pivot column of ProjMatrix for a fixed value of $D$ and extract the corresponding column of ProjMatrix. This element, called Q in the code, is proportional to the invariant tensor we were looking for.
    \begin{tcolorbox}[breakable, size=fbox, boxrule=1pt, pad at break*=1mm,colback=cellbackground, colframe=cellborder]
\prompt{In}{incolor}{8}{\boxspacing}
\begin{Verbatim}[commandchars=\\\{\}]
\PY{c+c1}{\PYZsh{}\PYZsh{} Now produce the matrix corresponding to left action of LeftProj and right action of RightProj}
\PY{n}{ProjMatrix} \PY{o}{=} \PY{n}{LeftProj}\PY{o}{.}\PY{n}{to\PYZus{}matrix}\PY{p}{(}\PY{n}{side}\PY{o}{=}\PY{l+s+s1}{\PYZsq{}}\PY{l+s+s1}{left}\PY{l+s+s1}{\PYZsq{}}\PY{p}{)}\PY{o}{*}\PY{n}{RightProj}\PY{o}{.}\PY{n}{to\PYZus{}matrix}\PY{p}{(}\PY{n}{side}\PY{o}{=}\PY{l+s+s1}{\PYZsq{}}\PY{l+s+s1}{right}\PY{l+s+s1}{\PYZsq{}}\PY{p}{)}
\PY{c+c1}{\PYZsh{}\PYZsh{} Compute the pivot columns of ProjMatrix for D=10}
\PY{n}{pivot} \PY{o}{=} \PY{n}{ProjMatrix}\PY{o}{.}\PY{n}{subs}\PY{p}{(}\PY{n}{D}\PY{o}{=}\PY{l+m+mi}{10}\PY{p}{)}\PY{o}{.}\PY{n}{pivots}\PY{p}{(}\PY{p}{)}
\PY{c+c1}{\PYZsh{}\PYZsh{} As long as ProjMatrix has rank 1, extract the pivot column,}
\PY{c+c1}{\PYZsh{}\PYZsh{} otherwise print a warning that G\PYZus{}A\PYZca{}\PYZbs{}Lambda G\PYZus{}B\PYZca{}\PYZbs{}Lambda dont define a non\PYZhy{}zero Q}
\PY{k}{if} \PY{n+nb}{len}\PY{p}{(}\PY{n}{pivot}\PY{p}{)} \PY{o}{\PYZgt{}} \PY{l+m+mi}{0}\PY{p}{:}
    \PY{n}{Q} \PY{o}{=} \PY{n}{ProjMatrix}\PY{o}{.}\PY{n}{column}\PY{p}{(}\PY{n}{pivot}\PY{p}{[}\PY{l+m+mi}{0}\PY{p}{]}\PY{p}{)}
\PY{k}{else}\PY{p}{:}
    \PY{n+nb}{print}\PY{p}{(}\PY{l+s+s1}{\PYZsq{}}\PY{l+s+s1}{Q does not exist for the chosen G\PYZus{}A\PYZca{}}\PY{l+s+se}{\PYZbs{}\PYZbs{}}\PY{l+s+s1}{Lambda and G\PYZus{}B\PYZca{}}\PY{l+s+se}{\PYZbs{}\PYZbs{}}\PY{l+s+s1}{Lambda}\PY{l+s+s1}{\PYZsq{}}\PY{p}{)}
\end{Verbatim}
\end{tcolorbox}
\end{appendices}

\cleardoublepage
\phantomsection
\addcontentsline{toc}{section}{\refname}
\bibliography{bibliography}

\end{document}